\documentclass[traditabstract]{aa} 

\usepackage{graphicx}
\usepackage{placeins}
\usepackage{txfonts}
\usepackage{longtable}
\usepackage{lscape}
\usepackage{hyperref}
\usepackage{soul}
\hypersetup{colorlinks=true, citecolor=blue}

\newcommand{\hi}{{\sc H\,i}}
\newcommand{\hii}{{H\,\sc{ii}}}
\newcommand{\nii}{{N\,\sc{ii}}}
\newcommand{\oiii}{{O\,\sc{iii}}}
\newcommand{\oii}{{O\,\sc{ii}}}
\newcommand{\sii}{{s\,\sc{ii}}}

\begin{document}

   \title{A Virgo Environmental Survey Tracing Ionised Gas Emission (VESTIGE)\\ XVII. Statistical properties of individual {\hii} regions in unperturbed systems\thanks{Based on observations obtained with
   MegaPrime/MegaCam, a joint project of CFHT and CEA/DAPNIA, at the Canada-French-Hawaii Telescope
   (CFHT) which is operated by the National Research Council (NRC) of Canada, the Institut National
   des Sciences de l'Univers of the Centre National de la Recherche Scientifique (CNRS) of France and
   the University of Hawaii.}
      }
   \subtitle{}
  \author{A. Boselli\inst{1,**}, 
 	  M. Fossati\inst{2,3},
	  Y. Roehlly\inst{1},
	  P. Amram\inst{1},
	  S. Boissier\inst{1},
	  M. Boquien\inst{4}, 
	  J. Braine\inst{5},
          P. C{\^o}t{\'e}\inst{6},
    	  J.C. Cuillandre\inst{7},
    	  L. Ferrarese\inst{6},
	  G. Gavazzi\inst{2},
    	  S. Gwyn\inst{6},
	  G. Hensler\inst{8},
	  G. Trinchieri\inst{9},
	  A. Zavagno\inst{1,10}
       }

\institute{     
                Aix Marseille Univ, CNRS, CNES, LAM, Marseille, France\thanks{Scientific associate INAF - Osservatorio Astronomico di Cagliari, Via della Scienza 5, 09047 Selargius (CA), Italy}
                \email{alessandro.boselli@lam.fr}
        \and
        	Universit\'a di Milano-Bicocca, piazza della scienza 3, 20100 Milano, Italy
	\and
		INAF - Osservatorio Astronomico di Brera, via Brera 28, 20121 Milano, Italy
	\and
		Universit\'e C\^ote d'Azur, Observatoire de la C\^ote d'Azur, CNRS, Laboratoire Lagrange, 06000, Nice, France
	\and
        	Laboratoire d'Astrophysique de Bordeaux, Univ. Bordeaux, CNRS, B18N, all\'ee Geoffroy Saint-Hilaire, 33615 Pessac, France	
	\and
        	National Research Council of Canada, Herzberg Astronomy and Astrophysics, 5071 West Saanich Road, Victoria, BC, V9E 2E7, Canada
	\and
       	 	AIM, CEA, CNRS, Universit\'e Paris-Saclay, Universit\'e Paris Diderot, Sorbonne Paris Cit\'e, Observatoire de Paris, PSL University, F-91191 Gif-sur-Yvette Cedex, France
	\and
		Department of Astrophysics, University of Vienna, T\"urkenschanzstrasse 17, 1180 Vienna, Austria
    	\and
		INAF - Osservatorio Astronomico di Brera, via Brera 28, 20121 Milano, Italy
	\and 
		Institut Universitaire de France, 1 rue Descartes, 75005 Paris, France
                 }

\authorrunning{Boselli et al.}
\titlerunning{VESTIGE: statistical properties of {\hii} regions }

   \date{}

 
  \abstract  
{The Virgo Environmental Survey Tracing Ionised Gas Emission (VESTIGE) is a blind narrow-band H$\alpha$+[{\nii}] imaging survey of the Virgo cluster carried 
out with MegaCam at the Canada-French-Hawaii telescope. The survey provides deep narrow-band images for 385 galaxies hosting star forming {\hii} regions. We 
identify individual {\hii} regions and measure their main physical properties such as H$\alpha$ luminosity, equivalent diameter, and electron density
with the purpose of deriving standard relations as reference for future local and high-$z$ studies of {\hii} regions in star forming systems in different environments.
For this purpose we use a complete sample of $\simeq$ 13.000 {\hii} regions of luminosity $L(H\alpha)$ $\geq$ 10$^{37}$ erg s$^{-1}$ to derive
the main statistical properties of {\hii} regions in unperturbed systems, identified as those galaxies with a normal {\hi} gas content (64 objects). These are 
the composite H$\alpha$ luminosity function, equivalent diameter and 
electron density distribution, and luminosity-size relation. We also derive the main scaling relations between several
parameters representative of the {\hii} regions properties (total number, luminosity of the first ranked regions, fraction of the diffuse component, 
best fit parameters of the Schechter luminosity function measured for individual galaxies) and those characterising the properties of the host galaxies
(stellar mass, star formation rate and specific star formation rate, stellar mass and star formation rate surface density, metallicity, molecular-to-atomic gas ratio,
total gas-to-dust mass ratio). We briefly discuss the results of this analysis and their implications in the study of the star formation process in galaxy discs.
}
   {}
   {}
   {}
   {}
   {}


   \maketitle
%

\section{Introduction}

The process responsible for the transformation of the cold gas distributed within the interstellar medium (ISM) of galaxies into new stars is 
one of the most studied phenomena in astronomy. Its importance resides in the fact that this process governs the transformation of the primordial baryonic 
component, the atomic hydrogen, into galaxies, and is thus at the origin of galaxy evolution. If at kpc scales the transformation of the gas along the disc of galaxies
mainly follows the Schmidt relation, where the star formation surface density is proportionally related to the cold gas column density  
(e.g., Schmidt 1959, Kennicutt 1998, Bigiel et al. 2008, Kennicutt \& Evans 2012), at smaller scales the formation of 
stars takes place within giant molecular clouds (GMC). Stars are here formed within compact regions generally called {\hii} regions,
where the hard stellar radiation produced by the youngest (O, early-B) and most massive ($\gtrsim$ 10 M$_{\odot}$, e.g., Kennicutt 1998) stars
is sufficiently energetic to ionise the surrounding gas. These star forming regions are easily identified thanks to their emission in several hydrogen lines,
the strongest of which visible in the optical domain is H$\alpha$ at 6563\AA\ (Balmer line). 

Although the star formation process at the scale of GMC has been the topic of hundreds of past and recent studies, many open questions remain unanswered.
The main difficulty in understanding this phenomenon resides in the fact that several competitive mechanisms (gravity, turbulence, different kinds of instabilities, 
gas cooling and heating in a multiphase
and mixed medium, confinement in magnetic fields, external pressure, shocks, etc.) act simultaneously. Still unclear are the initial conditions 
necessary for the gas to collapse into molecular clouds, in particular within those located along filaments, how the molecular clouds collapse and fragment into clumps
where stars are formed with the typical distribution characterised by the initial stellar mass function (IMF), 
and with which efficiency this occurs (e.g. ,Shu et al. 1987, McKee \& Ostriker 2007, Dobbs et al. 2014, Inutsuka et al. 2015, Chevance et al. 2020a, 2020b). 
It is also clear that the star formation process depends on several factors acting 
at different scales, from gas infall from the intergalactic medium into the disc (Mpc scales) down to the fragmentation of the molecular gas within GMC
into clumps and cores (1-0.1 pc scales, e.g., Kennicutt \& Evans 2012). At the scale of galactic discs, the process of star formation can depend on several factors, 
as the aforementioned gas column density (Schmidt law), possibly modulated by the rotational velocity of the disc (Wyse \& Silk 1989, Boissier et al. 2001), the stellar mass density 
of the disc (Shi et al. 2018), the dynamical time (Silk 1997, Elmegreen 1997), 
or the free-fall time on the cloud (Krumholz et al. 2012), the presence of
spiral density waves, the metallicity of the gas and its molecular-to-atomic fraction, the dust content and distribution (e.g., Dey et al. 2019). All these parameters are known to 
scale with the intrinsic size/mass of the host galaxies (scaling relations, e.g., Gavazzi et al. 1996, Boselli et al. 2001, 2014a, Tremonti et al. 2004, Cortese et al. 2012, 
Saintonge \& Catinella 2022), it is thus interesting to investigate how parameters representative of the star formation process at scales of individual {\hii} 
regions scale as a function of other indicative of global galaxy properties such as its morphological type, stellar mass, star formation rate, metallicity, gas and dust content, etc. 
The same properties might also change in hostile environments such as clusters and groups, where different kind of perturbations are known to modify the star 
formation process at large scales (star formation quenching, Boselli \& Gavazzi 2006, Peng et al. 2010).

A Virgo Environmental Survey Tracing Ionised Gas Emission (VESTIGE) is a blind H$\alpha$ narrow-band (NB) imaging survey of the Virgo cluster carried on at the 
Canada-Frence-Hawaii Telescope (CFHT) at Maunakea (Boselli et al. 2018a). Designed to cover the full cluster up to its virial radius at unprecedented depth 
and angular resolution, the survey has detected almost four hundred star forming galaxies, some of which still unperturbed by the hostile cluster environment
(Boselli et al. 2023a). The spectacular image quality of the data and their sensitivity allow us to identify thousands of individual {\hii} regions in most of these objects,
all located at about the same distance, and thus study in detail their statistical properties on a complete and homogeneous sample. The properties of the {\hii} regions 
identified in a few, representative objects have been already presented in dedicated works (the perturbed galaxy IC 3476, Boselli et al. 2021; lenticular 
galaxies, Boselli et al. 2022a). Those of a statistically significant sample of mainly perturbed systems have been discussed in Boselli et al. (2020). The main purpose of this work is 
to derive the main statistical properties (composite distributions, scaling relations) of a large sample of unperturbed systems mainly located at the 
periphery of the cluster. Most of these relations have been already derived, studied, and interpreted in dedicated works based on very local samples
including only a handful of galaxies (Kennicutt et al. 1989a, 1989b), other composed by more distant objects but with a limited quality of the data (Caldwell et al. 1991, 
Banfi et al. 1993, Rozas et al. 1996a, Elmegreen \& Salzer 1999, Beckman et al. 2000, Thilker et al. 2002, Helmboldt et al. 2005, Bradley et al. 2006). The most recent uses
spectacular NB imaging data in the near-IR domain gathered with the Hubble Space telescope (HST, Liu et al. 2013), where the exquisite angular resolution allows the authors to resolve {\hii} regions down to scales
of $\sim$ 10-20 pc. Others are based on the Very Large Telescope (VLT) with the Multi Unit Spectroscopic Explorer (MUSE) data obtained during the 
Physics at High Angular Resolution in Nearby Galaxies (PHANGS) survey and have the advantage to be based on integral field unit (IFU) spectroscopic data, of primordial 
importance to infer the physical properties of the identified star forming regions (Santoro et al. 2022). Both works, however, are limited in statistics, 
the former including only 12 objects, the latter 19, and the sampling of the galaxy parameter space which is mainly limited to massive systems ($M_{star}$ $\gtrsim$ 10$^{9.4-10}$ M$_{\odot}$).
The sample analysed in this work includes 64 galaxies, spans a wide range in morphological type (dE, S0, spirals, Magellanic irregulars, blue
compact dwarfs (BCD)) and stellar mass (10$^7$ $\lesssim$ $M_{star}$ $\lesssim$ 10$^{11}$ M$_{\odot}$), and has 
$\simeq$ 13.000 {\hii} regions of luminosity $L(H\alpha)$ $\geq$ 10$^{37}$ erg s$^{-1}$. The spectacular imaging quality of the data gathered at the CFHT 
($FWHM$ $\simeq$ 0.7\arcsec) allows us to resolve {\hii} regions down to $\simeq$ 60 pc. It is thus ideally designed to trace the main statistical properties 
of {\hii} regions located in normal, unperturbed star forming galaxies in the local Universe, and thus provide a benchmark reference for future local and high redshift studies. 
As designed, the VESTIGE survey also covers hundreds of perturbed systems located within the inner regions of the cluster,
ideal targets to study the effects of the different kinds of perturbations on the star formation process down to the scale of individual {\hii} regions.
The sample analysed in this work and based on the same set of data will also be an ideal reference for a comparative study 
between perturbed and unperturbed objects which will be presented in a 
forthcoming publication. 

The paper is structured as follows: in Sec. 2 we describe the sample, in Sec. 3 the full set of multifrequency data used for the analysis, which is presented 
in Sec. 4. This includes a detail description of the composite H$\alpha$ luminosity function, diameter, and electron density distribution of the detected {\hii} regions,
as well as the main scaling relations. The results are compared to those already available in the literature and discussed in Sec. 5, while the conclusions are given in Sec.6.

\section{Sample}

The sample of galaxies analysed in this work is composed by the star forming objects detected during the VESTIGE survey.
Out of the 385 galaxies detected in H$\alpha$, however, we exclude a few objects where the ionised gas emission is
diffuse and not associated to star forming episodes, for example in M87 (Boselli et al. 2019) or in a few bright 
lenticular galaxies (Boselli et al. 2022a). We also exclude nearly edge-on systems ($b/a$ $<$ 0.25,
$\sim$ corresponding to an inclination $i$ $\gtrsim$ 75 degrees, 
where $a$ and $b$ are the major and minor axes measured on the $i$-band Next Generation Virgo Survey (NGVS) images 
of Ferrarese et al. (2012) and available for all the targets), since
the morphology prevents an accurate identification of the {\hii} regions located along the disc. Finally, 
we also exclude the few {\hii} regions located outside the optical disc (see Sec. 3.4), as observed in several systems 
(e.g., IC 3418, Hester et al. 2010, Fumagalli et al. 2011;
NGC 4254, Boselli et al. 2018b; AGC226178, Junais et al. 2021, Jones et al. 2022; IC 3476, Boselli et al. 2021). These regions have been probably produced during 
the interaction of galaxies with their surrounding environment and will be studied in a dedicated work. Their exact number is still
unknown given the difficulty in identifying and disentangling them from other sources (planetary nebulae, background line emitters)
with consistent and homogeneous criteria on such a large region of the sky. Their number, however, is limited when compared to the number of {\hii}
within the disc of massive spirals (in NGC 4254 is $\simeq$ 0.1\%\ at the limiting luminosity of $L(H\alpha)$ $\geq$ 10$^{37}$ erg s$^{-1}$). 
It might be more important in dwarf, perturbed systems. In IC 3418 all {\hii} regions are located outside the stellar disc 
(Hester et al. 2010, Fumagalli et al. 2011), and are thus not analysed in this work.  

We then consider only those objects hosting {\hii} regions
brighter than the completeness limit of $L(H\alpha)$ $\geq$ 10$^{37}$ erg s$^{-1}$ (see Sec. 3.5).
Throughout the paper, we will refer to this sample as to the \textit{parent sample}.
The parent sample is thus composed of 322 star forming galaxies and is, at present, the largest sample ever studied in the literature
spanning a wide range in morphological type (from giant spirals and lenticulars to dwarf ellipticals, Magellanic
irregulars, and blue compact dwarfs (BCDs)) and stellar mass (10$^7$ $\lesssim$ $M_{star}$ $\lesssim$ 10$^{11}$ M$_{\odot}$) 
with an homogeneous set of narrow-band imaging data suitable for this analysis. 
The exceptional quality of this sample is also related to the fact that it is composed of galaxies all at a similar distance,
thus minimising distance related completeness biases. Indeed, the distance of each galaxy is here taken as the mean distance of the 
Virgo cluster subgroup to which the galaxy belongs. The membership to the different cluster substructures 
has been described in Boselli et al. (2023b). It depends on the relative distance of galaxies to the core of the different
clouds and on their recessional velocity. With respect to previous works of this series, however, we assume the updated mean distance 
to the different substructure recently published by Cantiello et al. (2024), i.e. 16.5 Mpc for galaxies belonging to the main body of the 
cluster associated to M87 (cluster A), and to the low velocity cloud (LVC), 15.8 Mpc for galaxies in cluster B (M49) and cluster C (M60)\footnote{The unperturbed 
galaxy sample does not include any galaxy belonging to cluster C.}, 23 Mpc
in the W' cloud, and 32 Mpc for objects in the W and M clouds (see Boselli et al. 2023b for details).

\begin{figure}
\centering
\includegraphics[width=0.49\textwidth]{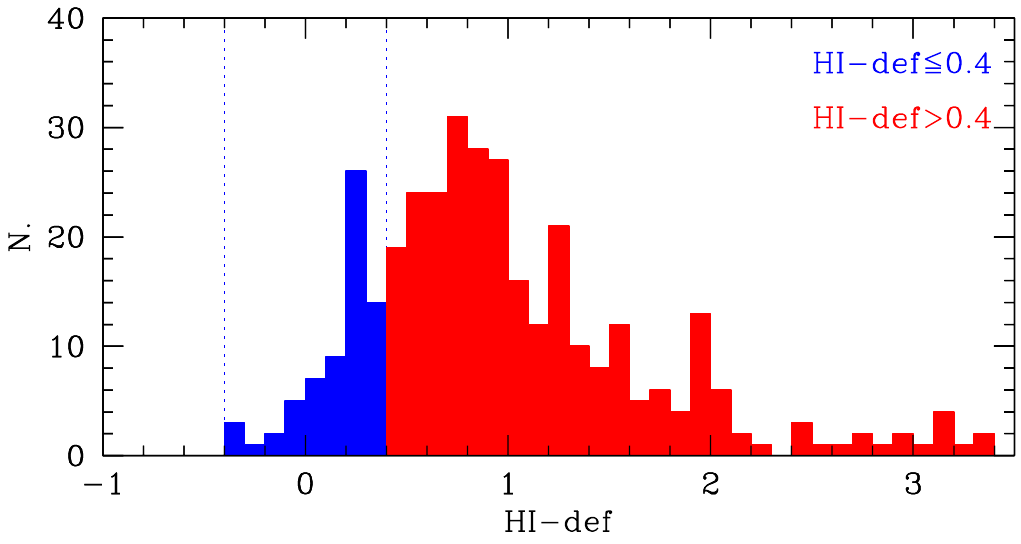}
\caption{Distribution of the {\hi}-deficiency parameter for the parent sample galaxies. The blue histogram
shows the distribution of the unperturbed systems with {\hi}$-def$ $\leq$ 0.4 and $b/a$ $\geq$ 0.25 analysed in this work (64 objects, unperturbed sample), 
the red histogram the one of perturbed galaxies with  {\hi}$-def$ $>$ 0.4 and $b/a$ $\geq$ 0.25 (258 objects, perturbed sample). The blue dotted vertical lines 
show the 1$\sigma$ range distribution of the {\hi}-deficiency parameter observed in isolated systems.}
\label{HIdefdist}%
\end{figure}

To separate normal galaxies mainly following a secular evolution from those perturbed by their surrounding environment 
we use the {\hi}-deficiency parameter (e.g., Boselli \& Gavazzi 2006, Cortese et al. 2021, Boselli et al. 2022b). 
This parameter is defined as the difference 
(in logarithmic scale) between the expected and the observed atomic gas mass of the targeted galaxies (Haynes \& Giovanelli 1984),
where the expected {\hi} mass of each object is here derived using the scaling relations of gas-rich field galaxies recently 
calibrated by Cattorini et al. (2023). This calibration is perfectly indicated for this sample because it includes dwarf systems
such as those detected during the VESTIGE survey. {\hi} masses or upper limits are available for all the 
targets (see Boselli et al. 2023a for details). In the following analysis, we consider as unperturbed systems those with
an {\hi}-deficiency parameter {\hi}$-def$ $\leq$ 0.4, in line with what was done in all previous VESTIGE publications (e.g., Boselli et al. 2020, 2023a, 2023b).
With all these criteria the parent sample is composed of 322 objects, out of which 64 unperturbed systems that will be used in this work\footnote{Releasing the condition 
that the luminosity of the {\hii} regions must be $L(H\alpha)$ $\geq$ 10$^{37}$ erg s$^{-1}$, the sample increases to 66 objects, implying that 
this luminosity condition does not bias the sample against systems with bright {\hii} regions. }. We will refer to this sample of 64 objects 
as to the \textit{unperturbed sample}.
Those possibly suffering an interaction with their surrounding environment (258 objects), will be the targets of a comparative
analysis which will be presented in a companion paper (\textit{perturbed sample}; see Table \ref{Tabgal}).

\begin{table}
\caption{The sample of galaxies
}
\label{Tabgal}
{
\[
\begin{tabular}{ccc}
\hline
\noalign{\smallskip}
\hline
Mass range						& {\hi}$-def$ $\leq$ 0.4&  {\hi}$-def$ $>$ 0.4  \\
							& Unperturbed		& Perturbed		\\
							& N.objects		& N.objects		\\
\hline
All							& 64			& 258			\\
$M_{star}$ $>$ 10$^{9.5}$ M$_{\odot}$			& 13			& 62			\\
10$^8$ $<$ $M_{star}$ $\leq$ 10$^{9.5}$ M$_{\odot}$	& 22			& 131			\\
$M_{star}$ $\leq$ 10$^{8}$ M$_{\odot}$			& 29			& 65			\\
\hline
\end{tabular}
\]
}
\end{table}

\section{Data}

\subsection{Narrow-band H$\alpha$ imaging data}

The data analysed in this work are extracted from the deep NB H$\alpha$ images 
gathered during the VESTIGE survey of the Virgo cluster undertaken with MegaCam at the CFHT.
The survey, which covers the Virgo cluster up to its virial radius (104 deg.$^2$), is extensively described in Boselli et al. (2018a).
The data were gathered in two filters, the NB filter MP9603 centred on the H$\alpha$ line ($\lambda_c$ = 6591 \AA;
$\Delta\lambda$ = 106 \AA), and the broad-band $r$ filter necessary for the subtraction of the stellar continuum (Boselli et al. 2019). 
The NB filter, which includes the emission of the H$\alpha$ ($\lambda$ = 6563 \AA) and of the two [{\nii}] lines ($\lambda$ = 6548, 6583 \AA), 
has its peak transmissivity in the velocity range -1140 $\leq$ $v_{hel}$ $\leq$ 3700 km s$^{-1}$ (see Boselli et al. (2018a). It is thus 
perfectly suited for the observations of all the galaxies at the distance of the cluster, whose redshift is included in the velocity range
-300 $\leq$ $v_{hel}$ $\leq$ 3000 km s$^{-1}$ (Boselli et al. 2014c). The data were collected thanks to several exposures using a large dither pattern 
to minimise any unwanted gradient in the sky background due to the reflection of bright stars possibly present in the fields.
The total integration time was 2 hours in the NB filter and 12 min in the broad $r$-band filter. The survey is now 76\%\ complete. 
Full sensitivity is reached almost everywhere within the VESTIGE footprint. A few shorter (a factor of $\sim$ 2) 
exposures are present at the far edges of the footprint. With these integration times the sensitivity of the survey is $f(H\alpha)$ $\simeq$ 4 $\times$ 10$^{-17}$
erg s$^{-1}$ cm$^{-2}$ (5$\sigma$) for point sources and $\Sigma(H\alpha)$ $\simeq$ 2 $\times$ 10$^{-18}$ erg s$^{-1}$ cm$^{-2}$ 
arcsec$^{-2}$ (1$\sigma$ after smoothing the data to $\simeq$ 3" resolution) for extended sources. The sensitivity might drop 
by a factor $\simeq$ 1.5 at the periphery of the cluster where full depth was not always reached.

The data were reduced using Elixir-LSB (Ferrarese et al. 2012), a pipeline expressly designed to optimise the 
detection of extended low surface brightness features such as those produced during the interaction of galaxies 
with their surrounding environment. The photometric calibration ($\leq$ 0.02-0.03 mag) and the astrometric corrections
($\leq$ 100 mas) in both the NB H$\alpha$ and broad-band 
$r$ filters was secured using the same standard procedures described in Gwyn (2008) and here optimised for these filters. 
This is done after cross-matching several thousand of unsaturated stars in the observed frames
with those available from the Panoramic Survey Telescope and Rapid Response System (Pan-STARRS). The photometric calibration 
in the NB H$\alpha$ filter was derived after applying a colour correction to unsaturated stars in each field, a
standard technique now commonly applied to wide field imaging data. This correction 
has been calibrated by convolving the spectra of a large sample of $\simeq$50000 stars of different type with available spectra in Sloian digital sky survey (SDSS) 
with the transmissivity curve of different CFHT filters, including the NB H$\alpha$ one, and fitted with a polynomial relation
as described in Gwyn (2008).

\begin{figure}
\centering
\includegraphics[width=0.49\textwidth]{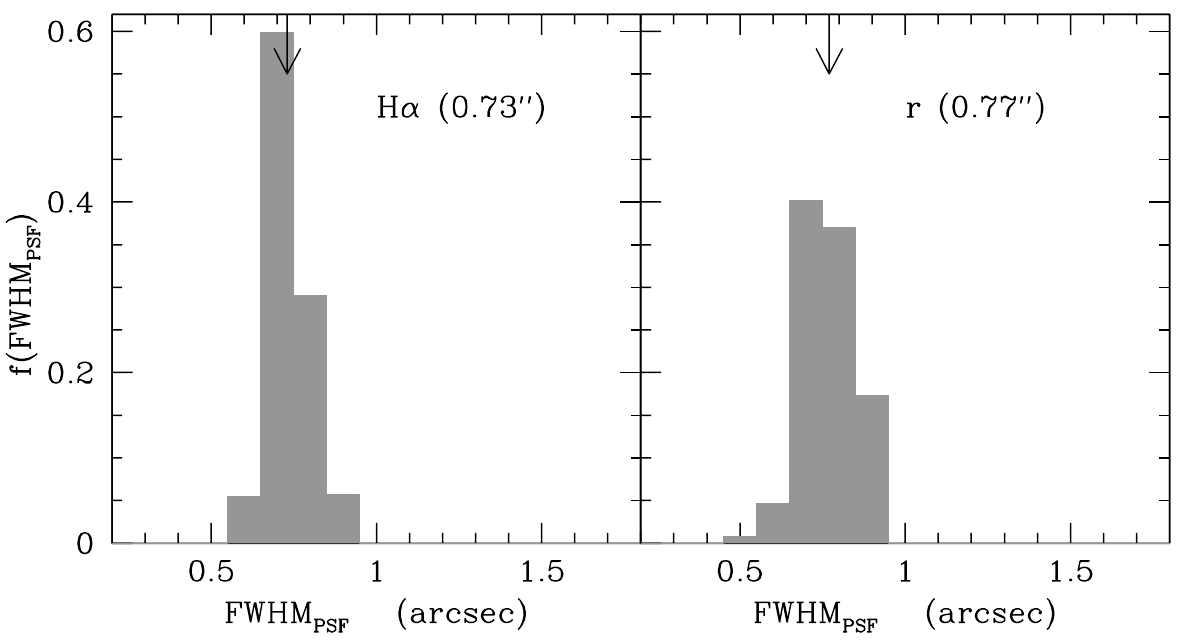}
\caption{Normalised seeing distribution in the H$\alpha$ NB (upper panel) and $r$-band (lower panel).
At the distance of the Virgo cluster (16.5 Mpc), the mean values (indicated by a vertical arrow 
and given in parenthesis) correspond to 58 pc and 62 pc in the H$\alpha$ NB and $r$-band, respectively.
}
\label{seeing}%
\end{figure}

We stress that the sensitivity of the survey for point sources at the mean distance of the Virgo cluster (16.5 Mpc, Gavazzi et al. 1999, 
Mei et al. 2007, Blakeslee et al. 2009, Cantiello et al. 2018, see below) is $L(H\alpha)$ $\geq$ 1.3 $\times$ 10$^{36}$ erg s$^{-1}$. 
This luminosity is lower than the H$\alpha$ luminosity expected for the emission of a single ionising O star and is comparable to that 
of a single early-B star (Sternberg et al. 2003). This unique set of data is thus potentially able to detect all the
ionising emission of star forming regions in the selected galaxies. As we will see in Sec. 3.4, however,
the detection of all {\hii} regions might be hampered by other factors such as confusion, the presence of contaminating stars,
dust attenuation which might affect in different ways regions of different age, and 
the correct determination of the diffuse underlying emission. The exceptional image quality of the data, which have a mean
seeing of $FWHM_{PSF}$ = 0.73" in the NB filter and $FWHM_{PSF}$ = 0.77" in the $r$-band (see Fig. \ref{seeing}), allows us to resolve {\hii} regions 
of diameter $\gtrsim$ 60 pc at the mean distance of the cluster (16.5 Mpc). 

\subsection{Multifrequency data}

As discussed in Boselli et al. (2023a), a complete set of multifrequency data from the far-UV to the far-IR, including NGVS $u,g,i,z$ broad-band 
imaging (Ferrarese et al. 2012), is available for the target galaxies.
Stellar masses are derived using a spectral energy distribution (SED) fitting analysis based on the CIGALE SED fitting code (Boquien et al. 2019) using integrated photometry. 
This is done assuming a Chabrier (2003) initial mass function (IMF)\footnote{We do not consider any possible variation in the IMF}. 
The same SED fitting code provides dust masses whenever the galaxies have been detected in a sufficient 
number of photometric mid- and far-IR bands. 
The dust masses are determined fitting the dust spectral energy distribution from 100~$\mu$m to 500$\mu$m with the Draine et al. (2007, 2014) 
models. The PAH fraction $q_\mathrm{PAH}$ is fixed at 2.5\%. 
We let the other parameters cover the full range: $U_\mathrm{min}$ (minimum radiation field) from 0.1 to 50, 
$\alpha$ (defined as $dU/dM\propto U^{-\alpha}$), and $\gamma$ (the fraction of dust illuminated by a star-forming region) 
from $10^{-3}$ to $10^{-0.3}$. Finally, the dust mass and the related uncertainties are computed from likelihood-weighted means 
and standard deviations over the full parameter space.

Total H$\alpha$ luminosities are extracted from the images by integrating the flux emission within elliptical
regions expressly defined to include all the emitting sources associated to each galaxy. These apertures have been manually optimised
to include all the diffuse emission and the {\hii} regions located within the optical disc of each galaxy and identified on the continuum-subtracted 
H$\alpha$ image, and at the same time minimise the sky contribution within the aperture to reduce the uncertainty on the measure. The {\hii} regions
selected within these elliptical apertures are located within the stellar extension of the galaxies as determined at the limiting surface brightness of 
the deep $r$-band images. At this surface brightness limit, the $r$-band extension of all galaxies 
is slightly larger than the one measured by Binggeli et al. (1985) at the $B$-band 25.5 mag arcsec$^{-2}$ isophote, in particular in dwarf irregular
objects with complex morphologies. 

In NGC 4567, where the optical disc of the galaxy partly overlaps with the one of the nearby NGC 4568, the elliptical aperture has been 
manually defined to maximise the contribution of NGC 4567 and minimise that of the companion. In this particular system, both the H$\alpha$ total flux and the identified {\hii} regions 
might be partly contaminated by those of the companion. The same elliptical apertures are used 
to estimate the sky background in regions randomly placed around the galaxy, as described in Boselli et al. (2023a).
Total H$\alpha$ luminosities are corrected for [{\nii}] contamination and dust attenuation using the spectroscopic data available for the parent sample,
as extensively described in Boselli et al. (2023a). This includes MUSE IFU or high-quality long slit spectroscopy 
(Fossati et al. 2018, Boselli et al. 2018c, Boselli et al. 2021, Boselli et al. 2022a; 12 objects), 
integrated long slit spectroscopy gathered by drifting the slit of the telescope
over the disc of the target galaxies (Gavazzi et al. 2004, Boselli et al. 2013, 2015; 116 objects), SDSS spectroscopy of the inner regions
(114 objects), or scaling relations for the remaining (mostly dwarf) systems (142 objects), where the [{\nii}] contamination is generally (if present) 
minimal (the [{\nii}]/H$\alpha$ ratio\footnote{Along the text, [{\nii}]/H$\alpha$ is defined as [{\nii}]/H$\alpha$
= [{\nii}]$\lambda$6548 + [{\nii}]$\lambda$6583/H$\alpha$.} 
is $<$0.2 in 60\%\ of this subsample) and dust attenuation $A(H\alpha)$ $\simeq$ 0 (see Boselli et al. 2023a). While Balmer decrement measurements take into account 
any effect related to the inclination of the galaxy on the plane of the sky, the adopted scaling relation does not. We expect, however,
that any second order effect related to inclination is marginal in these low-metallicity, dust-poor dwarf systems.
Luminosities are then converted to star formation rates assuming a Chabrier IMF.
The same set of spectroscopic data are also used to derive gas metallicities. For those galaxies with available integrated spectroscopy, metallicities are 
derived by Hughes et al. (2013) and calibrated on the Pettini \& Pagel (2004) relation based on the [{\oiii}]$\lambda$5007\AA\ and [{\nii}]$\lambda$6583\AA\ calibrators.
For those galaxies with only SDSS spectra, metallicities have been derived consistently using both Pettini \& Pagel (2004) calibrations. 
Metallicities are available for 39/64 of the unperturbed sample galaxies.

Finally, we derived the total gas content, defined as $M_{gas}$ = 1.3$\times$($M_{HI}$ $+$ $M_{H_2}$) (the factor 1.3 is to take into account the
Helium contribution), and the atomic-to-molecular gas ratio $M_{HI}/M_{H_2}$ for those galaxies with available high-quality molecular gas data. These have been taken from 
the Virgo environment traced in CO survey (VERTICO) when available (Brown et al. 2021), or from the Herschel Reference Survey (HRS, Boselli et al. 2014b). Both sets of data have been corrected to
a common and constant $X_{CO}$ conversion factor of $X_{CO}$ = 2.3 $\times$ 10$^{20}$ cm$^{-2}$/(K km s$^{-1}$) (Strong et al. 1988). As mentioned above, 
{\hi} data are available for the whole parent sample 
(see Boselli et al. 2023a for details), molecular masses for 15/64, and dust masses for 30/64. Gas-to-dust masses ($G/D$) are integrated values defined as
$G/D$ = $M_{gas}$/$M_{dust}$, and are available for 15/64 objects.

\subsection{Subtraction of the stellar continuum}

An accurate and unbiased subtraction of the stellar continuum is critical to measure the emission of the 
ionised gas in the H$\alpha$+[{\nii}] lines. This is done as extensively discussed in previous VESTIGE papers 
(e.g., Boselli et al. 2018a; 2019). In brief, the flux density of
of the stellar continuum emission at the effective wavelength of the NB filter is derived from the $r-$ band image 
using the relation (Boselli et al. 2019):

\begin{equation}
{\frac{r}{H\alpha+[NII]} = r - 0.1713 \times (g-r) + 0.0717}
\end{equation}

\noindent
where the emission in the $r$, $g$, and NB H$\alpha$+[{\nii}] filters is expressed in AB magnitudes, and where all 
the images have been taken in similar and excellent seeing conditions.
This colour correction has been 
calibrated using several thousands Milky Way stars with spectra available in the SDSS, and colours in the range
0.15 $\leq$ $g-r$ $\leq$ 1.0 mag. The residual bias after this correction is consistent with zero with a random 
scatter of 0.03 mag. This colour-dependent calibration is necessary to take into account the galaxy spectral slope 
between the effective wavelengths of the NB and $r-$ band filters. This procedure reduces systematic biases on the 
H$\alpha$+[{\nii}] flux (Spector et al. 2012). 
The residual scatter in the relation is taken into account in our error budget and combines spectral calibration 
uncertainties as well as effects related to the age of the stars, to their metallicity, and reddening. The lack of 
an accurate estimate of these parameters in the calibrating stars and in the target galaxies, in particular on scales 
of individual {\hii} regions, prevent us to make a further refinement in the subtraction of the stellar continuum. 
Other (mostly minor) effects such as the contribution of the emission lines in the broad-band $r$ filter or the presence 
of an underlying absorption at the H$\alpha$ line can also be present and cannot be accounted for with the available 
set of data. Those related to the underlying absorption in the H$\alpha$ line 
are certainly negligible given that typical equivalent width of the absorption line of individual {\hii} regions 
(H$\alpha$E.W.$_{abs}$ $\lesssim$ 5 \AA\ for observed regions, e.g., Diaz 1988, or H$\alpha$E.W.$_{abs}$ $\lesssim$ 2.5 \AA\
when measured using an updated version of the high spectral resolution Bruzual \& Charlot (2003) models for 
a single stellar population
after 10 Myr for metallicities $Z$ = $Z_{\odot}$ and $Z$ = $Z_{\odot}/5$) is small compared to their emission 
(H$\alpha$E.W.$_{em}$ $\gtrsim$ 100 \AA\ for
{\hii} regions younger than $\sim$ 7 Myr; Diaz 1988; Kennicutt et al. 1989b; Bresolin \& Kennicutt 1997; 
Cedr{\'e}s et al. 2005). We recall that imaging (and spectroscopic) data of normal galaxies such as those analysed in this
work do not allow to separate the stellar continuum emission of the disc from that of individual {\hii} regions. 
Their combined underlying Balmer absorption has been measured in a large variety of sources, and again it is generally $\lesssim$ 2 \AA
(e.g., Gavazzi et al. 2004; Moustakas \& Kennicutt 2006; Boselli et al. 2013),
thus negligible with respect to the H$\alpha$ line emission of the associated {\hii} region H$\alpha$E.W. $\gtrsim$ 100 \AA\, 
for the continuum and {\hii} region combined emission, Bresolin \& Kennicutt (1997).
The comparison with IFU spectroscopic data (see Sec. 3.6), however,
suggests that all these effects, if present, are minor compared to the dynamic range of the relations and distributions 
analysed here, making the results presented in this work robust. 
Despite these possible sources of uncertainty, we stress that the colour-dependent continuum subtraction procedure used in this work 
is rarely applied in the literature since very time consuming during the observations (need of observations in three different photometric bands). 
Therefore, the accuracy in the H$\alpha$+[{\nii}] flux emission reached in this work ($\sim$ 5-10\%\ ) is significantly better 
than the one obtained in most, if not all, previous similar statistical studies based on narrow-band imaging data 
(uncertainty on the flux $\sim$ 10-20\%\ ), including those where the data have been gathered with HST (e.g., Liu et al. 2013; $\sim$ 10\%\ ). 
IFU spectroscopic data do not suffer this potential problem.
The comparison of the integrated H$\alpha$+[{\nii}] fluxes measured in two VESTIGE star forming galaxies as those analysed in this work and observed 
also with MUSE (NGC 4424, IC 3476, Boselli et al. 2018c, 2021)
give consistent results within 3.5\%\ , proving once again the excellent quality of the present set of data.
For these reasons the VESTIGE survey is providing us with the best statistically significant sample of narrow-band imaging data suitable for
the purpose of this work.

\subsection{Identification of {\hii} regions}

\begin{figure}
\centering
\includegraphics[width=0.49\textwidth]{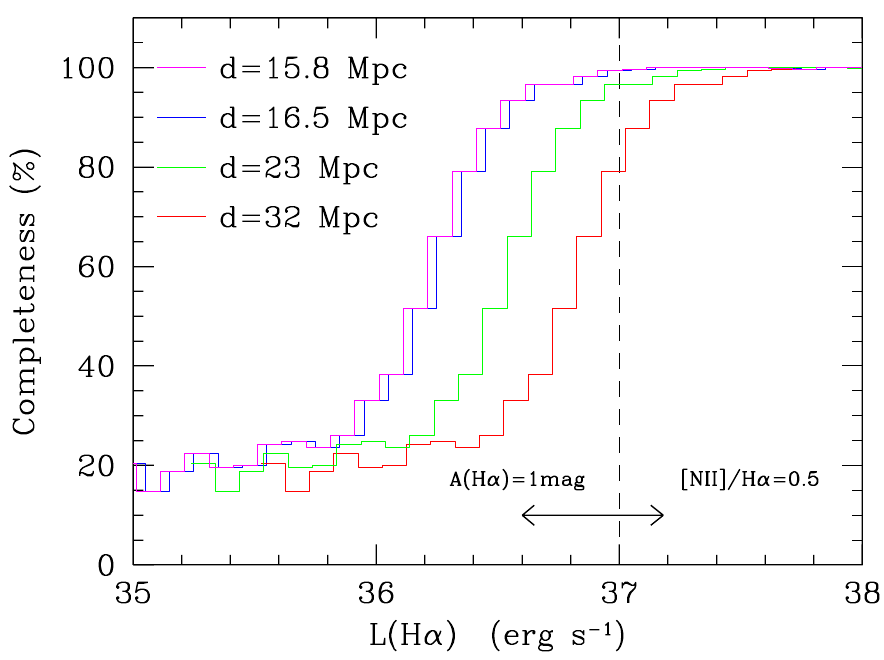}\\
\caption{Completeness function in luminosity of the {\hii} regions detected by \textsc{HIIphot} for galaxies belonging to the different substructures 
of the cluster, located at 15.8 (cluster B, C; magenta), 16.5 (cluster A, LVC; blue), 23 (W' cloud; green), and 32 Mpc (W and M clouds; red), respectively.
The vertical dashed line shows the limit in H$\alpha$ luminosity adopted to define a complete sample where statistical properties are derived. The black arrows 
show how the position of a limiting H$\alpha$ luminosity $L(H\alpha)$ = 10$^{37}$ erg s$^{-1}$ should be shifted for a dust attenuation 
of $A(H\alpha)$ = 1 mag or a [{\nii}] contamination of [{\nii}]/H$\alpha$ = 0.5 (for reference, the measured values for NGC 4321 used for this
test are $A(H\alpha)$ = 0.85 mag and [{\nii}]/H$\alpha$ = 0.48).}
\label{completeness}%
\end{figure}

\begin{figure}
\centering
\includegraphics[width=0.49\textwidth]{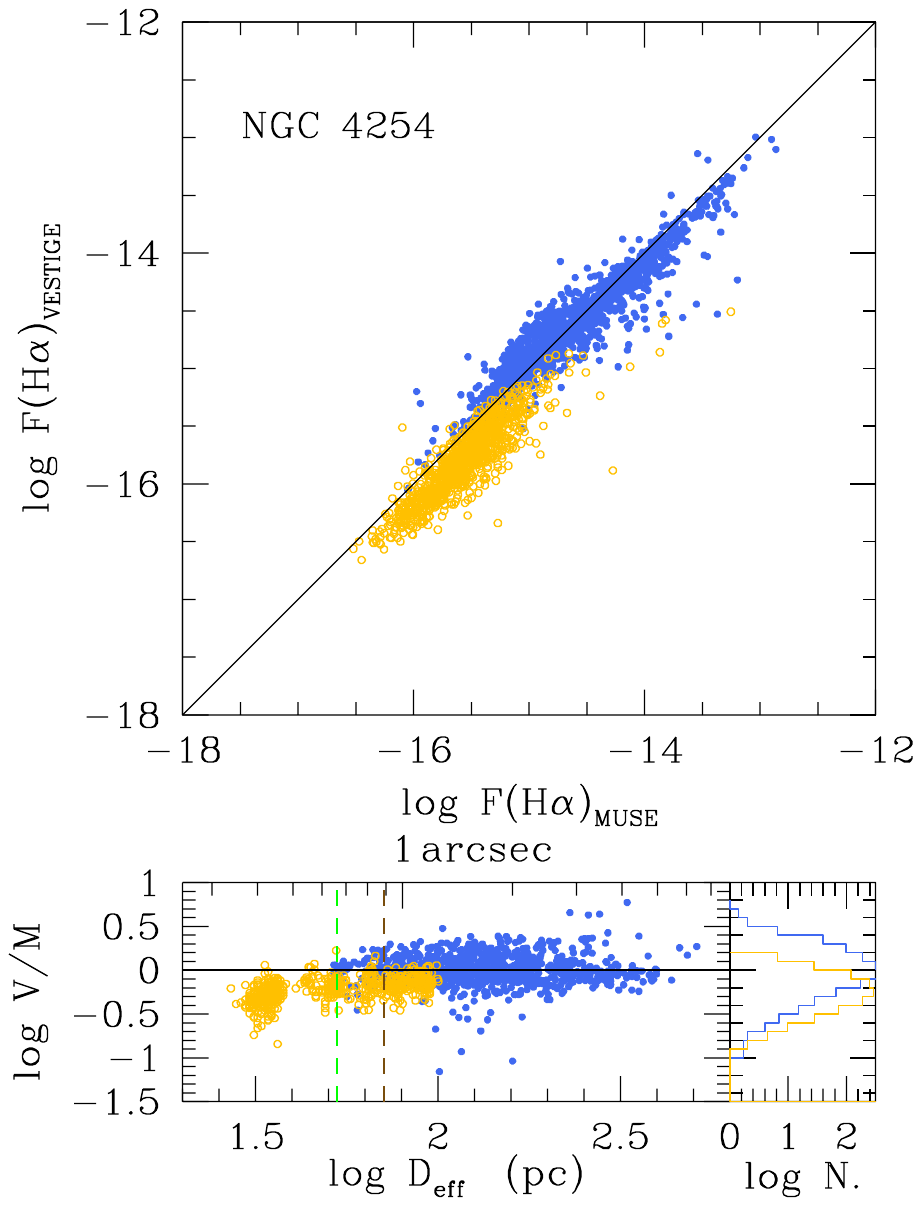}\\
\caption{Relationship between the flux of individual {\hii} regions of NGC 4254 extracted from the VESTIGE images and from the MUSE/PHANGS datacubes.
For consistency with the MUSE data, the VESTIGE data are the output of \textsc{HIIphot} uncorrected for the diffuse emission.
The lower panel shows how the ratio of the VESTIGE-to-PHANGS flux ratio changes as a function of the effective diameter corrected for seeing smearing.
The vertical green dashed line corresponds to the seeing of the VESTIGE image, the vertical brown dashed line to
the seeing of the MUSE/PHANGS dataset. Blue dots (orange circles) are for {\hii} regions with more (less) than 50 pixels in the VESTIGE image 
(regions with a typical radius of $\simeq$ 0.75\arcsec\ ). The distributions of their ratio is given in the lower right panel.
}
\label{4254}%
\end{figure}

\begin{figure}
\centering
\includegraphics[width=0.49\textwidth]{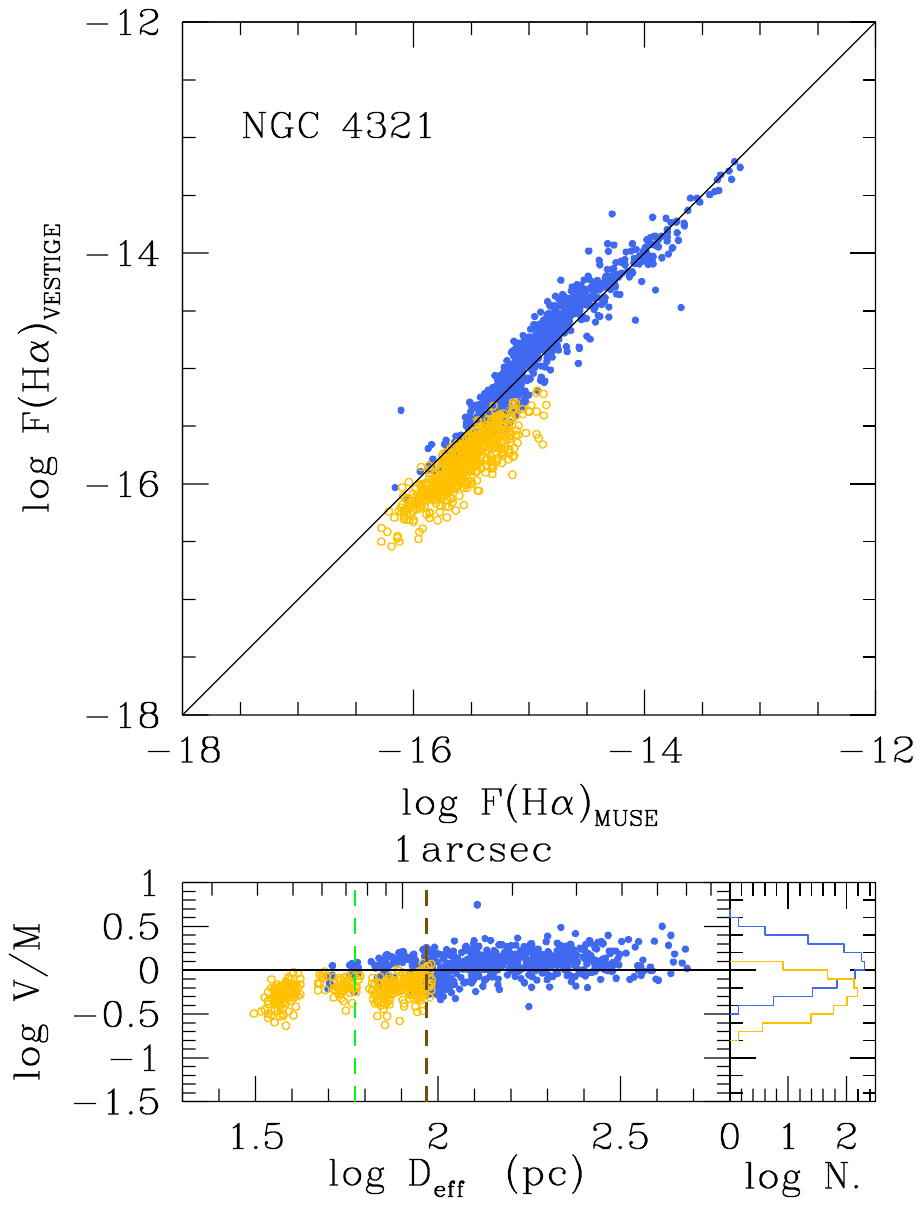}\\
\caption{Relationship between the flux of individual {\hii} regions of NGC 4321 extracted from the VESTIGE images and from the MUSE/PHANGS datacubes (see Fig. \ref{4254} for details).
}
\label{4321}%
\end{figure}

\begin{figure}
\centering
\includegraphics[width=0.49\textwidth]{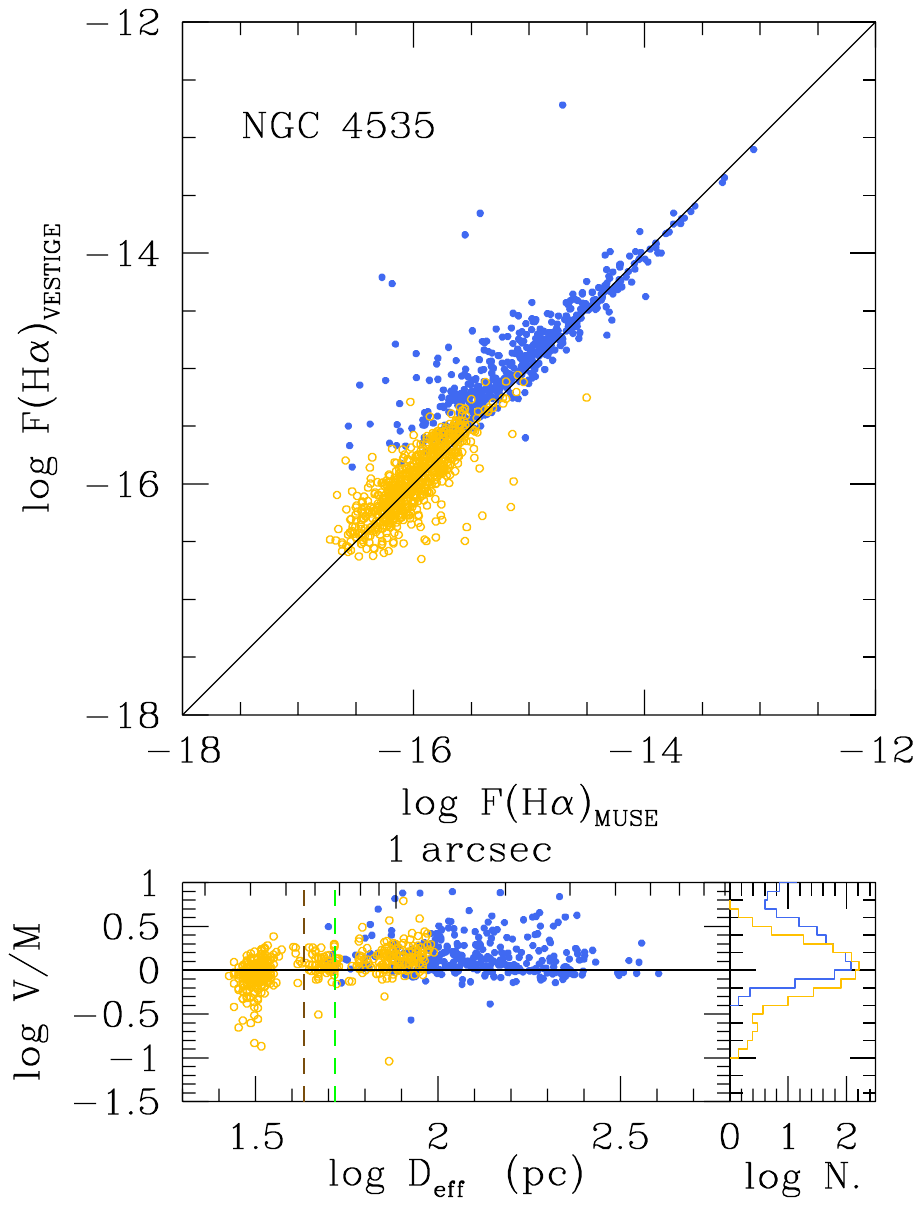}\\
\caption{Relationship between the flux of individual {\hii} regions of NGC 4535 extracted from the VESTIGE images and from the MUSE/PHANGS datacubes (see Fig. \ref{4254} for details).
}
\label{4535}%
\end{figure}

We identified the {\hii} regions and extracted their properties using the \textsc{HIIphot} data reduction pipeline (Thilker et al. 2000). 
This code has been expressly designed to extract the properties of these compact sources from NB imaging 
data as those analysed in this work. A detailed description of the code, of its qualities and possible limitations can be
found in Thilker et al. (2000), Scoville et al. (2001), Helmboldt et al. (2005), Azimlu et al. (2011), Lee et al. (2011), 
Liu et al. (2013), and Santoro et al. (2022). We already successfully applied this code on the VESTIGE data 
(Boselli et al. 2020, 2021, 2022a), and we did that here following the same methodology.

\textsc{HIIphot} uses a recognition technique based on an iterative growing procedure to identify individual {\hii} regions over the emission 
of a varying stellar continuum. The code is run on all the target galaxies after selecting adjacent regions necessary for the
determination of the sky background. These regions are selected well outside the stellar disc avoiding bright stars, ghosts, or unwanted
structures which might affect the determination of the background. The code requires for each target the seeing measured in the NB 
filter and the expected [{\nii}] contamination to the total H$\alpha$ emission. The seeing is directly measured
on the 1 deg.$^2$ stacked image to which the galaxy belongs. The distribution of the seeing in the stacked images is 
shown in Fig. \ref{seeing}. It has a mean value of $FWHM_{PSF}$ = 0.73" in the NB filter, and never exceeds 1". The contribution of the
two [{\nii}] lines to the total H$\alpha$ line emission is derived using spectroscopic data available for most of the targets, 
as described in Sec. 3.2. 

We fixed as termination gradient in the growth of the first detected "seed regions"  0.1 EM pc$^{-1}$ and a detection threshold over the background of 7$\sigma$.
This last number is significantly more aggressive than what usually used in the literature (Oey et al. 2007, Santoro et al. 2022)
but it is made possible thanks to the exceptional quality of the data in terms of sensitivity and angular resolution.
With respect to the typical MUSE/PHANGS data analysed in Santoro et al. (2022), whose sensitivity is comparable
to the one of VESTIGE, the seeing is generally better ($FWHM_{PSF}$ = 0.73" in VESTIGE vs. $FWHM_{PSF}$ = 1.2" in MUSE/PHANGS).
Adopting these parameters \textsc{HIIphot} detects {\hii} regions down to luminosities of $L(H\alpha)$ $\gtrsim$ 10$^{36}$ erg s$^{-1}$ and equivalent diameters 
$\gtrsim$ 50 pc (diameters of the circle of surface equivalent to the area of the detected {\hii} region down to a surface brightness limit of
$\Sigma(H\alpha)$ = 3 $\times$ 10$^{-17}$ erg s$^{-1}$ cm$^{-2}$ arcsec$^{-2}$, see eq. 1).
We recall that at these low emission levels where crowding becomes important, the code might suffer for incompleteness (e.g., Pleuss et al.
2000; Bradley et al. 2006; see Sec. 3.4). For this reason, although we use in several plots all {\hii} regions detected by \textsc{HIIphot},
we limit our analysis to those with $L(H\alpha)$ $\geq$ 10$^{37}$ erg s$^{-1}$ (see Sec. 3.4).

The code automatically removes the emission of unwanted stars in the field thanks to the combined and
simultaneous use of the three H$\alpha$ NB, $r$ broad-band, and H$\alpha$ continuum subtracted images. However, 
some spurious detections of other compact regions can occur in the presence of saturated stars and of their associated spikes
due to diffraction patterns of the spider of the secondary mirror of the telescope on the camera. To remove these possible contaminants, we visually inspected
the selected regions and removed any clear artefacts. Being based on a human selection, this cleaning might leave some
spurious source in the final catalogue of {\hii} regions. We consider, however, that the contribution of these unwanted objects
in the following analysis is very marginal for the following reasons: i) the density of saturated stars at the high Galactic latitude 
of the Virgo cluster is very low; ii) the number of possible contaminants increases with the angular extension of the target galaxies
on the plane of the sky. Very extended objects are rare compared to dwarf systems (Boselli et al. 2023b); iii) saturation occurs only
in bright stars, easily identifiable after comparing the broad-band $r$ image with the H$\alpha$ continuum subtracted image. 

The identification of the contaminants is thus relatively easy at bright luminosities, while might be more challenging at intermediate 
luminosities. In this luminosity range, however, the number of real {\hii} regions is significantly higher, thus the statistical weight
of the contaminants is almost negligible. We expressly decided to keep in the following analysis the compact regions detected in the
nucleus of some galaxies of the sample. This choice is justified by the fact that the number of possible AGNs identified using optical diagnostic diagrams
based on emission line ratios necessary to reject those nuclei 
where the H$\alpha$ emission is not due to a star forming episode in the whole sample
is extremely limited. The cross-match of the parent sample of H$\alpha$ detected galaxies in the VESTIGE survey with
the catalogue of Cattorini et al. (2023), which includes the most updated and homogenised nuclear classification of galaxies 
in this sky region, gives four objects classified as Seyfert using the Baldwin-Phillips-Telrevich (BPT) classification 
(Baldwin et al. 1980), eight using the H$\alpha$ equivalent width vs. {\nii}/H$\alpha$ (WHAN) classification (Cid Fernandes et al. 2011),
but only one, NGC 4388, classified as Seyfert using both diagnostics. Of these, only two galaxies (VCC 950 and AGC226326)
are included in the unperturbed sample analysed in this work. The number of low-ionisation nuclear emission-line region (LINER) in at least one of the two diagnostic diagrams is
36 in the parent sample, 4 in the unperturbed sample. We recall, however, that recent results indicate that the low ionisation emission lines observed in 
the nucleus and in the disc of nearby star-forming galaxies is mainly due to the underlying old stellar population and not necessarily to 
a nuclear activity (Belfiore et al. 2016). We thus do not consider them as active galaxies. 
Nuclear {\hii} regions are of primordial importance in environmental studies since they can be formed by the nuclear infall of gas from the disc 
induced by gravitational perturbations with other cluster members (high velocity flyby encounters, Henriksen \& Byrd 1996,
Moore et al. 1998, Lake et al. 1998, Mastropietro et al. 2005, Ellison et al. 2011). 
{\hii} regions with LINER-like 
spectra outside nuclear regions can be present in the sample (Belfiore et al. 2016). These regions cannot be identified 
in NB imaging data as those used in this work or in most of other similar works generally used in the study of the statistical properties of
extragalactic {\hii} regions.
Finally, we excluded all {\hii} regions located outside the stellar disc (see above).

The lack of IFU spectroscopic data for the whole sample prevents us from correcting the H$\alpha$ luminosities of individual {\hii} regions
for dust attenuation. Consistently with the correction for [{\nii}] contamination, we adopt here for the H$\alpha$ luminosity of individual {\hii} regions
the same correction for dust attenuation, this last measured for each galaxy using the mean 
Balmer decrement value derived from the available spectroscopic data (see Sec. 3.2), as extensively 
described in Boselli et al. (2023a). In Appendix \ref{appA} we analyse how these assumption can affect the results.

As in Boselli et al. (2021), equivalent diameters are corrected for the effects of the
point-spread function (PSF) following Helmboldt et al. (2005), where the corrected diameter $D_{eq}$ is given by the relation:

\begin{equation}
{D_{eq} = D_{HII}\frac{\sqrt{FWHM^2_{mod} - FWHM^2_{PSF}}}{FWHM_{mod}}}
\end{equation}

\noindent
where $D_{HII}$ is the output diameter of \textsc{HIIphot}, $FWHM_{mod}$ = $\sqrt{FWHM_{maj}FWHM_{min}}$ is the effective circular FWHM from
the Gaussian model fit of the 2D line intensity emission, and $FWHM_{PSF}$ is the seeing as obtained from fitting bright stars in the 
images with gaussian models. We recall that at the typical seeing of the survey ($FWHM_{PSF}$ = 0.73\arcsec in the NB filter),
the angular resolution of the images in physical scales is 56 pc at 15.8 Mpc (cluster B and C), 58 pc at 16.5 Mpc (cluster A and LVC), 
81 pc at 23 Mpc (W' cloud), and 112 pc at 32 Mpc (W and M clouds).
Finally, we derived electron densities $n_e$ of individual {\hii} regions following 
Scoville et al. (2001) adopting the relation (case B recombination, Osterbrock \& Ferland 2006):

\begin{equation}
n_\mathrm{e} = 43\bigg[\frac{(L_{cor}(H\alpha)/10^{37} ~\rm{erg~s^{-1}})(T/10^4 ~\rm{K})^{0.91}}{(D_{eq}/10 ~\rm{pc})^3}\bigg]^{1/2} ~~[\rm{cm^{-3}}]
\end{equation} 

\noindent
where $L_{cor}(H\alpha)$ is the H$\alpha$ luminosity of the individual {\hii} regions corrected for [{\nii}] contamination and dust attenuation, 
and $T$ the gas temperature (here assumed to be $T$ = 10000 K). To avoid large uncertainties in the adopted corrections,
equivalent diameters and electron densities 
are only derived for those regions where the correction for the effect of the PSF is less than 50\%. 
This choice might introduce a systematic bias in the physical properties of the analysed {\hii} regions, favouring regions with low electron densities.

\subsection{Completeness}

With decreasing luminosity, the increasing number of {\hii} regions populate more and more 
crowded structures such as spiral arms, and becomes thus more and more difficult to detect and measure them with automatic codes such as \textsc{HIIphot}.
To test the ability of \textsc{HIIphot} in detecting {\hii} regions of decreasing luminosity in crowded structures we created and injected mock {\hii} regions into 
real images, ran the code on these mock images, and counted the number of injected sources detected by the code. Mock {\hii} regions of different luminosity
have been created using the PSF measured on the image of the hosting galaxy. For this exercise we used the image of the bright galaxy NGC 4321 (M100),
a massive spiral galaxy seen face-on. Being one bright and massive unperturbed spiral galaxy of morphological type SAB(s)bc, NGC 4321 is one of
the galaxies within the parent sample hosting the largest number of {\hii} regions ($\sim$ 8000)\footnote{NGC 4321 is not included in the unperturbed sample 
because its {\hi}-deficiency parameter is {\hi}$-def$ $=$ 0.41. It is used in this test since it is the largest face-on galaxy of the parent sample, thus perfectly suited 
for this purpose. We do not expect that its possible moderate interaction with the surrounding cluster environment might affect this completeness test.}. 
It thus represents an extreme case where crowding
is particularly severe. The mock {\hii} regions have been randomly injected within the disc of NGC 4321, but avoiding those regions where 
{\hii} regions of H$\alpha$ luminosity $L(H\alpha)$ $\geq$ 10$^{37}$ erg s$^{-1}$ have been already detected by \textsc{HIIphot}. This lower limit in 
H$\alpha$ luminosity was necessary to grant enough positions over the disc of the galaxy where {\hii} regions could be injected 
given that at fainter luminosity all the disc is hosting {\hii} regions. The luminosity of the injected regions was uniformly distributed in log space over the 
range 10$^{35}$ $\leq$ $L(H\alpha)$ $\leq$ 10$^{38.5}$ erg s$^{-1}$. The counts injected in the image include the [{\nii}]
contribution, thus the detection performed by \textsc{HIIphot} includes H$\alpha$+[{\nii}] as for the real sources. The [{\nii}] contribution is then 
removed by \textsc{HIIphot} at the photometry step. We do not include the dust attenuation in these mocks, however,
we note that the putative detection limit is well below the luminosity limit we use for our analysis, therefore our results do not change appreciably even
if the completeness limit is shifted to higher luminosity by reasonable dust
extinction values. We checked that the physical size of the injected {\hii} regions is
comparable to that of the real {\hii} regions in different bins of luminosity.
Each injection was done using $\sim$ 300 mock {\hii} regions, and
this operation was repeated 30 times to grant sufficient statistics. The injection of these {\hii} regions was done on both the H$\alpha$ 
narrow-band and on the H$\alpha$ continuum-subtracted images, while not in the broad-band $r$ image where the contribution of these line emitters
to the continuum is only marginal (a few per cent).

We then run the \textsc{HIIphot} code on the images including the injected mock {\hii} regions using all the same parameters of the code as those used on real images. 
The fraction of detected HII regions in the mocks is purely defined by a match (within 3 pixels) between the injected position and the recovered
position. We have verified that the detected luminosities are in agreement ($\lesssim$ 0.2 dex for $L(H\alpha)$ $\geq$ 10$^{36}$ erg s$^{-1}$,
dropping to $\lesssim$ 0.12 dex for $L(H\alpha)$ $\geq$ 10$^{37}$ erg s$^{-1}$, the limiting luminosity used in the following analysis) with the injected ones,
 confirming the quality of the \textsc{HIIphot} photometry. 
Figure \ref{completeness} shows the completeness function, i.e. 
the fraction of detected/injected mock {\hii} regions as a function of their luminosity assuming that the galaxy is located at 
the typical distance of the different Virgo cluster substructures. Figure \ref{completeness} clearly shows that, when the galaxy is located 
within the main body of the cluster (16.5 Mpc), most of the mock 
{\hii} regions of luminosity $L(H\alpha)$ $\geq$ 10$^{37}$ erg s$^{-1}$ are detected by \textsc{HIIphot}. The completeness function is 99\%\ at 
$L(H\alpha)$ $=$ 10$^{37}$ erg s$^{-1}$, $\simeq$ 80\%\ at $L(H\alpha)$ $=$ 10$^{36.5}$ erg s$^{-1}$, dropping to $\simeq$ 30\%\ 
at $L(H\alpha)$ $=$ 10$^{36}$ erg s$^{-1}$. The completeness function at $L(H\alpha)$ $=$ 10$^{37}$ erg s$^{-1}$ drops to 97\%\ for galaxies at 23 Mpc, 
and to 80\%\ at 32 Mpc, but becoming 87\%\ complete at $L(H\alpha)$ $\geq$ 10$^{37.05}$ erg s$^{-1}$ at this distance. Since 1) NGC 4321 used for this 
completeness test is one of the most extreme galaxies in terms of crowdness, thus the present completeness function should be 
considered as a lower limit for the typical galaxies of the sample; 2) the number of objects belonging to the M and W clouds at 32 Mpc is
only 14/64, 3) 13/14 of them have a very limited number of {\hii} regions satisfying this selection ($\leq$ 237), with only one object (VCC 89) hosting 654
regions, and 4) their total number (1263)
is $\leq$ 10\%\ of those of the entire unperturbed sample (13278)
we decided to consider in the following analysis only {\hii} regions with H$\alpha$ luminosity 
$L(H\alpha)$ $\geq$ 10$^{37}$ erg s$^{-1}$ when deriving their statistical properties. We will discuss, whenever appropriate, the possible impact
of this distance-related bias in the following analysis.
This completeness test also confirms that the 
aggressive thresholds used in the identification of {\hii} regions in \textsc{HIIphot} (see Sec. 3.4) do not affect the results once 
a limiting H$\alpha$ luminosity of $L(H\alpha)$ $\geq$ 10$^{37}$ erg s$^{-1}$ is adopted.

\subsection{Comparison with the literature}

We checked the accuracy of the derived parameters by comparing them to those extracted from an independent set of {\hii} regions extracted from MUSE data
gathered during the PHANGS survey of nearby galaxies (Emsellem et al. 2022) and recently published and analysed in Santoro et al. (2022).
The two sets of data have three massive galaxies in common, NGC 4254, NGC 4321, and NGC 4535. \footnote{NGC 4321 is not included in 
the unperturbed sample because its {\hi}-deficiency parameter is {\hi}$-def$ $=$ 0.41. We include it in this comparison which it would be otherwise  
limited to only two objects. We do not expect that the comparison of fluxes measured using these two different methods would be 
affected by any external perturbation, which in any case, if present, is minimal in this particular object.}
We recall that the parameters of the {\hii} regions of
these galaxies in the PHANGS data have been extracted using a modified version of the \textsc{HIIphot} pipeline optimised to work on IFU
spectroscopic data. The main differences with the one used on the VESTIGE data are: i) the PHANGS H$\alpha$ data are corrected for dust attenuation 
and do not include any [{\nii}] contamination; ii) the PHANGS data are not corrected for the emission line flux of the background, i.e.
the diffuse ionised gas (DIG) component permeating star-forming discs (Haffner et al. 2009). As stressed in Santoro et al. (2022), 
this diffuse emission becomes relevant only in the fainter {\hii} regions; iii) the PHANGS data cover just a fraction of the 
disc of all the three targets, where the contribution of the stellar continuum emission and of the DIG are important.

\begin{table*}
\caption{Comparison VESTIGE vs. PHANGS/MUSE 
}
\label{TabPHANGS}
{
\[
\begin{tabular}{ccccc}
\hline
\noalign{\smallskip}
\hline
Galaxy	& FWHM$_{VESTIGE}$	& FWHM$_{MUSE}$	& log $F(H\alpha)_{VESTIGE}/F(H\alpha)_{MUSE}$	& log $F(H\alpha)_{VESTIGE}/F(H\alpha)_{MUSE}$ \\
	& 			& 			& all {\hii} regions					& resolved {\hii} regions				\\
\hline
NGC4254	& 0.66\arcsec		& 0.89\arcsec		& -0.11$\pm$0.21					& -0.02$\pm$0.18 \\
NGC4321	& 0.74\arcsec		& 1.16\arcsec		& -0.07$\pm$0.21					& 0.07$\pm$0.15	 \\
NGC4535 & 0.68\arcsec		& 0.56\arcsec		& 0.09$\pm$0.25						& 0.16$\pm$0.25  \\
\hline
\end{tabular}
\]
}
\end{table*}

The comparison of the fluxes extracted by \textsc{HIIphot} on the two sets of data is shown in Figures \ref{4254}, \ref{4321}, and \ref{4535}. 
For consistency, the comparison is done using VESTIGE observed fluxes not corrected for the DIG emission. 
The comparison of the two sets of data shows that a higher number of {\hii} regions is detected in the 
VESTIGE NB imaging data than in the MUSE/PHANGS IFU spectroscopic data. This is probably due to the fact that
a more aggressive surface brightness threshold is adopted in this work compared to the one used on the PHANGS data. 
Table \ref{TabPHANGS} gives the difference in logarithmic scale between the fluxes in the H$\alpha$+[{\nii}] 
emission line uncorrected for DIG emission and measured between the two instruments. The 
agreement between the two sets of data is very good (systematic difference of $\sim$ 0.10 dex with a dispersion of $\sim$ 0.20 dex), 
and this despite the use of a different \textsc{HIIphot} extraction code on a completely different set of data.
The difference in the two sets of data is reduced in those {\hii} regions extended over more than 50 pixels 
(blue symbols), the most extended regions resolved by the two instruments.   
This result suggests that the seeing can slightly affect the output of the code and stresses once again the importance of using
a uniform set of high-quality imaging data for such an analysis. VESTIGE is providing this ideal sample
given that all images have been gathered during excellent and very constant seeing conditions (see Fig. \ref{seeing}).
Figures \ref{4254}, \ref{4321}, and \ref{4535} do not show any increase of the systematic difference between the two sets of data
in the brightest {\hii} regions, where the contribution of the line emission in the broad-band $r$ filter is expected to be more
important, suggesting that this effect is negligible. The available set of data does not allow us to test whether
the scatter in the relations is correlated to any other physical parameter such as the metallicity or dust attenuation in the {\hii} region
or to the underlying Balmer absorption in the H$\alpha$ line. Considering that the typical instrumental uncertainty in the MUSE data is of the order of 
$\sim$ 5\%\ (from the instrument documentation), that on the H$\alpha$ measurement of another $\lesssim$ 2\%\ (Santoro et al. 2022), and that 
part of the scatter is due to the different apertures identified by \textsc{HIIphot} in the IFU MUSE data and in the narrow-band 
VESTIGE imaging data, we can conclude that the uncertainty in the measured fluxes ($\simeq$ 0.15 dex if we consider 
that part of the dispersion in the Figures \ref{4254}, \ref{4321}, and \ref{4535} is due also to the MUSE data) is small compared to the dynamic range
of the relations and distributions analysed in this work. We thus expect that the impact of the third-order effects mentioned here and in Sec. 3.3
are negligible in the following analysis. 
In \ref{appA} we show how the use of a single correction for dust attenuation and [{\nii}] contamination derived using integrated
measurements for each individual galaxy does not have major systematic effects in the derived parameters.

\subsection{Sample of {\hii} regions in the unperturbed sample}

Adopting the above criteria, we end up with a sample of 27.330 {\hii} regions in 64 gas-rich ({\hi}$-def$ $\leq$ 0.4), 
unperturbed systems. Of these, 13.278 have a H$\alpha$ luminosity corrected 
for dust attenuation and [{\nii}] contamination $L(H\alpha)$ $\geq$ 10$^{37}$ 
erg s$^{-1}$. This is the final sample of {\hii} regions analysed in this work.
Those with an effective diameter derived assuming a correction $\leq$ 50\%\ (see eq. 2) are 11293, and 6520 when
limited to $L(H\alpha)$ $\geq$ 10$^{37}$ erg s$^{-1}$. 
The galaxies analysed in this work (unperturbed sample) are listed in Table \ref{gal} along with their parameters.
Table \ref{Tabds9} gives the parameters of the elliptical apertures used for the identification of the {\hii} regions 
and for the extraction of the total H$\alpha$ flux. Tables \ref{gal} and \ref{Tabds9} are given in \ref{appB}.

\section{Analysis}

In this section we analyse the statistical properties of the {\hii} regions in unperturbed cluster galaxies.
We do that first by deriving the composite distribution of the whole unperturbed sample (luminosity function, effective
diameter, and electron density distribution) and the luminosity-diameter relation.
Composite luminosity functions and distributions are important to derive the statistical properties of a complete sample 
of {\hii} regions in the Virgo cluster. Although these distributions might marginally change when derived in the different cluster substructures, 
as they do from galaxy to galaxy, they composite shapes can be used as reference for comparative studies for samples of galaxies located in different environments 
(voids, field, groups, other clusters etc). They also provide sufficient statistics to analyse dwarf systems, 
otherwise impossible for their very limited number of {\hii} regions.
We then study the main scaling relations between several representative properties of individual {\hii} regions and those 
of their parent galaxy. All this analysis is made possible thanks to the excellent statistic,
the large dynamic range in the parameter space of galaxies properties, and the uniform quality of the data
which minimises any possible bias.

\subsection{Composite distributions}

\subsubsection{Composite luminosity function}

\begin{figure}
\centering
\includegraphics[width=0.49\textwidth]{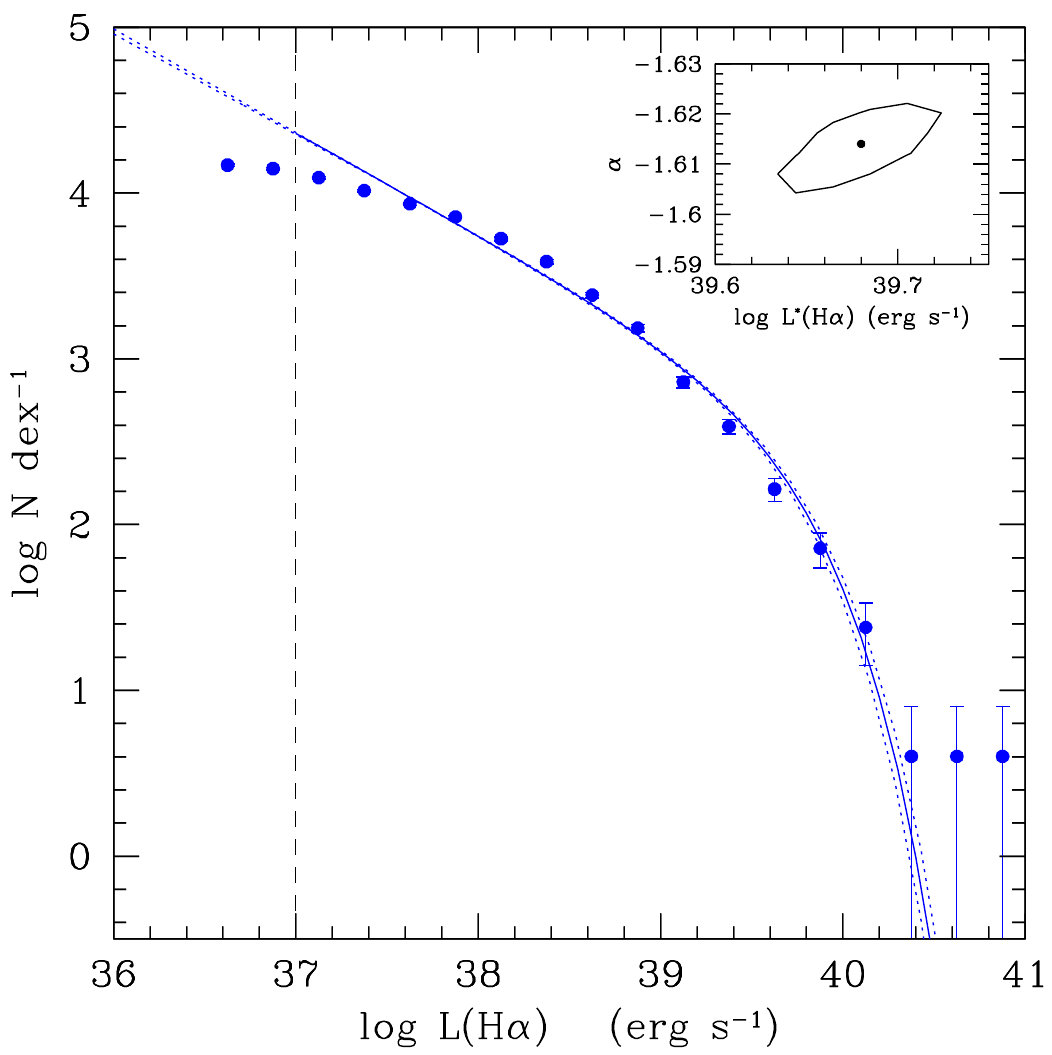}\\
\caption{Composite luminosity function of the {\hii} regions detected by \textsc{HIIphot} on the selected galaxies. The H$\alpha$ luminosities of 
individual {\hii} regions are corrected for dust attenuation and [{\nii}] contamination as described in Sec. 3.2. The solid and dotted lines
indicate the best fit and 1$\sigma$ confidence regions for the Schechter luminosity function parametrisation. The vertical dashed line 
shows the completeness of the survey. The small panel in the top right corner indicates the 1$\sigma$ probability distribution of the 
fitted Schechter function parameters.}
\label{LFHIInondef}%
\end{figure}

Figure \ref{LFHIInondef} shows the composite H$\alpha$ luminosity function of all the {\hii} regions of the unperturbed sample.
The H$\alpha$ luminosities are corrected for [{\nii}] contamination and dust attenuation as mentioned above.
This function is constructed by counting the number of {\hii} regions per bin of H$\alpha$ luminosity (0.25 dex in log scale; differential 
luminosity function).
The distribution shows a steep rise of the number of {\hii} regions from $L(H\alpha)$ $\simeq$ 10$^{40.5}$ erg s$^{-1}$
to $L(H\alpha)$ $\simeq$ 10$^{39-39.5}$ erg s$^{-1}$, a relative flattening up to $L(H\alpha)$ $\simeq$ 10$^{36.5}$ erg s$^{-1}$,
with a moderate (up to $L(H\alpha)$ $\simeq$ 10$^{36}$ erg s$^{-1}$) then an abrupt decrease below this luminosity due to completeness (see Sec. 3.5 and Fig. \ref{completeness}).

\begin{table}
\caption{Parameters of the Schechter function
}
\label{TabLF}
{
\[
\resizebox{\columnwidth}{!}{\begin{tabular}{ccccc}
\hline
\noalign{\smallskip}
\hline
Sample					& N.objects		&  $\alpha$     & log$L^*(H\alpha)$     & $\Phi^*$ \\
					&			&		& erg s$^{-1}$		&		\\
\hline
Composite				& 13278			& -1.61$\pm$0.01& 39.68$\pm$0.04	& 226$\pm$21	\\
Mean$^a$				& 27			& -1.66$\pm$0.43& 39.57$\pm$0.44	& -		\\
$R$ $\geq$ $R_{eff}$			& 9084			& -1.66$\pm$0.01& 39.61$\pm$0.05	& 136$\pm$17	\\
$R$ $<$ $R_{eff}$			& 4194			& -1.50$\pm$0.02& 39.70$\pm$0.06	& 113$\pm$14	\\
cluster A+LVC				& 9348			& -1.55$\pm$0.02& 39.59$\pm$0.03	& 245$\pm$24	\\
cluster B				& 2344			& -1.78$\pm$0.03& 39.56$\pm$0.13	& 23.4$\pm$6.2\\
W' cloud				& 323			& -1.74$\pm$0.05& 40.35$\pm$0.38	& 0.91$\pm$2.00	\\
W+M clouds				& 1263			& -1.78$\pm$0.05& 40.42$\pm$0.32	& 2.53$\pm$4.00	\\
\hline
\end{tabular}}
\]
Notes: all fits are done for {\hii} regions with $L(H\alpha)\geq10^{37}$ erg s$^{-1}$. $a$): mean values derived for 27 galaxies having more than 20 individual {\hii} regions.
Uncertainties given here are the dispersion in the parameter distribution. 
}
\end{table}

Following the same procedure described in Mehta et al. (2015), Fossati et al. (2021) and Boselli et al. (2023b), 
we tentatively fit the distribution with a Schechter (1976) function of the form of ${\mathcal L}$ = $\log_{10}(L)$ :


\begin{equation}
{\rm{\Phi}({\mathcal L})d{\mathcal L} = \rm{ln(10)\Phi^*10}^{({\mathcal L}-{\mathcal L}^*)(1+\alpha)}\rm{exp}(-10^{({\mathcal L}-{\mathcal L}^*)})d{\mathcal L}}
\end{equation}

\noindent
with $d{\mathcal L}$ = $d\log_{10}(L)$, ${\mathcal L}^*$=$\log_{10}L^*$ (in units of erg s$^{-1}$), 
where we derived the posterior 
distribution and the best-fit parameters using the MULTINEST Bayesian algorithm
(Feroz \& Hobson 2008, Feroz et al. 2019). $L^*$, $\Phi^*$, and $\alpha$ are the characteristic luminosity at the knee
of the distribution, the number of objects at $L^*$, and the slope of the distribution at the faint end, respectively.
We limit the fit of the Schechter function to $L(H\alpha)$ $\geq$ 10$^{37}$ erg s$^{-1}$, where the sample is almost complete (see Sec. 3.5).
The adopted function reproduces the data, although with a systematic 
underestimate of the number of bright {\hii} regions for $L(H\alpha)$ $\gtrsim$ 10$^{40.5}$ erg s$^{-1}$, where the statistic is poor, 
and overestimating it at the faint end ($L(H\alpha)$ $\lesssim$ 10$^{37.5}$ erg s$^{-1}$), where incompleteness might start to be present.
Despite these systematic differences, a Schechter function is here more indicated than a single or double power law, as often used 
in the literature (e.g., Kennicutt et al. 1989a; Beckman et al. 2000; Helmboldt et al. 2005; Bradley et al. 2006; 
Liu et al. 2013; Cook et al. 2016; Santoro et al. 2022), because the data have a smooth distribution, steep at the bright end and flat at the faint end. 
The parameters of the best fit for individual galaxies are given in Table \ref{galLF} in \ref{appB}, while those
for the composite luminosity functions in Table \ref{TabLF} (median of the posterior samples from the nested
sampling solver, with uncertainties). They are given for the composite luminosity function (first row)
or the mean of the parameters with the dispersion of their distribution derived for those galaxies having more than 20 {\hii} 
regions\footnote{The limiting number of 20 is here arbitrary taken to grant at the same time a
robust derivation of the best fit parameters and to have a sufficient number of objects necessary for a statistical analysis. } of 
luminosity $L(H\alpha)$ $\geq$ 10$^{37}$ erg s$^{-1}$ (27 objects, second row). These are all galaxies with a stellar mass 
$M_{star}$ $\geq$ 10$^8$ M$_{\odot}$ (see Sec. 4.2.2 and Table \ref{gal}). The individual luminosity functions of these 27 galaxies are shown in 
Fig. \ref{LFindividual} given in \ref{appE}. 
The fitting method used in this work is optimised to treat distributions with different statistical numbers and uncertainties on
the sampled range, as is the case for a luminosity function from the faint to the bright end, since it uses individual points rather than binned data 
(e.g., Mehta et al. 2015). The outputs of the fit might thus systematically differ from those gathered using other fitting techniques as done in previous works.
The comparison with other works, however, is made difficult by other factors such as the choice of different parametric functions (e.g., power 
low vs. Schechter function), dust attenuation and [{\nii}] contamination corrections, completeness, or sampled H$\alpha$ luminosity range. Despite the 
adopted Schechter function only approximately fits the observed distributions, the homogeneity of the sample in terms of completeness and the use of the same 
fitting function within the same luminosity range for all targets secures a robust comparative analysis of the output parameters.

\begin{figure}
\centering
\includegraphics[width=0.49\textwidth]{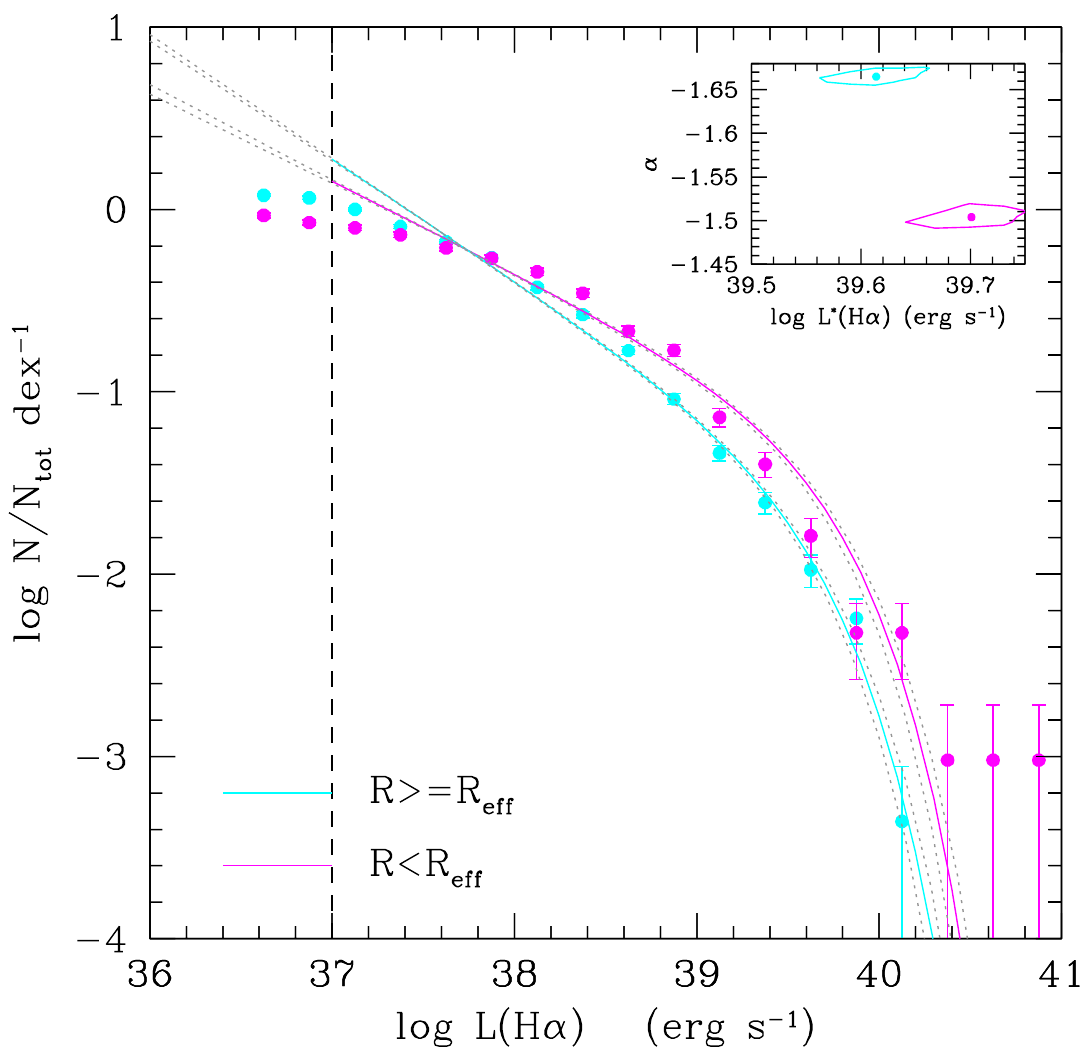}\\
\caption{Composite luminosity function of the {\hii} regions detected within (magenta) and outside (cyan) 
the $i$-band effective radius of the target galaxies normalised to the total number of {\hii} regions with $L(H\alpha)$ $\geq$ 10$^{37}$ erg s$^{-1}$
within ($N_{tot}$=4194) and outside ($N_{tot}$=9084) the effective radius. The vertical dashed line 
shows the completeness of the survey. The small panel in the top right corner indicates the 1$\sigma$ probability distribution of the 
fitted Schechter function parameters.}
\label{LFHIIReffnondef}%
\end{figure}

\begin{figure}
\centering
\includegraphics[width=0.49\textwidth]{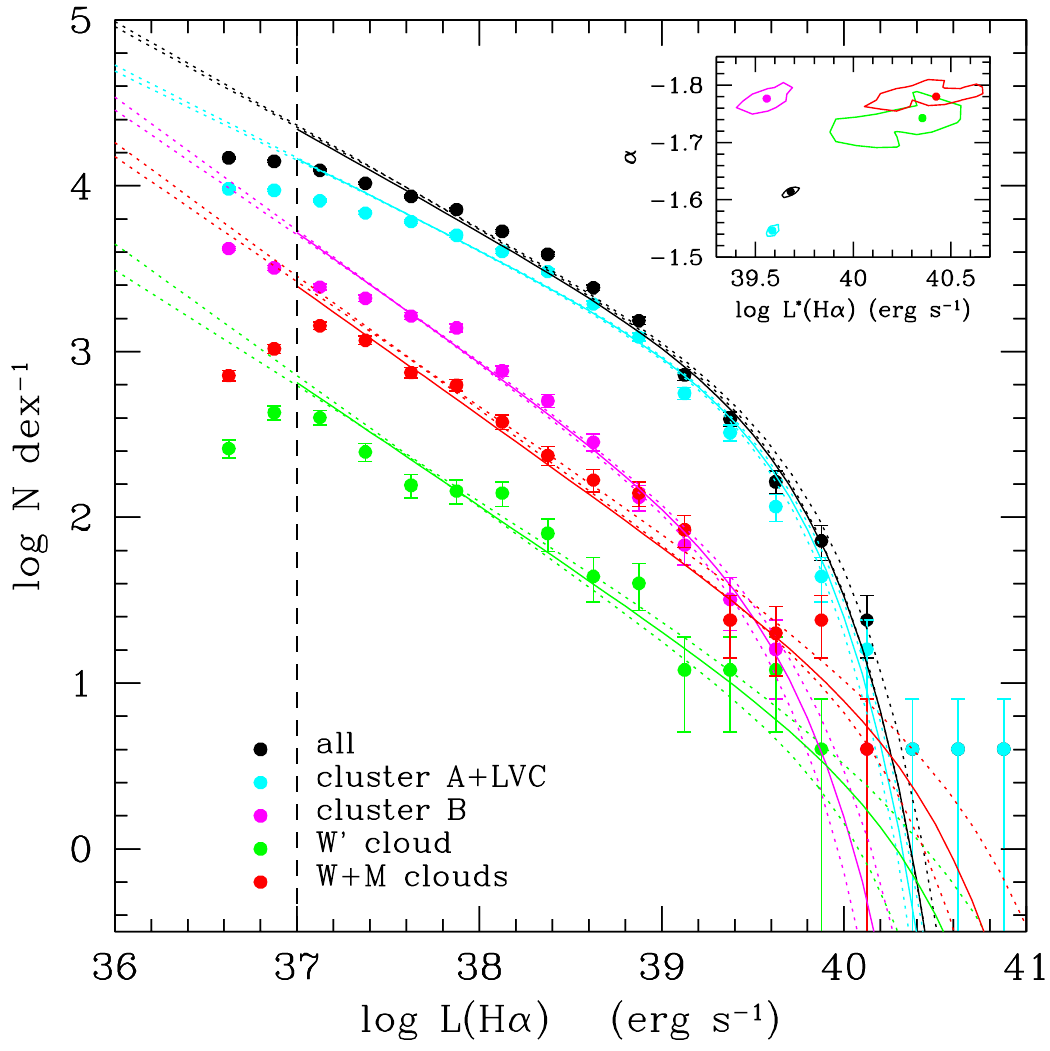}\\
\caption{Composite luminosity function of the {\hii} regions detected by \textsc{HIIphot} on the selected galaxies: different colours are used for galaxies belonging to 
different cluster substructures (cyan for cluster A and LVC, 16.5 Mpc; magenta for cluster B, 15.8 Mpc;
green for W' cloud, 23 Mpc; red for W and M clouds, 32 Mpc). The H$\alpha$ luminosities of 
individual {\hii} regions are corrected for dust attenuation and [{\nii}] contamination as described in Sec. 3.2. The solid and dotted lines
indicate the best fit and 1$\sigma$ confidence regions for the Schechter luminosity function parametrisation. The vertical dashed line 
shows the completeness of the survey. The small panel in the top right corner indicates the 1$\sigma$ probability distribution of the 
fitted Schechter function parameters.}
\label{LFHIInondefdist}%
\end{figure}

\begin{figure}
\centering
\includegraphics[width=0.49\textwidth]{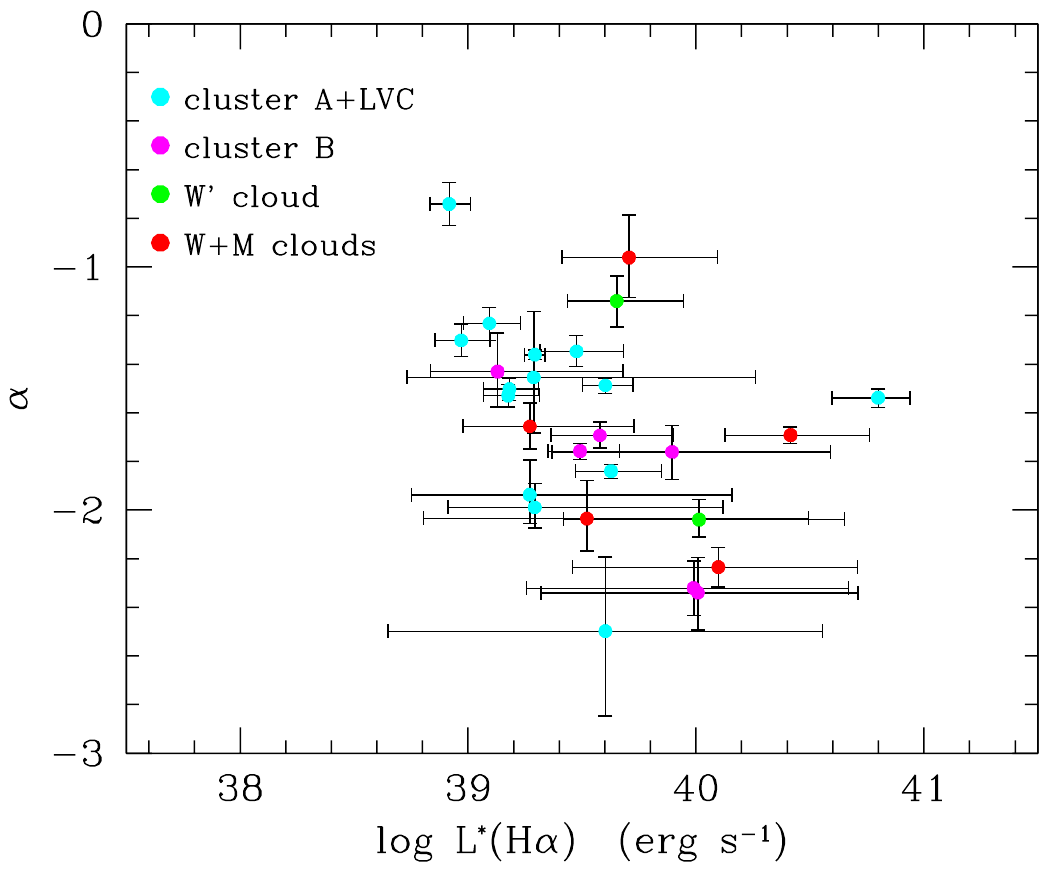}\\
\caption{Relation between the faint end slope $\alpha$ of the luminosity function and the characteristic luminosity
of in dividual galaxies with more than 20 {\hii} regions brighter than $L(H\alpha)$ $\geq$ 10$^{37}$ erg s$^{-1}$. Different colours are used for galaxies belonging to 
different cluster substructures (cyan for cluster A and LVC, 16.5 Mpc; magenta for cluster B, 15.8 Mpc;
green for W' cloud, 23 Mpc; red for W and M clouds, 32 Mpc). The H$\alpha$ luminosities of 
individual {\hii} regions are corrected for dust attenuation and [{\nii}] contamination as described in Sec. 3.2. }
\label{individualparameters}%
\end{figure}

Figure \ref{LFHIIReffnondef} shows the normalised luminosity function derived for {\hii} regions located inside or outside the deprojected effective radius
measured in the $i$-band, available to the NGVS team, and measured as described in Ferrarese et al. (2020). 
The total number of {\hii} regions with $L(H\alpha)$ $\geq$ 10$^{37}$ erg s$^{-1}$
located outside the effective radius (9084) exceeds by a factor of $\sim$ 2 the one located in the inner regions (4194).
Despite that, the brightest {\hii} regions are preferentially located in the inner disc, some of which at less than 3 arcsec from the nucleus (12/64),
as indicated by the different shape of the two functions (a Kolmogorov-Smirnov test indicates that the two distributions are statistically different, 
p-value = 1.3 $\times$ 10$^{-19}$). 
The faint-end slope of the distribution in the inner regions is flatter than the one observed in the outer discs. The difference in the 
distributions observed below $L(H\alpha)$ $\lesssim$ 10$^{37.5}$ erg s$^{-1}$, however, should be considered with caution since it
might result form possible biases in the \textsc{HIIphot} pipeline in detecting {\hii} regions of low luminosity located
in crowded regions, in particular where the stellar continuum might be important (inner disc, see Pleuss et al. 2000, Bradley et al. 2006).
Indeed, it is likely that incompleteness, if present, is more important in the inner regions, where crowding is also more important.
Finally, to quantify any possible effect related to the different distance of galaxies,
we construct and compare four independent composite luminosity functions for galaxies located in the different substructures of the cluster 
(Fig \ref{LFHIInondefdist}). Although this test is limited by poor statistics for galaxies located in the W' (23 Mpc), W and M clouds (32 Mpc), 
we do not observe any strong dependence of the best fit parameters with distance, nor a clear sign of incompleteness in the lowest luminosity bins
in the distribution for galaxies located in the far W', W, and M clouds, suggesting that the luminosity function shown in Fig. \ref{LFHIInondef}
well represents the distribution of luminosities in unperturbed, gas-rich systems.
Figure \ref{individualparameters} shows the relation between the faint end slope and the characteristic luminosity measured for all the individual galaxies with
more than 20 {\hii} regions of H$\alpha$ luminosity corrected for dust attenuation and [{\nii}] contamination $L(H\alpha)$ $\geq$ 10$^{37}$ erg$^{-1}$.
The two parameters are not correlated, and no trend with distance is observable.

\subsubsection{Composite diameter distribution}

\begin{figure}
\centering
\includegraphics[width=0.49\textwidth]{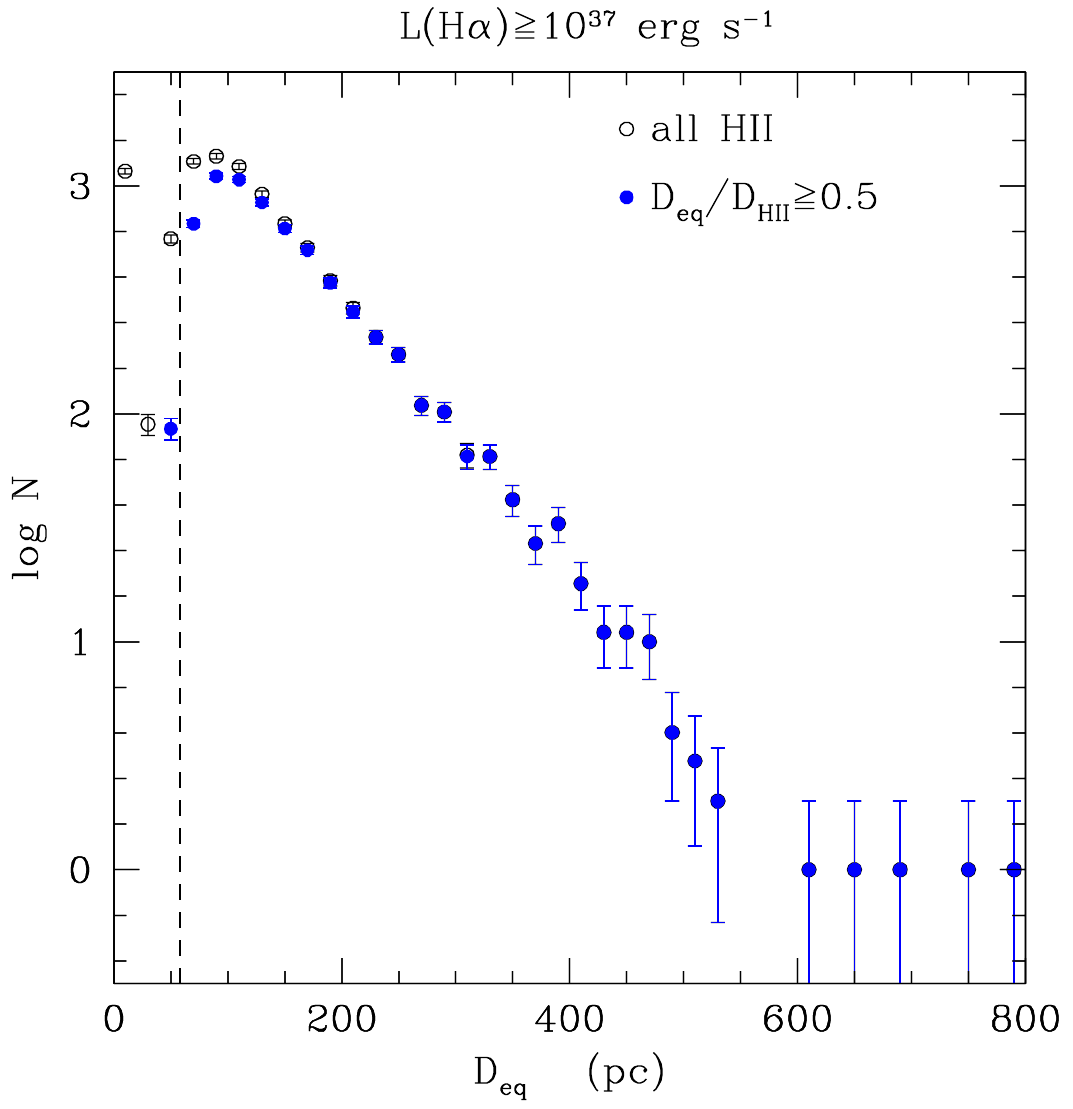}\\
\caption{Distribution of the observed (black dots) and corrected for seeing effects (blue dots) equivalent diameters 
for {\hii} regions with $L(H\alpha)$ $\geq$ 10$^{37}$ erg s$^{-1}$. The dashed vertical line shows the equivalent diameter corresponding
to the mean seeing of the survey at the distance of the cluster (16.5 Mpc).}
\label{Deffdist_nondef_lin}%
\end{figure}

\begin{figure}
\centering
\includegraphics[width=0.49\textwidth]{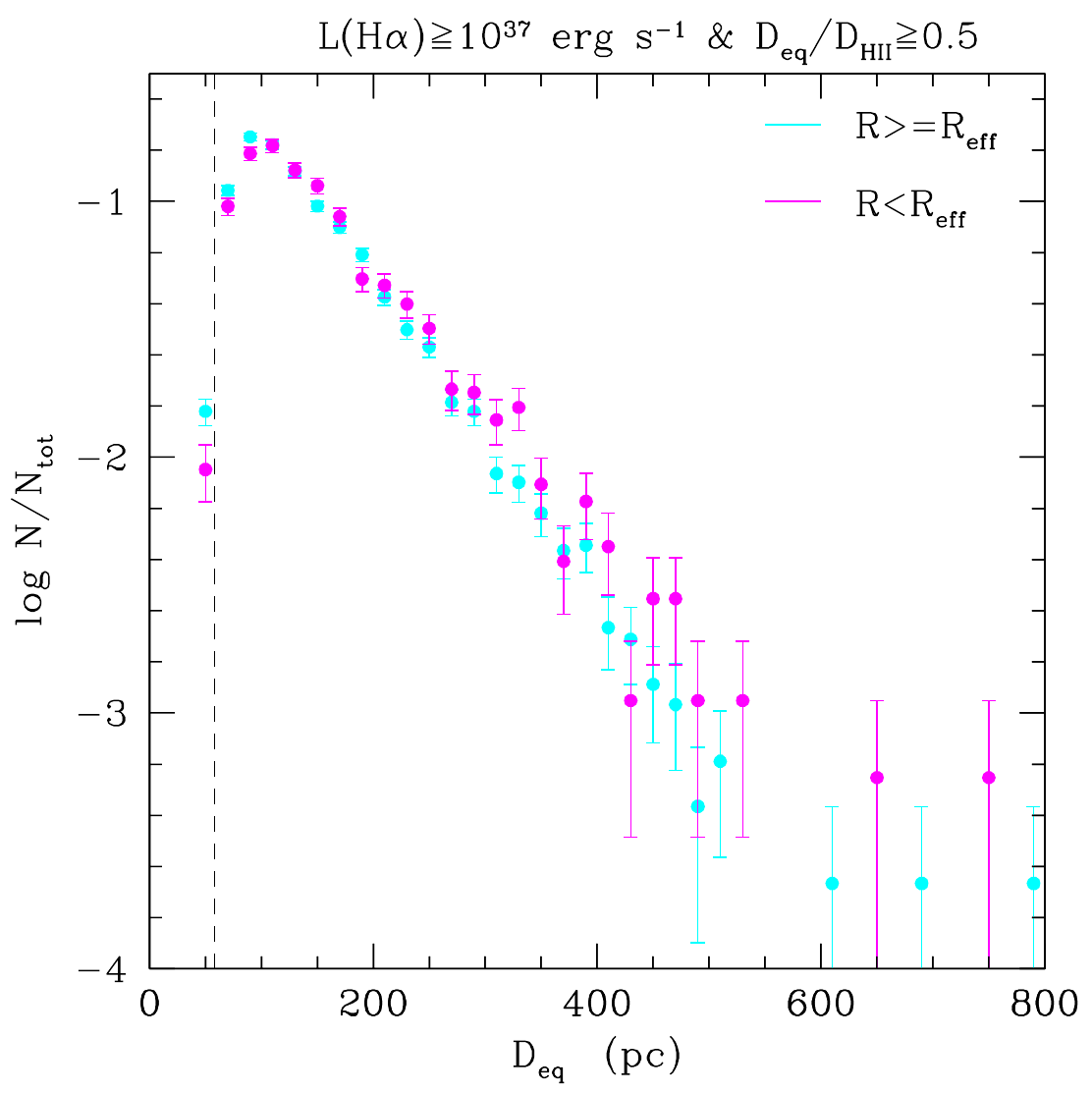}\\
\caption{Normalised distributions of the equivalent diameters corrected for seeing effects measured within (magenta) and outside (cyan) 
the $i$-band effective radius of the selected galaxies for {\hii} regions with $L(H\alpha)$ $\geq$ 10$^{37}$ erg s$^{-1}$. 
The dashed vertical line shows the equivalent diameter corresponding
to the mean seeing of the survey at the distance of the cluster (16.5 Mpc). }
\label{DeffdistReff_nondef_lin}%
\end{figure}

\begin{figure}
\centering
\includegraphics[width=0.49\textwidth]{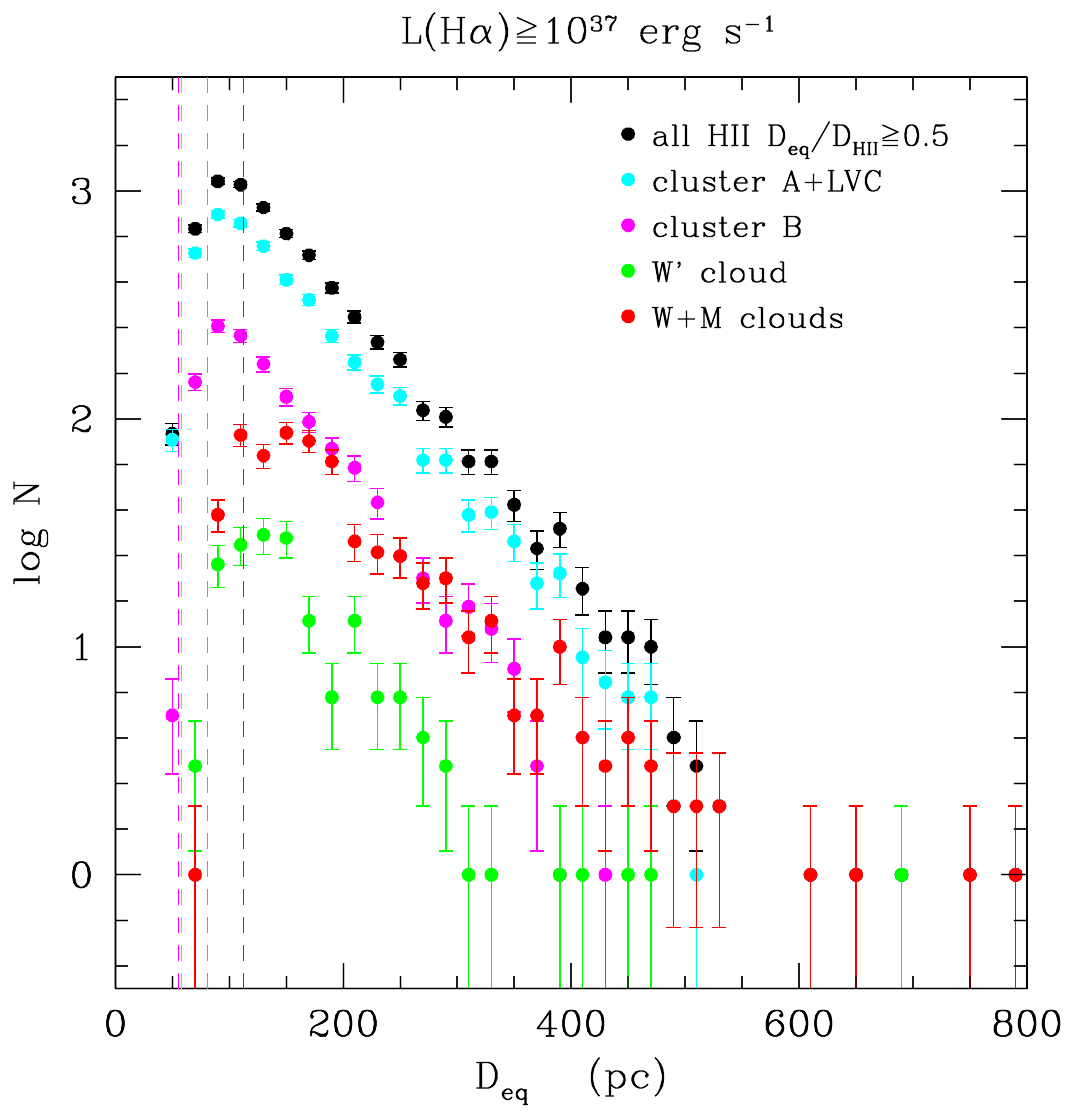}\\
\caption{Distribution of the equivalent diameters corrected for seeing effects equivalent diameters 
for {\hii} regions with $L(H\alpha)$ $\geq$ 10$^{37}$ erg s$^{-1}$, with different colours for galaxies belonging to 
different cluster substructures (cyan for cluster A and LVC, 16.5 Mpc; magenta for cluster B, 15.8 Mpc;
green for W' cloud, 23 Mpc; red for W and M clouds, 32 Mpc). The dashed vertical lines show the equivalent diameter corresponding
to the mean seeing of the survey at the distance of the different cluster substructures.}
\label{Deffdist_nondef_lin_dist}%
\end{figure}

Figure \ref{Deffdist_nondef_lin} shows the combined distribution of the observed and corrected for the seeing (6520 regions) 
equivalent diameters of {\hii} regions 
with an H$\alpha$ luminosity $L(H\alpha)$ $\geq$ 10$^{37}$ erg s$^{-1}$. The distribution shows a constant increase from 
$D_{eq}$ $\simeq$ 600 to 100 pc with a steep decrease below this limit. 
The difference between the observed and corrected for seeing effects distributions becomes significant below $D_{eq}$ $\lesssim$ 100 pc,
with an expected lower number of corrected measurements when the sample includes only moderately corrected values 
($D_{eq}/D_{HII}$ $\geq$ 0.5).
Figure \ref{DeffdistReff_nondef_lin} shows the distribution of the equivalent diameters corrected for seeing effects for {\hii} regions 
of luminosity  $L(H\alpha)$ $\geq$ 10$^{37}$ erg s$^{-1}$ located within (1834 objects) and outside (4686) the $i$-band 
effective radius of the galaxy. A Kolmogorov-Smirnov test indicates that the two distributions are statistically different (p-value = 3.7 $\times$ 10$^{-5}$). 
We observe a mild excess of extended regions
in the inner galaxy discs than in the outer regions. These differences, as the observed difference at small diameters ($D_{eq}$ $\lesssim$ 100 pc)
might be related to confusion and incompleteness more important in the crowded inner regions of the stellar disc. 
As for the luminosity function, we do not observe any evident variation of the equivalent diameter 
distribution with the distance of galaxies (Fig. \ref{Deffdist_nondef_lin_dist}).

\subsubsection{Composite electron density distribution}

\begin{figure}
\centering
\includegraphics[width=0.49\textwidth]{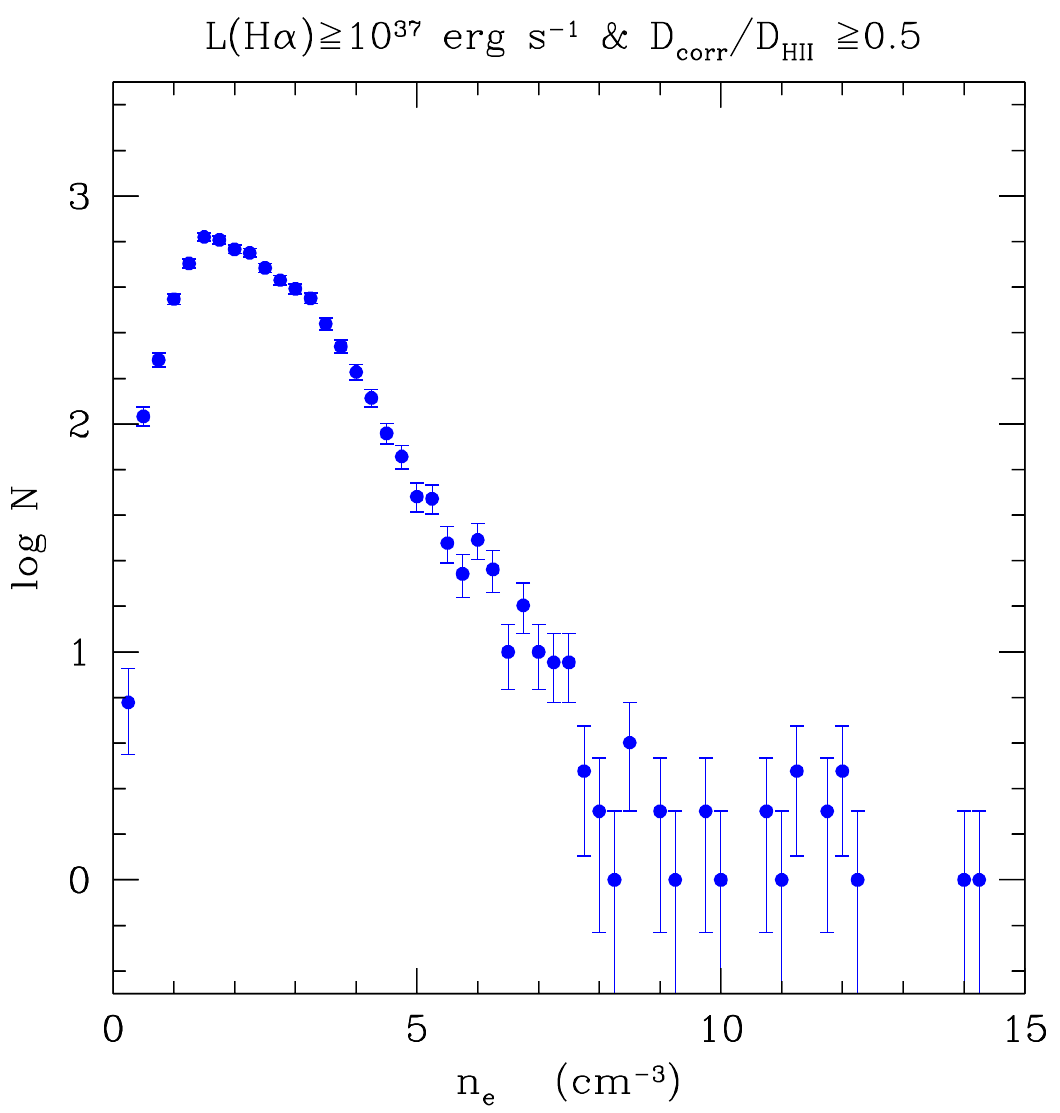}\\
\caption{Distribution of the electron density derived using equivalent diameters corrected for seeing effects 
for {\hii} regions with $L(H\alpha)$ $\geq$ 10$^{37}$ erg s$^{-1}$ and a diameter correction factor $\leq$ 50\%. }
\label{nedist_nondef}%
\end{figure}

\begin{figure}
\centering
\includegraphics[width=0.49\textwidth]{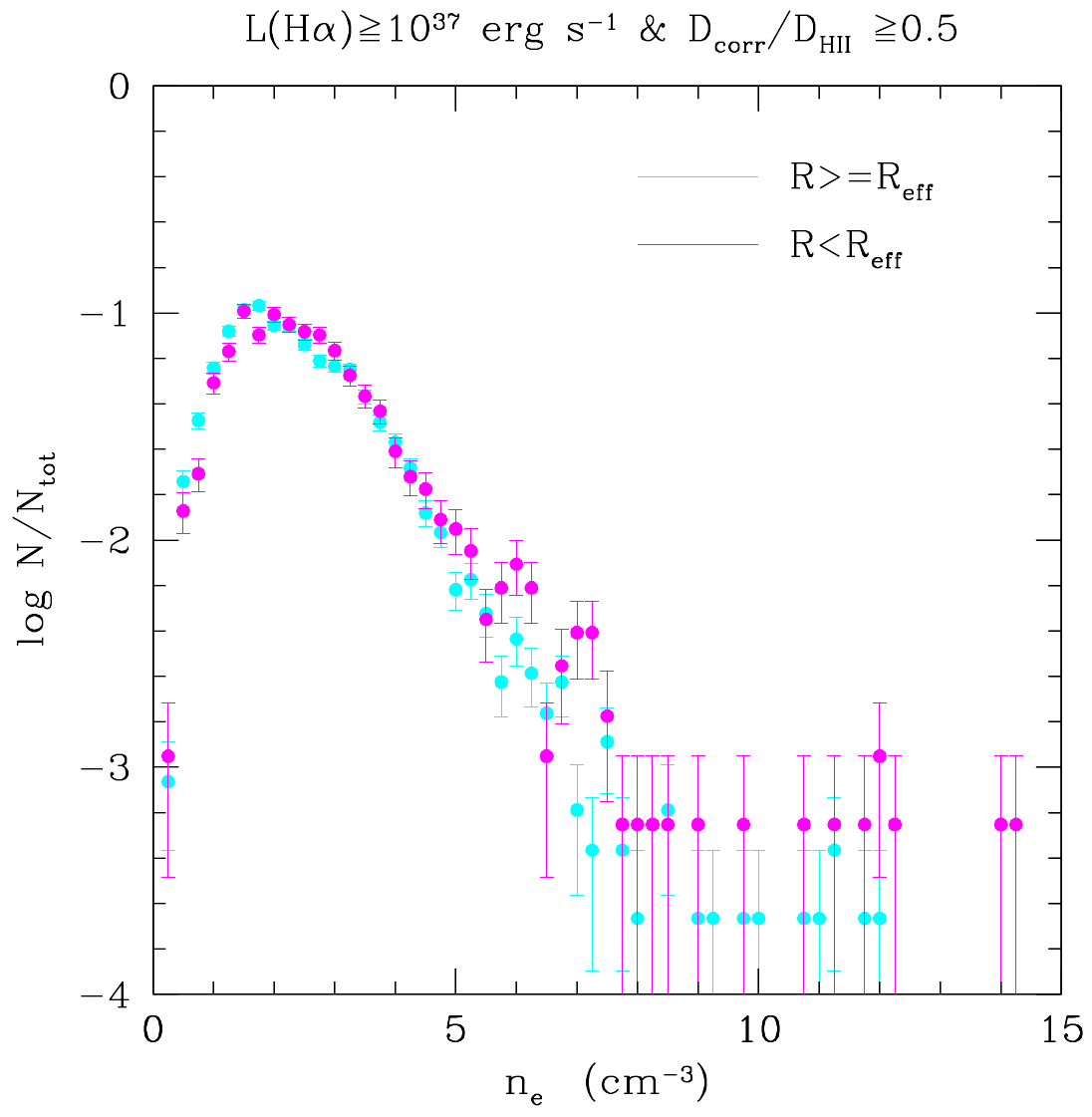}\\
\caption{Normalised distributions of the electron density derived using equivalent diameters corrected for seeing effects 
for {\hii} regions with $L(H\alpha)$ $\geq$ 10$^{37}$ erg s$^{-1}$ for {\hii} regions located inside (magenta) and
outside (cyan) the $i$-band effective radius. The two distributions are normalised to the total number of regions
located within and outside the effective radius. The number of {\hii} regions with densities $n_e$ $\geq$ 7.5 cm$^{-3}$ for the two samples
are most one per bin but are plotted at different positions because of the different normalisation.}
\label{nedistReff_nondef}%
\end{figure}

\begin{figure}
\centering
\includegraphics[width=0.49\textwidth]{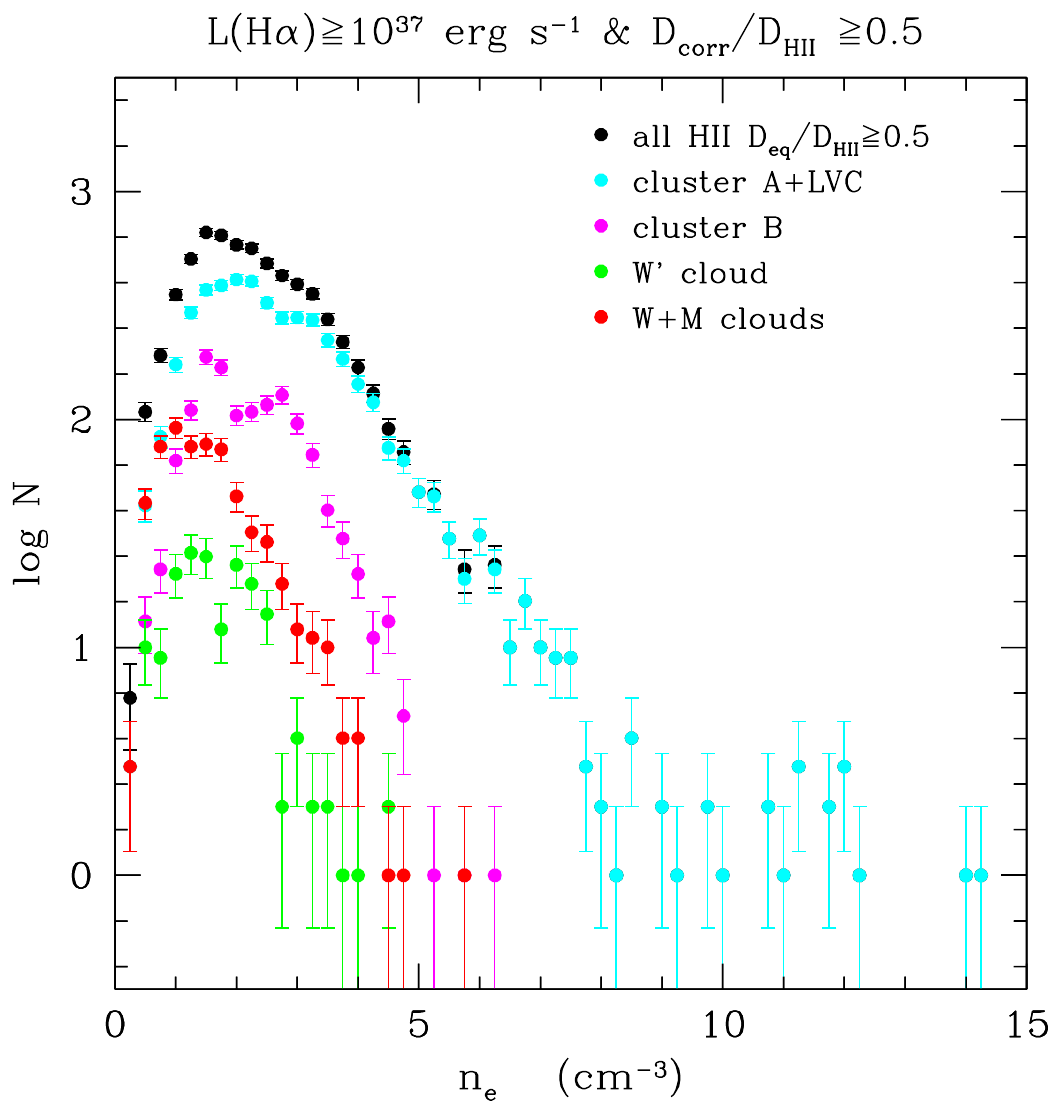}\\
\caption{Distribution of the electron density derived using equivalent diameters corrected for seeing effects 
for {\hii} regions with $L(H\alpha)$ $\geq$ 10$^{37}$ erg s$^{-1}$ and a diameter correction factor $\leq$ 50\%. 
Different colours are for galaxies belonging to the different Virgo cluster substructures (cyan for cluster A and LVC, 16.5 Mpc; magenta for cluster B, 15.8 Mpc;
green for W' cloud, 23 Mpc; red for W and M clouds, 32 Mpc).}
\label{nedist_nondef_dist}%
\end{figure}

Figure \ref{nedist_nondef} shows the distribution of the electron density derived using eq. (3) for {\hii} regions with 
$L(H\alpha)$ $\geq$ 10$^{37}$ erg s$^{-1}$ and equivalent diameters corrected for seeing effects with corrections not exceeding
50\%\ . The distribution is peaked at $n_e$ $\simeq$ 2 cm$^{-3}$, and sharply decreases up to  $n_e$ $\simeq$ 7-8 cm$^{-3}$,
where it reaches its minimum. There are a very few regions with densities $n_e$ $>$ 8 cm$^{-3}$.
Figure \ref{nedistReff_nondef} shows the same distribution derived for {\hii} regions located within and outside
the $i$-band effective radius of the parent galaxies. Although the two distributions look very similar,
a Kolmogorov-Smirnov test indicates that they are drown by statistically different populations ($p$-value = 10$^{-7}$). 
Again, we do not observe any clear dependence of the shape of the composite electron density distribution 
with the distance of galaxies located in the different cluster substructures (Fig. \ref{nedist_nondef_dist}).
We notice, however, that the distributions of $n_e$ look steeper in cluster B and 
W' than in the W and M clouds or in cluster A and LVC. We also notice that in the different subclusters the distributions 
are very noisy even at their peak where the statistic is sufficiently high, while the comparison at $n_e$ $\gtrsim$ 4 cm$^{-3}$ is impossible given 
the very limited number of {\hii} regions.

\subsubsection{Luminosity - diameter relation}

\begin{figure}
\centering
\includegraphics[width=0.49\textwidth]{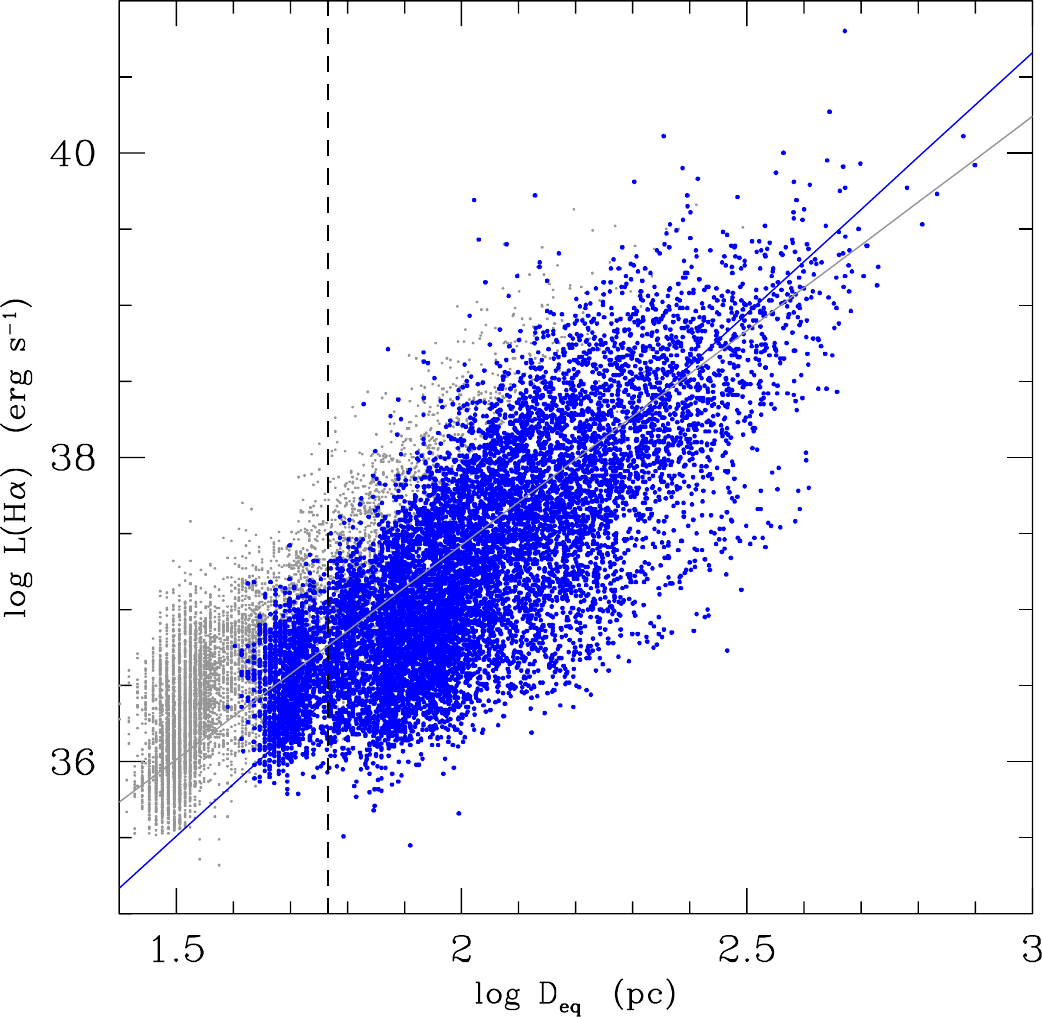}\\
\caption{Relation between the H$\alpha$ luminosity of individual {\hii} regions corrected for [{\nii}] contamination and dust 
attenuation and the equivalent diameter corrected for seeing effects. Blue filled dots are for {\hii} regions  
with a diameter correction factor $\leq$ 50\%, grey dots for
all {\hii} regions with no limits in diameter correction. The vertical dashed line shows the mean FWHM of the survey
assuming galaxies at the distance of the main body of the cluster (16.5 Mpc). The blue solid line gives the bisector fit (Isobe et al. 1990) of the relation
derived for {\hii} regions with a diameter correction factor $\leq$ 50\%\ (blue dots), the grey solid line
for {\hii} regions with no limits in the diameter correction.
}
\label{sizelum}%
\end{figure}

\begin{figure}
\centering
\includegraphics[width=0.49\textwidth]{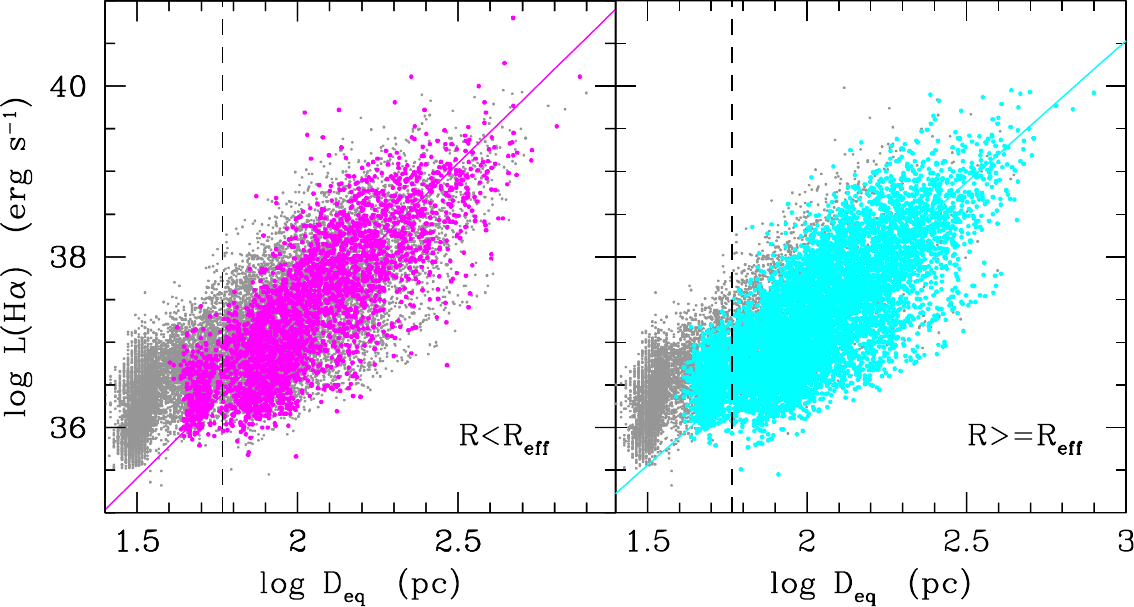}\\
\caption{Relation between the H$\alpha$ luminosity of individual {\hii} regions corrected for [{\nii}] contamination and dust 
attenuation and the equivalent diameter corrected for seeing effects for regions located within (right, filled magenta dots) and outside (left, filled cyan dots)
the $i$-band effective radius of the parent galaxy, where the correction is less than 50\%\ . Grey dots show
all {\hii} regions with no limits in luminosity and diameter correction. The vertical dashed line shows the mean FWHM of the survey
assuming galaxies at the distance of the main body of the cluster (16.5 Mpc). The magenta and cyan solid lines give the bisector fit of the relations
derived for galaxies with a diameter correction factor $\leq$ 50\%\ (magenta and cyan filled dots).
}
\label{sizelumReff}%
\end{figure}

\begin{figure}
\centering
\includegraphics[width=0.49\textwidth]{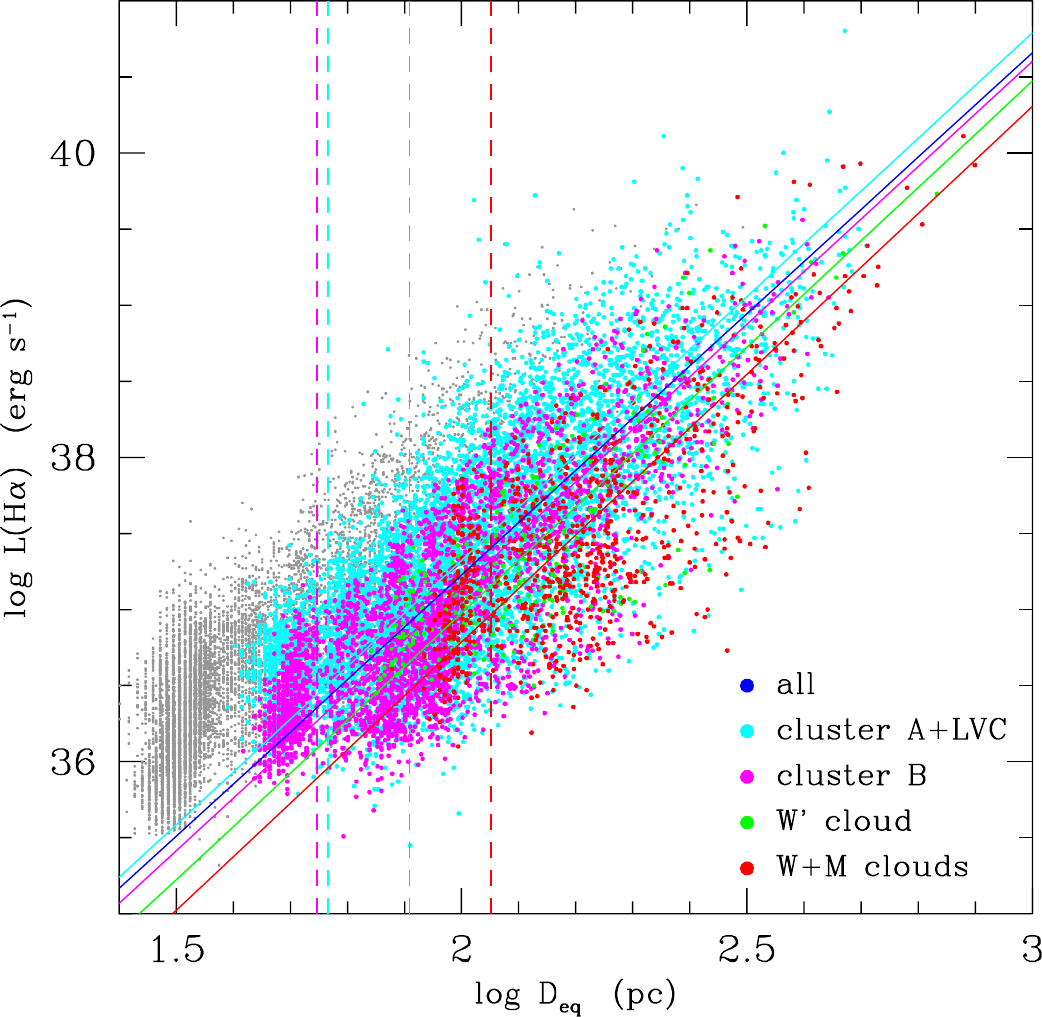}\\
\caption{Relation between the H$\alpha$ luminosity of individual {\hii} regions corrected for [{\nii}] contamination and dust 
attenuation and the equivalent diameter corrected for seeing effects. Coloured (cyan for cluster A and LVC, 16.5 Mpc; magenta for cluster B, 15.8 Mpc;
green for W' cloud, 23 Mpc; red for W and M clouds, 32 Mpc) dots are for {\hii} regions  
with a diameter correction factor $\leq$ 50\%, grey dots for
all {\hii} regions with no limits in diameter correction. The vertical dashed lines show the mean FWHM of the survey
assuming for the different substructures of the cluster. The coloured solid lines give the bisector fit of the relation
derived for galaxies with a diameter correction factor $\leq$ 50\%\ .
}
\label{sizelum_dist}%
\end{figure}

Figure \ref{sizelum} shows the relation between H$\alpha$ luminosity of individual {\hii} regions corrected for [{\nii}] contamination and dust 
attenuation and the equivalent diameter corrected for seeing effects. The two variables are tightly correlated, with the H$\alpha$ luminosity of the 
individual {\hii} regions increasing with the size. Figure \ref{sizelumReff} shows the same relation for {\hii} regions located within 
and outside the $i$-band effective radius of the parent galaxy. The best fit parameters are given in Table \ref{Tablumsize}. The relation is
flatter when no limits on the diameter correction are applied.
The difference between the two relations in the inner and outer disc regions seems statistically significant, with a steeper slope 
observed in the inner regions, although the two best fits are within the 1$\sigma$ dispersion of the two distributions. 
The slope of the relations is fairly constant while the intercepts slightly change when measured on the subsample of galaxies 
located at different distances (Fig. \ref{sizelum_dist}). The decrease of the intercept with increasing distance observed between 
cluster A+LVC, W', and W+M is expected given the slope of the relation ($\simeq$3.4-3.5) and that at further distances only 
the largest {\hii} regions are resolved, thus reducing the dynamic range of the relation. This, however, does not explain the observed difference between
cluster A+LVC and cluster B. We recall, however, that the difference between the two intercepts is just marginal ($\sim$ 2 sigma), as the difference in distance 
of the two subclusters A and B (0.7 Mpc) as measured by Cantiello et al. (2024) using mainly early-type galaxies. 
The distance of the star-forming systems hosting the {\hii} regions analysed in this work might not exactly match that of early-types since most of them 
are now infalling for the first time into the cluster (e.g., Gavazzi et al. 1999). Any interpretation of the observed difference in the intercept of the $L(H\alpha)$ vs. $D_{eq}$ relation
between cluster A and B in terms of distance of galaxies should thus be done with extreme caution.

\begin{table}
\caption{Best fit parameters for the luminosity-size relation (log $L(H\alpha)$ = $a$ $\times$ log $D_{eq}$ + $b$)
}
\label{Tablumsize}
{
\[
\begin{tabular}{cccccc}
\hline
\noalign{\smallskip}
\hline
Sample					& N.objects		&  $a$  	& $b$   	& $\rho$	& $\sigma$ 	\\
					&			&		&		&	&		\\
\hline
All$^a$					& 18474			& 2.82$\pm$0.01	& 31.79$\pm$0.04& 0.79	& 0.17   	 \\		 
Diam. corr.$<$50\%\ 			& 11263			& 3.43$\pm$0.02	& 30.36$\pm$0.05& 0.76	& 0.14   	 \\
$R$ $\geq$ $R_{eff}$			& 8268			& 3.32$\pm$0.02 & 30.58$\pm$0.06& 0.74	& 0.14   	 \\
$R$ $<$ $R_{eff}$			& 2995			& 3.69$\pm$0.02	& 29.88$\pm$0.10& 0.81	& 0.13   	 \\
cluster A+LVC				& 6970                  & 3.47$\pm$0.02 & 30.38$\pm$0.07& 0.76	& 0.14   	 \\
cluster B				& 3261			& 3.46$\pm$0.02 & 30.23$\pm$0.09& 0.77	& 0.12	 	 \\
W' cloud				& 239			& 3.50$\pm$0.08 & 29.98$\pm$0.37& 0.70	& 0.11	 	 \\
W+M clouds				& 793			& 3.52$\pm$0.05 & 29.74$\pm$0.23& 0.61	& 0.14	 	 \\
\hline
\end{tabular}
\]
Notes: fits are done for all {\hii} regions with a diameter correction factor $\leq$ 50\%\ , unless otherwise stated. No limits on the H$\alpha$
luminosity is taken.
$\rho$ is the Spearman correlation coefficient, $\sigma$ the dispersion perpendicular to the fitted relation.\\
$a$ : no limits in the diameter correction.
}
\end{table}

\subsection{Scaling relations}

We analyse in this section the main scaling relations between several representative properties of the {\hii} regions
and the main properties of their parent galaxies for a statistically significant sample of unperturbed systems spanning a wide range 
in stellar mass and morphological type. H$\alpha$ luminosities are corrected for [{\nii}] contamination and dust extinction as 
indicated in Sec. 3.2.

\begin{figure*}
\centering
\includegraphics[width=0.49\textwidth]{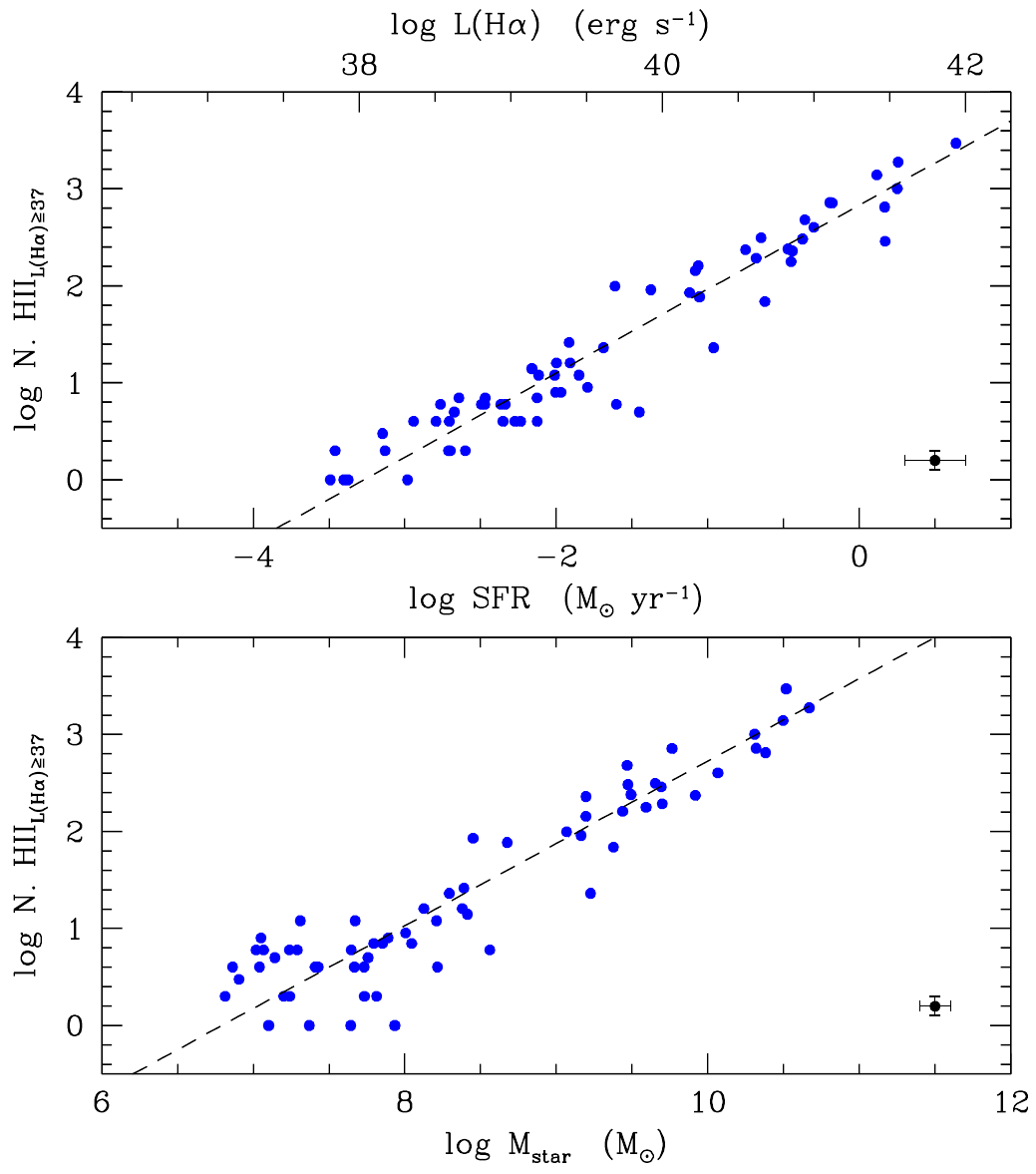}
\includegraphics[width=0.49\textwidth]{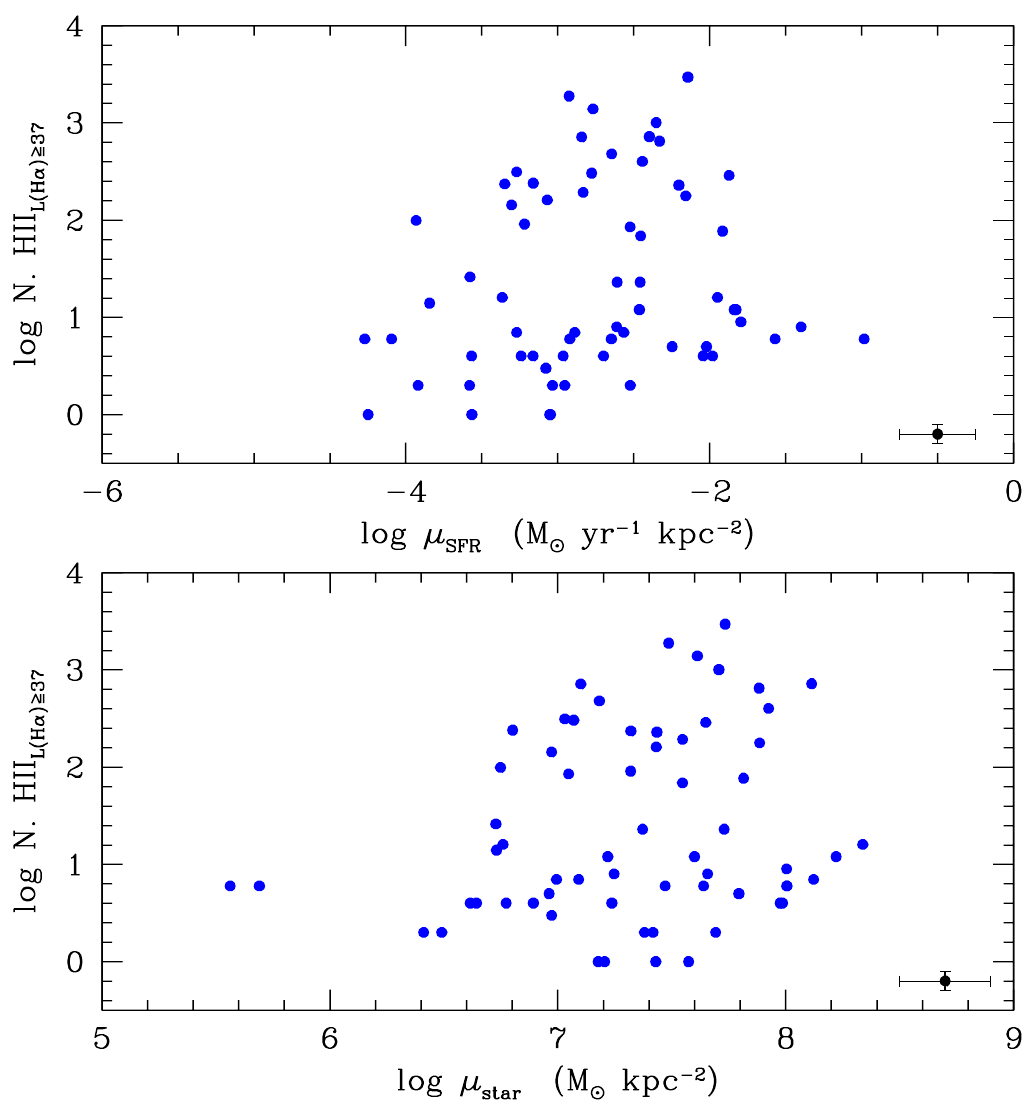}\\
\caption{Relation between the total number of {\hii} regions detected by \textsc{HIIphot} with $L(H\alpha)$ $\geq$ 10$^{37}$ erg s$^{-1}$ and the total star
formation rate (upper left panel), total stellar mass (lower left), mean star
formation rate surface density (upper right), and mean stellar mass surface density (lower right) of the host galaxies. Star formation rates (lower axis) have been derived from 
H$\alpha$ luminosities (upper axis) corrected for dust attenuation and [{\nii}] contamination assuming a Chabrier IMF.
The black dashed line shows the best fit to the data (bisector fit). The dot in the lower right corner shows the typical error bar in the data.}
\label{scaling1N37}%
\end{figure*}

Figure \ref{scaling1N37} shows the relation between the total number of {\hii} regions with $L(H\alpha)$ $\geq$ 10$^{37}$ erg s$^{-1}$
and the total star formation rate (upper left panel), stellar mass (lower left), mean star formation rate (upper right) and 
stellar mass density (lower right) within the $i$-band effective radius of the parent galaxies. As expected, 
the number of {\hii} regions linearly increases with the total star formation rate and the stellar mass since 
'bigger galaxies have more of everything" (Kennicutt 1990). The parameters of the best fit of Fig. \ref{scaling1N37}
and of the following Fig. \ref{scaling1Lup}-\ref{SSFR} are shown in the Figures and given in Table \ref{Tabscalingfit} 
only whenever the probability $P$ that the two variables are correlated with $P$ $\geq$ 95\%\ ($p$-value $<$0.05).
All the following figures relative to the scaling relations discussed in this and in the next sections are given in Appendix \ref{appC}.

A scattered diagram is observed whenever the total number of {\hii} regions with $L(H\alpha)$ $\geq$ 10$^{37}$ erg s$^{-1}$ is plotted 
vs. the logarithm of the star formation and of the stellar mass surface densities, where these two normalised entities are defined as the mean star formation rate
and stellar mass within the deprojected effective radius $R_{eff}$ here measured in the $i$-band (radius containing 50\% of the $i$-band light)
as described in Ferrarese et al. (2020), and available to the NGVS team; see Fig. \ref{scaling1N37}):

\begin{equation}
{\mu_{SFR} = \frac{SFR}{2 \pi R_{eff}^2}  ~~~~~~~~ \rm{[M_{\odot} yr^{-1} kpc^{-2}]}}
\end{equation}

\noindent
and

\begin{equation}
{\mu_{star} = \frac{M_{star}}{2 \pi R_{eff}^2}  ~~~~~~~~ \rm{[M_{\odot} kpc^{-2}]}}
\end{equation}

\noindent
We stress that the mean surface densities defined in eq. (5) and (6) uses $i$-band effective radii and not radii
measured on the H$\alpha$ and stellar mass 2D distributions. While the $i$-band distribution well traces that of stellar mass, 
this is not necessary the case for the distribution of star forming regions. Indeed, Equation (5) is an hybrid definition roughly measuring
the star formation activity normalised to the size of the stellar disc as defined by the old stellar population, and should thus be considered with caution.
We adopted this hybrid definition since more indicated than the standard mean star formation rate surface density definition (star formation
rate divided by the size of the star forming disc measured in H$\alpha$) to quantify the truncation of the star forming disc
with respect to that of the stellar disc occurring in ram pressure stripped galaxies (Boselli et al. 2022), something which
will be fully analysed in a future VESTIGE paper.



Figure \ref{scaling1Lup} shows the relations between the H$\alpha$ 
luminosity of the brightest {\hii} region and the total star formation rate, 
stellar mass (and their surface densities) of the parent galaxies. The same figures done using the mean luminosity of the first three ranked {\hii} regions
and the same galaxy parameters are also shown in Appendix \ref{appC}.
The H$\alpha$ luminosity of the brightest {\hii} region is
tightly related to the star formation rate, stellar mass, mean star formation rate and stellar mass surface density of their host.
The brightest {\hii} regions with luminosities $L(H\alpha)$ $\simeq$ 10$^{40}$ erg s$^{-1}$ are preferentially located in those systems 
with the highest star formation rate.

Figure \ref{scaling1L37} shows the relation between $L(H\alpha)_{\geq 37}/L(H\alpha)$, defined as 
the ratio between the sum of the H$\alpha$ luminosity of all the {\hii} regions 
with luminosity $L(H\alpha)$ $\geq$ 10$^{37}$ erg s$^{-1}$ and the integrated H$\alpha$ luminosity of the galaxy, and the total star formation rate, 
the total stellar mass and their surface densities of the host galaxies. The ratio $L(H\alpha)_{\geq 37}/L(H\alpha)$ roughly corresponds to 
the fraction of H$\alpha$ emission due to compact {\hii} regions, and is thus a direct estimate of the DIG in these normal, star forming systems.
This ratio is always in the range -0.7 $\lesssim$ log$L(H\alpha)_{\geq 37}/L(H\alpha)$ $\lesssim$ 0 and does not changes with stellar mass and star formation rate,
but rather increases with increasing star formation rate and stellar mass mean surface densities, suggesting that the contribution of the DIG
increases in those galaxies where the density of stars is low (i.e. low surface brightness systems), which are also objects where 
the star formation rate surface density is low given that the two variables are also correlated (see Appendix \ref{appD}).


Figure \ref{scaling1L37physical} shows the dependence of the ratio on three different parameters characterising the physical properties of the ISM
where stars are formed: the metallicity of the gas, the fraction of molecular-to-atomic gas, and the total gas-to-dust ratio. With the only exception 
of the metallicity, which is available for 39/64 galaxies of the unperturbed sample, the two other parameters are limited to the most massive galaxies
where both the molecular gas (CO) and dust masses are measured. This limits the sampled parameter space. Figure \ref{scaling1L37physical}
shows that the $L(H\alpha)_{\geq 37}/L(H\alpha)$ ratio is fairly constant in galaxies having a molecular-to-atomic gas ratio in the range
0.03 $\lesssim$ $M_{H_2}/M_{HI}$  $\lesssim$ 1.6 and a total gas-to-dust ratio 100 $\lesssim$ $G/D$ $\lesssim$ 650. On the contrary,
the ratio decreases with increasing metallicity, from $L(H\alpha)_{\geq 37}/L(H\alpha)$ $\simeq$ 0.9 at 12$+$ log O/H $\simeq$ 8.1 to
$L(H\alpha)_{\geq 37}/L(H\alpha)$ $\simeq$ 0.4 at 12$+$ log O/H $\simeq$ 8.7.

Figure \ref{scaling2Lup} shows the relation between $L(H\alpha)_{1}/L(H\alpha)$, the ratio between the luminosity of the brightest
{\hii} region and the total H$\alpha$ luminosity of the galaxy, and the integrated star formation rate and stellar mass of the host.
We recall that out of the 66 galaxies with $HI-def$ $\leq$ 0.4, only two have $L(H\alpha)_{1}$ $<$ 10$^{37}$ erg s$^{-1}$
and are thus excluded from the unperturbed sample. As defined, the sample is thus not biased to those galaxies with bright {\hii} regions.
Similar figures with the contribution of the brightest three {\hii} regions to the total H$\alpha$ luminosity of the galaxy are shown in Appendix \ref{appC}. 
Figure \ref{scaling2Lup} indicates that the contribution of the brightest
{\hii} regions to the total H$\alpha$ emission of galaxies strongly increases from $\simeq$ 1-10\%\ in star forming ($SFR$ $\simeq$ 1 M$_{\odot}$ yr$^{-1}$)
and massive ($M_{star}$ $\simeq$ 10$^{10}$ M$_{\odot}$) objects to $\simeq$ 50-100\%\ in dwarf systems ($SFR$ $\simeq$ 10$^{-3}$ M$_{\odot}$ yr$^{-1}$;
$M_{star}$ $\simeq$ 10$^{7}$ M$_{\odot}$).


Figure \ref{scaling1Lupphysical} shows the relation $L(H\alpha)_{1}/L(H\alpha)$ and the metallicity, the molecular-to-atomic gas ratio, and the
total gas-to-dust ratio. The poor statistic (15 objects) prevents us to see whether $L(H\alpha)_{1}/L(H\alpha)$ is related to $M_{H_2}/M_{HI}$ and $M_{gas}/M_{dust}$,
but Fig. \ref{scaling1Lupphysical} clearly shows an anti-correlation with the gas metallicity.



Figure \ref{SSFR} shows the relation between the ratio of the total H$\alpha$ luminosity of {\hii} regions with $L(H\alpha)$ $\geq$ 10$^{37}$ erg s$^{-1}$  
($L(H\alpha)_{\geq 37}/L(H\alpha)$), of the brightest {\hii} region ($L(H\alpha)_{1}/L(H\alpha)$), and of the sum of the H$\alpha$ luminosity of the brightest 
three {\hii} regions ($L(H\alpha)_{3}/L(H\alpha)$, measured only for those galaxies with at least 3 {\hii} regions with $L(H\alpha)$ $\geq$ 10$^{37}$ erg s$^{-1}$,
where $L(H\alpha)_{3}$ is the contribution of the three brightest {\hii} regions) 
and the integrated H$\alpha$ luminosity and the specific star formation rate of the parent galaxies ($SSFR$). 
The three variables are only marginally related to the specific star formation rate. While at low specific star formation rates (log$SSFR$ $\lesssim$ -10.5 yr$^{-1}$)
all ratios with -0.7 $\lesssim$ log$L(H\alpha)_{\geq 37}/L(H\alpha)$ $\lesssim$ 0, -2 $\lesssim$ log$L(H\alpha)_{1}/L(H\alpha)$ $\lesssim$ 0, 
and -1.6 $\lesssim$ log$L(H\alpha)_{3}/L(H\alpha)$ $\lesssim$ 0 are observed, in those systems with the highest specific star formation rate
only the highest ratios are present, suggesting that only in these extreme systems the contribution of all the individual {\hii} regions,
or that of the brightest ones, to the total ionising radiation of galaxies is dominant.

\subsubsection{Relations with morphological type}

We checked whether the main normalised properties of {\hii} regions change in different morphological classes, as previously stated in the literature (e.g., Kennicutt et al. 1989a;
see Fig. \ref{density37type} in Appendix \ref{appC}).
We do not observe any clear variation of the density of {\hii} regions of luminosity $L(H\alpha)$ $\geq$ 10$^{37}$ erg s$^{-1}$ per unit disc size with morphological type, where the size of the galaxy is measured 
using the optical diameter at the 25.5 isophote (extracted from Binggeli et al. 1985 whenever available or derived from the $g$-band NGVS effective radius
using a relation previously calibrated on objects with both measurements available). The only exception are BCD galaxies where the density of {\hii} regions is 
increased with respect to that observed in the other morphological classes. We also do not observe any variation of the number of {\hii} regions
of luminosity $L(H\alpha)$ $\geq$ 10$^{37}$ erg s$^{-1}$ per unit stellar mass with the morphological type. 


\subsubsection{Fit parameters of the luminosity function on individual galaxies}

We fitted the H$\alpha$ luminosity function of {\hii} regions in individual galaxies using the same parametric Schechter (1976) function given in eq. (4).
Because the observed shape of the non-parametric luminosity function abruptly decreases below $L(H\alpha)$ $\lesssim$ 10$^{37}$ erg s$^{-1}$ (see Sec. 3.5), 
we limited the fit to {\hii} regions of luminosity 
$L(H\alpha)$ $\geq$ 10$^{37}$ erg s$^{-1}$. We also limited the fit to those objects with at least 20 {\hii} regions detected above 
this luminosity limit (27 objects, all with $M_{star}$ $\gtrsim$ 10$^8$ M$_{\odot}$). Figure \ref{scaling1alfa} to \ref{scaling1phistar} (Appendix \ref{appC}) show how the fitted 
characteristic parameters $\alpha$, $L^*$, and $\Phi^*$
are related to the total star formation rate, stellar mass, mean star formation rate and stellar mass surface brightness.
The best fit parameters of the Schechter function are given in Table \ref{galLF}, while the fitted functions are compared to the
data in Fig. \ref{LFindividual} presented in Appendix \ref{appE}.

Figure \ref{scaling1alfa} shows that slope of the faint end of the luminosity function is only barely related to the 
total star formation rate of the host but not to their total stellar mass. On the contrary, the 
slope $\alpha$ is tightly related to the mean star formation and stellar mass 
surface densities of their host galaxies. Objects with a high surface density of star forming regions or evolved stars
have, on average, a flatter slope of the H$\alpha$ luminosity function of {\hii} regions at the faint end than those characterised by low densities of recently formed or evolved stars. 
On the contrary, the characteristic luminosity $L^*$ is not related to the star formation rate, stellar mass, and their mean surface densities (Fig.
\ref{scaling1Lstar}). The characteristic parameter $\Phi^*$, which gives the number of {\hii} regions at $L^*(H\alpha)$, 
is related, although with a significant dispersion, to
the total star formation rate, stellar mass, and mean star formation rate and stellar mass 
surface density of the host galaxies (Fig. \ref{scaling1phistar}). The relation with the stellar mass and the star formation rate 
is expected being $\Phi^*$ a scale parameter counting the number of {\hii} regions in each object. Out of these three parameters, only $\alpha$ is barely
related to the specific star formation rate of the host galaxies (Fig. \ref{SSFRLF}).
We also investigated the relationship between $\alpha$, $L^*$, and $\Phi^*$, and three parameters characterising the properties of the ISM, metallicity, 
molecular-to-atomic, and total gas-to-dust ratio, but we did not found any clear relation.


\begin{table*}
\caption{Coefficients of the scaling relations: y = ax + b (dashed lines in Figs. \ref{scaling1N37}  to \ref{SSFRLF})
}
\label{Tabscalingfit}
{
\[
\begin{tabular}{ccccccc}
\hline
\noalign{\smallskip}
\hline
x							&y					&  a     		& b     	& $\rho$ &$\sigma$ &	$P$-value	\\
\hline
log$SFR$ [M$_{\odot}$ yr$^{-1}$]			& log$N. HII_{\geq 37}$			& 0.87$\pm$0.04   	& 2.83$\pm$0.07	& 0.94  & 0.21 & $<$0.0005   \\ 		
							& log$L(H\alpha)_{1}$			& 0.74$\pm$0.06		& 40.03$\pm$1.48& 0.92	& 0.29 & $<$0.0005   \\ 		
							& log$\langle L(H\alpha)_3\rangle$	& 0.76$\pm$0.05		& 39.85$\pm$1.42& 0.94	& 0.24 & $<$0.0005   \\ 		
							& log$L(H\alpha)_{1}/L(H\alpha)$	&-0.48$\pm$0.11		& -1.63$\pm$0.13&-0.67	& 0.34 & $<$0.0005   \\ 		
							& log$L(H\alpha)_3/L(H\alpha)$		&-0.44$\pm$0.13		& -1.25$\pm$0.12&-0.64	& 0.30 & $<$0.0005   \\ 		
							& $\alpha$				& 0.75$\pm$0.20		& -1.22$\pm$0.29& 0.37	& 0.40 & 0.05        \\ 		
							& $\Phi^*$				& 1.76$\pm$0.17		& 1.49$\pm$0.16	& 0.72	& 0.44 & $<$0.0005   \\ 		
\hline
log$M_{star}$ [M$_{\odot}$]				& log$N. HII_{\geq 37}$			& 0.85$\pm$0.05  	& -5.78$\pm$0.21& 0.88  & 0.27 & $<$0.0005   \\ 		
							& log$L(H\alpha)_{1}$			& 0.73$\pm$0.09		& 32.62$\pm$2.66& 0.69	& 0.50 & $<$0.0005   \\ 		
							& log$\langle L(H\alpha)_3\rangle$	& 0.68$\pm$0.10		& 32.89$\pm$2.78& 0.76	& 0.43 & $<$0.0005   \\ 		
							& log$L(H\alpha)_{1}/L(H\alpha)$	&-0.46$\pm$0.10		&  3.02$\pm$0.21&-0.79	& 0.29 & $<$0.0005   \\ 		
							& log$L(H\alpha)_3/L(H\alpha)$		&-0.39$\pm$0.12		&  2.70$\pm$0.16&-0.80	& 0.25 & $<$0.0005   \\ 		
							& $\Phi^*$				& 1.63$\pm$0.20		&-15.14$\pm$1.52& 0.57	& 0.57 & 0.005       \\ 		
\hline
log$\mu_{SFR}$ [M$_{\odot}$ yr$^{-1}$ kpc$^{-2}$]	& log$L(H\alpha)_{1}$			& 1.13$\pm$0.11		& 41.92$\pm$1.92& 0.60	& 0.49 & $<$0.0005   \\ 		
							& log$\langle L(H\alpha)_3\rangle$	& 1.08$\pm$0.12		& 41.66$\pm$2.36& 0.43	& 0.55 & $<$0.0005   \\ 		
							& log$L(H\alpha)_{\geq 37}/L(H\alpha)$	& 0.28$\pm$0.21		& 0.55$\pm$0.10 & 0.66	& 0.16 & $<$0.0005   \\ 		
							& $\alpha$				& 0.84$\pm$0.13		& 0.63$\pm$0.29 & 0.76	& 0.23 & $<$0.0005   \\ 		
							& $\Phi^*$				& 2.15$\pm$0.21		& 6.33$\pm$0.53 & 0.61	& 0.43 & 0.001   \\ 		
\hline
log$\mu_{star}$ [M$_{\odot}$ kpc$^{-2}$]		& log$L(H\alpha)_{1}$			& 1.28$\pm$0.13		& 29.39$\pm$2.20& 0.39	& 0.53 & 0.002   \\ 		
							& log$\langle L(H\alpha)_3\rangle$	& 1.22$\pm$0.13		& 29.77$\pm$2.44& 0.37	& 0.51 & 0.002   \\ 		
							& log$L(H\alpha)_{\geq 37}/L(H\alpha)$	& 0.44$\pm$0.20		& -3.48$\pm$0.19& 0.39	& 0.22 & 0.002   \\ 		
							& $\alpha$				& 1.10$\pm$0.16		& -9.80$\pm$0.79& 0.63	& 0.26 & 0.001   \\ 		
							& $\Phi^*$				& 2.70$\pm$0.28		&-19.56$\pm$1.92& 0.61	& 0.36 & 0.001   \\ 		
\hline
\noalign{\smallskip}
log$SSFR$  [yr$^{-1}$]					& log$L(H\alpha)_{\geq 37}/L(H\alpha)$	& 0.46$\pm$0.18		&  4.37$\pm$0.26& 0.45	& 0.20 & $<$0.0005   \\ 		
							& log$L(H\alpha)_{1}/L(H\alpha)$	& 1.02$\pm$0.12		&  9.50$\pm$0.70& 0.27	& 0.43 & 0.023   \\ 		
							& log$L(H\alpha)_3/L(H\alpha)$		& 0.99$\pm$0.12		&  9.36$\pm$0.66& 0.46	& 0.32 & 0.001   \\ 		
							& $\alpha$				& 1.23$\pm$0.16		& 10.77$\pm$1.26& 0.49	& 0.27 & 0.01    \\ 		
\hline
\noalign{\smallskip}
12+log (O/H)						& log$L(H\alpha)_{\geq 37}/L(H\alpha)$	&-0.68$\pm$0.17		&  5.46$\pm$0.39&-0.53	& 0.14 & 0.015   \\  	       
							& log$L(H\alpha)_{1}/L(H\alpha)$	&-2.11$\pm$0.12		& 16.75$\pm$0.82&-0.85	& 0.13 & $<$0.0005   \\  	       
							& log$L(H\alpha)_3/L(H\alpha)$		&-1.81$\pm$0.11		& 14.53$\pm$0.72&-0.86	& 0.13 & $<$0.0005   \\  	       
\hline
\noalign{\smallskip}
log G/D							& log$L(H\alpha)_{\geq 37}/L(H\alpha)$	&0.58$\pm$0.28		&  -1.53$\pm$0.19& 0.59	& 0.09 & 0.04   \\ 		
\hline
\end{tabular}
\]
Notes: bisector fit. The best fit parameters are given here and shown in the Figures only whenever the probability that the 
variables are correlated is larger than 95\%\ ($p$-value of the null hypothesis $<$ 0.05).
$\rho$ gives the Spearman correlation coefficient, $\sigma$ the dispersion in the relations. Relations with $L(H\alpha)_{\geq 37}$, $L(H\alpha)_{1}$ are derived for the 
whole unperturbed sample of 64 objects, those with $L(H\alpha)_3$ for a sample of 55 objects, 
with metallicity 39 objects, with the $G/D$ ratio 15 objects, and with the output of the fit of the luminosity function 27 objects, respectively. $p$-values are derived using the Spearman's ranked
correlation coefficient (Scipy Python package). \\
}
\end{table*}

\subsection{Radial variations}

We also study the variation of the properties of {\hii} regions with their deprojected radial distance from the galaxy centre.
For this purpose, we divide {\hii} regions located inside and outside the effective radius measured in the $i$-band.
We do not observe any relation between the ratio of number of {\hii} regions with luminosity $L(H\alpha)$ $\geq$ 10$^{37}$ erg s$^{-1}$
located within and outside the $i$-band effective radius and the morphological type, star formation rate, and stellar mass (and their associated surface densities) of the host galaxies.
The ratio distribution has a mean value of log$\frac{N(HII)_{in}}{N(HII)_{out}}$ = -0.15$\pm$0.42, corresponding to $\frac{N(HII)_{in}}{N(HII)_{out}}$ = 0.7.

\section{Discussion}

The main purpose of this work is that of providing a reference for future studies. Most of the relations analysed in this work have been already studied in dedicated papers.
Their physical interpretation has been extensively discussed in these works, to which we address the interested readers.
Here we will briefly discuss the main results of this analysis in comparison with what already presented in the literature.

\subsection{Strengths and weaknesses of the sample}

The statistical analysis presented in this work is based on a sample of 64 unperturbed star-forming galaxies with 13278 {\hii} regions 
of luminosity $L(H\alpha)$ $\geq$ 10$^{37}$ erg s$^{-1}$ located at the periphery of the Virgo cluster. Despite this large statistic, the data used in this work
have several limitations which should be considered in the discussion. First of all, the H$\alpha$ luminosities of individual {\hii} regions are corrected for [{\nii}] contamination 
and dust attenuation using a single value which changes from galaxy to galaxy. This is not the case in similar studies based on samples of {\hii} regions observed with IFU spectroscopy
such as those recently conducted or under way using VLT/MUSE (PHANGS, Santoro et al. 2022, Groves et al. 2023; MAUVE), CFHT/SITELLE (SIGNALS, Rousseau-Nepton et al. 2018, 2019), 
or Calar Alto/PPAK (CALIFA, S{\'a}nchez et al. 2015, Espinosa-Ponce et al. 2020). IFU spectroscopy allows for an accurate determination of [{\nii}] contamination and dust attenuation
in each individual {\hii} region. IFU data are thus more suitable to trace variations within the disc of galaxies, where metallicity and dust attenuation gradients are present.

Our set of data, however, has several other advantages. First of all, it is composed of galaxies located all at almost the same distance with a unique set of data all of similar quality 
in terms of sensitivity and angular resolution. It includes objects spanning a wide range in morphological type (from lenticulars to spirals, Magellanic irregulars and BCDs)
and stellar mass (10$^{7}$ $\lesssim$ $M_{star}$ $\lesssim$ 10$^{11}$ M$_{\odot}$). It is thus ideally suited to study scaling relations and to become a reference for 
future comparative analyses. We recall that, at this sensitivity, the data allow us to detect a single early-B ionising star ($L(H\alpha)$ $\simeq$ 10$^{36}$ erg s$^{-1}$)
if located in uncrowded regions, and to resolve {\hii} regions of diameter $\gtrsim$ 60 pc. Up to now, these low luminosities have been reached on statistically significant samples
only in a handful of galaxies belonging to the Local Group (e.g., Kennicutt et al. 1989a, 1989b), or using 8 metre class telescopes on nearby
samples of massive galaxies ($M_{star}$ $\gtrsim$ 10$^{9.5}$ M$_{\odot}$, 
Santoro et al. 2022). Other statistical samples are generally limited in terms of sensitivity and angular resolution (Caldwell et al. 1991, Banfi et al. 1993,
Rozas et al. 1996a, Elmegreen \& Salzer 1999, Thilker et al. 2002, Helmboldt et al. 2005, Bradley et al. 2006). Furthermore, our sample includes galaxies located in a nearby cluster, and 
is thus suitable to study the effects of the environment on galaxy evolution down to the scale of individual {\hii} regions. This will be the topic of a forthcoming publication.

\subsection{Parameters of the luminosity function}

The data in our hand allow us to trace the statistical properties of this representative set of local galaxies. The composite H$\alpha$ luminosity function
of individual {\hii} regions can be fitted with a Schechter function of faint end slope $\alpha$ = -1.61$\pm$0.01 and characteristic luminosity $L^*(H\alpha)$ = 39.68$\pm$0.04 erg s$^{-1}$.
When fitted to individual galaxies having more than 20 {\hii} regions above the threshold of $L(H\alpha)$ $\geq$ 10$^{37}$ erg s$^{-1}$, 
$<\alpha>$ =  -1.66, with a dispersion $\sigma$ = 0.43, where $<\alpha>$ is the mean slope derived on a sample of 27 objects. We recall that the fit parameters might suffer for 
low-number statistics in some low-mass objects.
Schechter functions are not often used in the literature since data with the sensitivity and angular resolution comparable to those available in this work, necessary to sample 
a sufficient range in luminosity,
are rarely available. Power laws, or double power laws, are generally used. The slope of the power law can be compared to the faint 
end slope of the Schechter function derived in this work (see however Sec. 4.1.1). The value derived for the PHANGS sample composed of 19 nearby mostly massive galaxies 
($M_{star}$ $\gtrsim$ 10$^{9.4}$ M$_{\odot}$, Santoro et al. 2022) is $<\alpha>$ =  -1.79, with a dispersion $\sigma$ = 0.15. These parameters have been 
derived after fitting a power law 
above a limiting luminosity changing from galaxy to galaxy, with values close to the lower limit used in our analysis ($L(H\alpha)$ $\geq$ 10$^{37}$ erg s$^{-1}$). 
The two sets of data give consistent results despite the two samples 
cover a different range in galaxy stellar mass. This can be explained considering that the $\alpha$ parameter does not depend on the stellar mass of the host galaxies (see Sec. 4.2.2). 
The values given by Santoro et al. (2022) have been derived using H$\alpha$ luminosity of {\hii} regions individually corrected for [{\nii}] contamination and dust attenuation.  
These authors have shown that the use of data uncorrected for dust extinction might induce a slight steepening of the slope. Our set of data might suffer from this
effect since the dust extinction correction that we have used is just a rigid shift in the H$\alpha$ luminosity of all the {\hii} regions belonging to a galaxy. 
The difference observed at the bright end, where we see a drop of the number of {\hii} regions while Santoro et al. (2022) do not, is possibly partly related
to the fact that we include the nuclear region, while they do not. Figure \ref{LFHIIReffnondef} clearly shows that the brightest regions with $L(H\alpha)$ $\geq$ 10$^{40}$ erg s$^{-1}$
are all located within the effective radius of the galaxies, with 12/64 within 3 arcsec from the nucleus. Given the tight relation between the luminosity of the brightest region
$L(H\alpha)_{1}$ and the total H$\alpha$ luminosity of the host galaxy, in $\sim$ 20\%\ of the objects the bright end of the luminosity function is dominated by nuclear regions.
The faint-end slope of the fitted function found in this work is also consistent with those found in Local Group 
galaxies by Kennicutt et al. (1989a, $\alpha$ = -2.0$\pm$0.5) in a fairly similar luminosity range using uncorrected data,
or slightly steeper than those found in another sample of 12 nearby galaxies derived using the Pa$\alpha$ line at 1.8756 $\mu$m observed with NICMOS/HST ($\alpha$ $\simeq$ -1.0; Liu et al. 2013).
It is 
 comparable to that derived for star forming complexes in the UV band 
($\alpha$ $\simeq$ -1.75, Cook et al. 2016) and flatter than the luminosity function of the star clusters 
populations of nearby galaxies ($\alpha$ = -2.37$\pm$0.18, Whitmore et al. 2014; see also Larsen 2002).

A flattening of the distribution below $L(H\alpha)$ $\simeq$ 10$^{38.6}$ erg s$^{-1}$, has been already claimed in the past (e.g., Kennicutt et al. 1989a, Beckman et al. 2000, Bradley et al. 2006).
A break at this luminosity has been interpreted in the literature as due to the transition of {\hii} regions from ionisation bounding 
(the central star/star cluster only ionises the gas within the {\hii} region) at low luminosities
to density bounding (a large amount of Lyman continuum photons from the central
ionising source escapes the {\hii} region and ionises the diffuse surrounding medium) at higher luminosities (Beckman et al. 2000; see also Lee et al. 2011).
The shape of the luminosity function has been also simulated adopting a stellar mass distribution as predicted by an initial mass function 
and different star formation scenarios (Oey \& Clarke 1998).
The composite luminosity function derived in this work does not have any sharp discontinuity at the observed break luminosity of $L(H\alpha)$ $\simeq$ 10$^{38.6}$ erg s$^{-1}$,
but rather a smooth flattening towards low luminosities characterised by a log $L^*(H\alpha)$ = 39.68$\pm$0.04 erg s$^{-1}$ in the Schechter parametrisation.
A sharp discontinuity is also lacking in the luminosity functions of individual galaxies, in particular in the most massive objects unaffected by low-number statistics.
Our result is consistent with what derived from the analysis of the PHANGS data, which do not show any evident change of slope at these luminosities.
Since a further flattening of the luminosity distribution is seen in our data and in the PHANGS data at $L(H\alpha)$ $\simeq$ 10$^{37}$ erg s$^{-1}$,
which corresponds to the completeness limit of the survey, it is conceivable that the H$\alpha$ luminosity break at $L(H\alpha)$ $\simeq$ 10$^{38.6}$ erg s$^{-1}$
observed in previous works could be due to blending effects in low angular resolution data, as first claimed by Lee et al. (2011).

\subsection{Equivalent diameter and electron density distributions}

The size distribution of individual {\hii} regions observed in a few well resolved nearby spiral galaxies is generally represented by
a power law in the range 50 $\lesssim$ log $D$ $\lesssim$ 250 pc (Lee et al. 2011, Liu et al. 2013),
with the number of regions dropping for smaller size diameters. The limiting diameter at small sizes depends on the linear resolution of the data.
The composite distribution observed in our data shows a somewhat different
shape (see Fig. \ref{Deffdist_nondef_lin}, as observed in some other objects (Gutierrez et al. 2011, Liu et al. 2013,  Rousseau-Nepton et al. 2018). 
This gradual variation is visible for $D_{eq}$ $\gtrsim$ 100 pc,
where the correction for seeing effects are negligible. Interestingly, the size composite distribution is very similar to the one observed in NGC 628
by Rousseau-Nepton et al. (2018), although with a slightly different slope.
A distribution more similar to a power law is however observed 
in our data in the equivalent diameter range 60 $\lesssim$ $D_{eq}$ $\lesssim$ 300 pc when the brightest galaxies are considered individually. 
It is thus likely that the observed composite diameter distribution might result from the combination of data for galaxies of different mass/luminosity. 
Indeed, the scaling relations shown in Sec. 4.2, combined with the luminosity-size relation (Sec. 4.1.2), consistently suggest that galaxies of different mass
host regions of different luminosity and size, and thus might populate different regions in the composite equivalent diameter distribution.

The composite electron density distribution is regularly decreasing in the range 2 $\lesssim$ $n_e$ $\lesssim$ 7 cm$^{-3}$, it has a few objects 
above this threshold, while it sharply drops at lower densities. A similar distribution is also observed individually in the most massive galaxies
where the statistic is sufficiently high. In other galaxies the distribution is fairly symmetric and peaked at a given density 
(e.g., Liu et al. 2013), with slightly higher values (3 $\lesssim$ $n_e$ $\lesssim$ 30 cm$^{-3}$, Liu et al. 2013).
Notice that the range of electron densities derived using narrow-band imaging data ($n_e$ $\lesssim$ 30 cm$^{-3}$) is generally lower than the one that can be 
probed with optical emission lines ([{\sii}]$\lambda$6716/6731\AA\ and [{\oii}]$\lambda$3726/3729\AA\ line ratios; e.g., Kennicutt et al. 1989b) since these indicators 
are degenerate at these low-density levels (e.g., Osterbrock \& Ferland 2006). As extensively discussed in Cedr{\'e}s et al. (2013),
the higher densities given by emission line diagnostics than those measured assuming eq. 3 are due to the fact that the ionised gas is not uniformly
distributed within the {\hii} regions, and the line emission is dominated by the densest inner regions while the equivalent diameter measures the
extension of the low-density ionised gas.
The observed differences with other works might depend on several factors. First of all, the shape of the distribution and its mean value 
change according to the adopted [{\nii}] and dust attenuation correction. This is due to the fact that the slope of the luminosity-size relation is not three (see Table \ref{Tablumsize}).
Being the electron density defined as a ratio of these two variables (see eq. 3), the rigid corrections here applied to the luminosity has thus systematic effects on $n_e$.
Furthermore, the absolute values of $n_e$ strongly depend on the definition of equivalent diameter and on the adopted corrections for seeing effects.
These might change in different works, where the codes used to identify {\hii} regions might change, as the definition of the diameter might also change. 
Finally, as discussed in the next section, the luminosity-size relation is also strongly affected by observational biases.
The comparison with other samples should thus be considered with these caveats whenever these parameters are derived using non-homogeneous sets of data
or different tracers.

\subsection{Luminosity-size relation}

A luminosity-size relation in {\hii} regions in local galaxies has been already observed in the past (e.g., Rozas et al. 1996b, Liu et al. 2013). 
Liu et al. (2013) reported a slope for the relation of $\sim$ 2.5-3 when luminosities are derived using the Pa$\alpha$ emission line. Rozas et al. (1996b) 
finds a slope $\sim$ 3 using H$\alpha$ data. These slopes are consistent with the one found in this work 
for the composite relation ($a$ = 3.43$\pm$0.02). 
A slope of $\sim$ 3 suggests that the luminosity increases linearly with the volume of the region. Scoville et al. (2001) argued that a 
slope of $\sim$ 3 is expected if all {\hii} regions 
are resolved, while the slope should be closer to $\sim$ 2 if blending is important. Figure \ref{sizelum} and Table \ref{Tablumsize} show 
a flattening of the relation if all {\hii} regions, including those with a high uncertainty on the adopted correction, are considered.
Liu et al. (2013), however, questioned the idea that the flattening of the relation is due to blending with convincing arguments
and rather predicted an opposite effect (steepening of the relation with increasing blending). They have shown that the slope of the relation 
and its scatter depend on the algorithm used to identify and measure the parameters of individual {\hii} regions and on the noise in the data
(an increasing noise decreases the slope of the relation). It is thus clear that the slope of the relation is not only related to the intrinsic physical 
properties of the {\hii} regions but also strongly depends on the quality of the data and on the code adopted for their identification (e.g., Scoville et al. 2001, Liu et al. 2013).

\subsection{Scaling relations} 

\subsubsection{General trends}

Variations of some representative properties of the {\hii} regions with the properties of their host galaxies (morphological type, luminosity, ...)
have been already explored in the past (Kennicutt 1988, Kennicutt et al. 1989a, Banfi et al. 1993, Rozas et al. 1996b, 
Elmegreen \& Salzer 1999, Helmboldt et al. 2005, Santoro et al. 2022). 
Most of the results previously found in the literature are reproduced on strong statistical basis with our dataset. These include the relation 
between the luminosity of the first ranked {\hii} region (Kennicutt 1988), the shape of the luminosity function (Liu et al. 2013) 
with several other characteristic parameters of the host galaxies (Santoro et al. 2022). 
In particular, we find strong dependences of the total number of {\hii} regions (Fig. \ref{scaling1N37}), of the first ranked (Fig. \ref{scaling1Lup}) 
and of the three first ranked (Fig. \ref{scaling1Lup3}) {\hii} regions, 
with the total H$\alpha$ luminosity (or star formation rate) and stellar mass of the host galaxies. 

The first of these relations is due to a simple scaling factor (bigger galaxies have more of everything). 
Those relative to the first ranked {\hii} regions can also be understood when combined to the luminosity function.
For statistical reasons, the bright end of the luminosity function is populated only in galaxies hosting many {\hii} regions, thus in massive systems with a very 
strong star formation activity\footnote{There are only four galaxies with {\hii} regions of luminosity $L(H\alpha)$ $\geq$ 10$^{40}$ erg s$^{-1}$: NGC 4189, 4383, 4535, and
4654. Of these, NGC 4383, which is a strong starburst characterised by a prominent outflow (Watts et al. 2024), 
host six {\hii} regions brighter than this limit.  }. They also reflect what found for the massive young star clusters observed with HST in a few nearby galaxies, where the properties 
of the most massive one is tightly related to the star formation rate density over the galaxy disc (e.g., Larsen 2002, Johnson et al. 2017). 
Twelve over sixty-four of these bright {\hii} regions are located in the galaxy nucleus. Interesting to note is that the luminosity of the brightest {\hii} regions
of some well known galaxies of the Local Group, 30 Doradus (log $L(H\alpha)$ = 39.76 erg s$^{-1}$, Kennicutt et al. 1995) in the Large Magellanic Cloud\footnote{The 
LMC is interacting with the Milky Way, the comparison with 30 Doradus should thus be considered here with some caution. } 
(log $L(H\alpha)$ = 40.49 erg s$^{-1}$, Kennicutt et al. 2008) and NGC 604 (log $L(H\alpha)$ = 39.63 erg s$^{-1}$, Rela{\~n}o \& Kennicutt 2008) in M33
(log $L(H\alpha)$ = 40.51 erg s$^{-1}$, Kennicutt et al. 2008) also fall on the upper envelope
of the relation shown in Fig. \ref{scaling1Lup} and \ref{scaling2Lup}. 

The observed decrease of the $L(H\alpha)_{1}/L(H\alpha)$ ratio
with $L(H\alpha)$ (Fig. \ref{scaling2Lup}) is also expected given that the $L(H\alpha)$ variable is present on both axis. It is interesting, 
however, to see that the correlation of  $L(H\alpha)_{1}/L(H\alpha)$ with $M{star}$ is stronger than with $L(H\alpha)$, suggesting 
once again that stellar mass is a fundamental parameter in regulating the star formation process (e.g., Boselli et al. 2001) and thus in driving galaxy
evolution (e.g., Cowie et al. 1996; Gavazzi et al. 1996). The relation of $L(H\alpha)_{1}/L(H\alpha)$ with metallicity (Fig. \ref{scaling1Lupphysical}) can be easily understood
considering that $L(H\alpha)_{1}/L(H\alpha)$ is also related to stellar mass (Fig. \ref{scaling2Lup}), as does the gas metallicity (metallicity-mass relation, Tremonti et al. 2004).
The fact that the correlation is tighter (higher correlation coefficient, lower dispersion; see Table \ref{Tabscalingfit}) in the relation with metallicity than with stellar mass 
(this is also the case when the correlation with $M_{star}$ is measured only on the 39/64 galaxies with available metallicity measurements) lets us think
that metallicity might participate in regulating the star formation process. This is also suggested by the fact that $L(H\alpha)_{\geq 37}/L(H\alpha)$ is correlated with
metallicity (Fig. \ref{scaling1L37physical}) while not with stellar mass (Fig. \ref{scaling1L37}). Metallicity must thus be an important parameter regulating the 
star formation activity at the scale of individual {\hii} regions.

These well-defined scaling relations can be used to predict the expected number of {\hii} regions in typical unperturbed galaxies 
in the local Universe undergoing a secular evolution.
We do not confirm, however, the observed trends of the number density of {\hii} regions per unit stellar surface and stellar luminosity with morphological type
claimed by Kennicutt et al. (1989a).

Interesting is the dependence of several representative parameters (total number of {\hii} regions, luminosity of the first ranked {\hii} regions) 
with the star formation rate and stellar mass surface density. Dependencies with the surface brightness of the host galaxies have been already 
noticed by Helmboldt et al. (2005, 2009) and more recently by Santoro et al. (2022) in massive systems. These results have been interpreted 
as due to a different star formation mode in low-density regimes, where star formation is episodic and with a low efficiency (e.g., Boissier et al. 2008) 
while it is constant in high surface brightness regions where it generally occurs within spiral density waves (Helmboldt et al. 2005). 
It is important to remember that the two variables mean star formation and stellar mass surface densities are tightly related (see Appendix \ref{appD}).
Although part of this trend might be due to the use of the same normaliser for the two variables, this tight relation is
suggesting that the star formation process is triggered by the midplane pressure produced by the stellar gravity of the disc (Shi et al. 2018).
The scaling relations on the properties of {\hii} regions shown in Sec. 4.2 thus reflect the general star formation mode observed at the scale of galactic discs.
The relation with the specific star formation rate (Fig. \ref{SSFR}) is also very interesting since it indicates that in those systems with the highest present-to-mean past activity
the star formation is dominated by one or a few bright {\hii} regions. These are principally BCD galaxies, which can be considered as analogues of 
starburst at high redshift.

Another interesting result is that the first ranked {\hii} regions (the brightest or the three brightest ones) dominate the total H$\alpha$ emission of 
dwarfs. This result is very important since it implies that in these low-mass systems the activity of star formation is dominated by one or a few {\hii} regions. 
It cannot thus be considered constant over time. We recall that the stationarity condition of the star formation activity on scales as 
long as the mean age of the emitting stars is a fundamental prerequisite to transform H$\alpha$ or UV
luminosities into star formation rates using the simple prescriptions proposed in the literature (e.g., Kennicutt 1998, Boselli et al. 2009).
The implications of this results are major also for the study of the IMF. Indeed, it has been claimed that the observed decrease of the 
H$\alpha$ over FUV flux ratio with decreasing stellar mass was due to a variation of the IMF in dwarf systems (Meurer et al. 2009), a result 
that, if confirmed, would have had major implications for all studies of galaxy evolution. The analysis presented in Sec. 4.2 and the results shown in Figs.
\ref{scaling2Lup} and \ref{scaling2Lup3} disprove this result and rather confirm what proposed by Boselli et al. (2009), i.e. that the decrease of the 
H$\alpha$ over FUV flux ratio observed in dwarf systems is principally due to an age effect. The typical age of ionising stars on the main sequence 
(O and early-B stars, $\lesssim$ 10 Myr) is significantly shorter than that of the stars dominating the emission in the far-UV (A-F stars, $\lesssim$ 100 Myr).
In the case of a sporadic star formation activity as the one encountered in individual {\hii} regions, 
the H$\alpha$ emission decreases more rapidly than the FUV flux after the burst.

\subsubsection{Diffuse fraction}

The scaling relations analysed in Sec. 4.2 are also important to quantify the contribution of the diffuse ionising emission in galaxies (DIG)
of different stellar mass and star formation activity. Previous works on selected objects  mainly observed in narrow-band imaging 
(Scoville et al. 2001, Guti{\'e}rrez et al. 2011, Lee et al. 2011) 
or on statistical samples (Hoopes et al. 1996, 1999, Thilker et al. 2002, Helmboldt et al. 2005, Oey et al. 2007) have shown that the 
mean contribution of the DIG to the total ionised gas emission of galaxies is of $\sim$ 30-60\%, with an increasing importance in low surface brightness systems 
(Hoopes et al. 1996, 1999, Oey et al. 2007). The advent of IFU spectroscopy with wide field detectors allowed to extend these first study 
and identify the main source of ionisations in different regions within the disc of galaxies (e.g., Zhang et al. 2017, Lacerda et al. 2018, Belfiore et al. 2022).
The main radiation sources have been identified, and are probably ionising photons produced by massive OB stars leaking out from {\hii} regions, with
the possible contribution of other mechanisms such as shocks, turbulent mixing layers, cosmic rays, hot low-mass stars and weak emission lines in retired galaxies
(e.g., Haffner et al 2009, Belfiore et al. 2022). 

The analysis presented in Sec. 4.2 clearly shows that the contribution of {\hii} regions brighter than $L(H\alpha)$ $\simeq$ 10$^{37}$ erg s$^{-1}$
to the total H$\alpha$ emission of the unperturbed sample galaxies ranges from 20\%\ to 100\%\, implying that the contribution of the DIG is in 
between the 0\%\ $\lesssim$ DIG $\lesssim$ 80\%\ , in agreement with previous results (e.g., Oey et al. 2007). 
The fractions of DIG measured in this work would obviously change assuming different cuts in the luminosity function and/or in the 
threshold used to identify {\hii} regions, but overall indicate that the observed trends would not change given the homogeneous set of data. 
Figure \ref{scaling1L37}
clearly shows that the importance of the DIG does not change with the total mass or the star formation activity of galaxies. On the contrary, it increases 
with decreasing stellar surface density (Helmboldt et al. 2005, Oey et al. 2007), star formation surface density, and specific star formation rate (Fig. \ref{SSFR}).
This result has been explained as due to an important escape fraction of the Lyman continuum radiation in compact, starburst galaxies, where the ISM is shredded 
by pressure-driven supernova feedback and the cold gas is all consumed via star formation, ionised, or ejected by outflows (Oey et al. 2007).
Figures \ref{scaling1Lupphysical}, \ref{scaling1Lup3physical}, and \ref{scaling1L37physical} (although the trend observed in this last figure 
should be considered with caution given the large uncertainty in the data) add an interesting new information in this context.
They consistently show that the DIG is minimal in low metallicity environments where most of the ionising radiation is due to one or very few massive star forming regions.
Interesting, the relation between $L(H\alpha)_{\geq 37}/L(H\alpha)$ and metallicity is present despite the same ratio does not change with stellar mass
even within the subsample of galaxies with available metallicity. Given the strong 
metallicity-mass relation observed in star forming galaxies (e.g., Tremonti et al. 2004), this suggests that metallicity plays a role in regulating 
the ionised gas diffuse emission. These low metallicity objects are dominated by young stellar populations; thus the expected contribution
of hot low-mass evolved stars (HOLMES) here is probably lower than in massive, metal-rich systems. 
In low-metallicity objects the gas is mainly in its atomic phase since the molecular gas component, which is generally associated to dust and metals,
is efficiently dissociated by the harsh ionising radiation. It is thus likely that here a significant fraction of the ionising radiation leaking from these 
bright {\hii} regions can also escape the whole galaxy. The data in our hand do not allow us to further investigate this speculative scenario
mostly because molecular gas and dust masses are still difficult to measure in these metal poor dwarf systems. If confirmed
with tuned observations and simulations, however, it could be of fundamental importance in the study of the nature of the ionising sources in the epoch of reionisation
where metallicity was lower than in the local Universe.  

\subsubsection{Properties of the luminosity function}

The relations between the best fit parameters of the luminosity function with the properties of the host galaxies have been already studied in 
Santoro et al. (2022). They have shown that the faint-end slope of the luminosity function is related to the star formation rate, specific star formation rate, 
star formation rate density, offset from the resolved main sequence relation, Kennicutt-Schmidt relation, and molecular gas main sequence relation.
Our dataset does not allow us to test all these parameters but rather allows us to extend the parameter space in the input sample, which here includes 
also low-mass systems down to $M_{star}$  $\simeq$ 10$^{8.5}$ M$_{\odot}$ (about one order of mag lower than in Santoro et al. 2022). 
Our data do not show any relation between the faint-end slope of the luminosity function and the stellar mass and the star formation activity of 
the host galaxies, but it shows a tight relation with the star formation rate and stellar mass surface densities (Fig. \ref{scaling1alfa}, see Sec. 4.2.2).
The characteristic luminosity of the fitted Schechter function is not correlated with any of these galaxy parameters (Fig. \ref{scaling1Lstar}),
while $\Phi^*$ increases with the star formation rate, stellar mass, and their surface densities (Fig. \ref{scaling1phistar}).
Being $\Phi^*$ connected to the total number of {\hii} regions, these last relations can be understood as explained in Sec. 5.5.1.

\subsection{Radial variations}

Variations of the mean properties of {\hii} regions with galactocentric distance have been reported in the literature for regions
located outside the optical disc (Ferguson et al. 1998, Leli{\`e}vre \& Roy 2000), now generally referred to as the extended UV discs
(Thilker et al. 2007). Here, the activity of star formation might be episodic, thus the associated {\hii} regions might be on average
older than those located within the inner disc, principally within spiral arms, where star formation is self-propagating since it is triggered
by the density waves. In these extreme regions, the low level of star formation might just be related to the small number of 
ionising stars (Boissier et al. 2007).
Differences within the optical disc are also expected given the observed variation with galactocentric distance of the mass function 
of giant molecular clouds, where star formation takes place (Rosolowsky et al. 2007, Colombo et al. 2014, 
Rice et al. 2016, Faesi et al. 2018, Schruba et al. 2019). Galaxy discs are also characterised by metallicity gradients (e.g., Zaritsky et al. 1994). 
These variations are probably due to the changing of the external pressure 
from molecular gas dominated regions (inner disc) to atomic hydrogen dominated ones (outer disc) (e.g., Schruba et al. 2019). 
Analysing a sample of 19 massive local galaxies observed during the PHANGS survey with MUSE, Santoro et al. (2022) did not find any significant variation of the slope
of the luminosity function with galactocentric distance. Our analysis shows that the number of {\hii} regions located within the deprojected $i$-band effective radius
is $\sim$ 70\%\ lower than that in the outer disc. This difference is not surprising given that the effective radius is measured in the $i$-band (old stellar population)
which is more peaked in the inner regions because of the presence of the nucleus, bulges, and the underlying stellar disc, than the young stellar population, 
as indeed suggested by the observed radial colour gradients in spiral discs.

As mentioned in Sec. 3, we expressly excluded from the analysis all {\hii} regions located outside the optical disc. 
Being all galaxies located in a cluster environment, these external {\hii} regions might have been formed during the interaction of galaxies
with their surrounding environment, as indeed observed in several Virgo cluster objects (e.g., Hester et al. 2010, Fumagalli et al. 2011, Arrigoni Battaia et al. 2012, 
Kenney et al. 2014, Fossati et al. 2018, Boselli et al. 2018b, 2021, 2023c, Cramer et al. 2019). 
Their origin is thus significantly different than the one observed in extended UV discs, 
where they have been probably formed within the fresh gas recently accreted on the disc (e.g., Thilker et al. 2007).
The properties of the composite luminosity function shown in Fig. \ref{LFHIIReffnondef} and measured within and outside the $i$-band effective radius
are significantly different: those located within the inner disc have a flatter distribution at the faint end compared to those in the outer disc.
The inner disc also includes the brightest {\hii} regions. We caution, however, in interpreting these differences as due to physical reasons. 
Although the unperturbed sample analysed in this work spans a wide range in morphology and stellar mass and is dominated by dwarf systems of irregular shape
(Im, BCD), the inner discs are generally characterised by a higher density of {\hii} regions. It is thus likely that blending is here more important
than in the outer discs. These arguments should also be considered when interpreting the observed steepening of the 
luminosity-size relation shown in Fig. \ref{sizelumReff}. As discussed in Sec. 5.4, simulations indicate that a steepening of the relation is indeed expected 
when blending is important (Liu et al. 2013). What should be however stressed is the fact that the brightest {\hii} regions are generally 
located within the effective radius, with 10/51 in the nucleus.

\section{Conclusion}

This work presents a complete statistical study of an unperturbed sample of $\simeq$ 27.000 {\hii} regions, out of which 13.000 of luminosity $L(H\alpha)$ $\geq$ 10$^{37}$ erg s$^{-1}$ 
located in 64 galaxies of different morphological type (dE, S0, spirals, Magellanic irregulars, BCDs) spanning a wide range in stellar mass
(10$^7$ $\lesssim$ $M_{star}$ $\lesssim$ 10$^{11}$ M$_{\odot}$). The uniform set of data, gathered during a NB imaging survey of the Virgo cluster,
are of excellent image quality, $FWHM$ $\simeq$ 0.7\arcsec, corresponding to $\simeq$ 60 pc at the distance of the sample. This unique set of data is used to
trace the composite luminosity function, equivalent diameter and electron density distribution of the detected regions, and drive the main 
scaling relations linking representative quantities of the {\hii} regions (total number, first ranked and sum of the first three ranked luminosities, best fit parameters of
the fitted Schechter luminosity function) to those of their host galaxies (morphological type, stellar mass, star formation rate and specific star formation rate,
metallicity, molecular-to-atomic gas fraction, and total gas-to-dust ratio).
The derived distributions and scaling relations are in line with those already presented in the literature on smaller and heterogeneous samples
of local galaxies, and can be summarised as follows: \\
1) Several properties such as the total number of {\hii} regions, the H$\alpha$ luminosity
of the first ranked and the mean luminosity of the first three ranked {\hii} regions, and their relative contribution to the total H$\alpha$ emission of the galaxy, 
the $\Phi^*$ parameter of the fitted Schechter function, are strongly correlated to the stellar mass 
and the total star formation activity of their host galaxies. \\
2) Some of these parameters are also tightly connected to the star formation and stellar mass surface density of the host galaxies.\\
3) The faint end slope of the H$\alpha$ luminosity function of {\hii} regions increases with decreasing star formation, stellar mass surface density, and specific star formation rate.\\
4) There is also a clear anti-correlation between the contribution of all {\hii} regions with $L(H\alpha)$ $\geq$ 10$^{37}$ erg s$^{-1}$, of the first ranked {\hii} regions and 
of the three first ranked {\hii} regions to the total H$\alpha$ emission of these galaxies and their metallicity, with the emission of the DIG minimal 
in low metallicity environments. 

Thanks to its homogeneity, sensitivity, and complete coverage of the galaxy's parameter space, this unperturbed sample is an excellent reference for future local and high redshift studies.
It will be perfectly suited to study the perturbations induced by the cluster environment on galaxy evolution down to the scale of individual {\hii} regions, as well as 
in the study of the star formation process in individual {\hii} regions and its evolution with cosmic time when compared to the spectacular set of narrow-band imaging 
data that JWST is now providing for galaxies at different redshift.

\begin{acknowledgements}

We thank the anonymous referee for a carefully reading of the manuscript, and for precious and constructive comments and suggestions which helped improving its quality.
We are grateful to the whole CFHT team who assisted us in the preparation and in the execution of the observations and in the calibration and data reduction: 
Todd Burdullis, Daniel Devost, Bill Mahoney, Nadine Manset, Andreea Petric, Simon Prunet, Kanoa Withington.
This work was supported by the Programme National Cosmology et Galaxies (PNCG) of CNRS/INSU with INP and IN2P3, co-funded by CEA and CNES.
This research has made use of the NASA/IPAC Extragalactic Database (NED) 
which is operated by the Jet Propulsion Laboratory, California Institute of 
Technology, under contract with the National Aeronautics and Space Administration
and of the GOLDMine database (http://goldmine.mib.infn.it/) (Gavazzi et al. 2003).

\end{acknowledgements}

\newpage
\clearpage

\begin{appendix}
\onecolumn

\section{Radial variation of [{\nii}]/H$\alpha$ and $A(H\alpha)$}
\label{appA}

\begin{figure*}[h!]
\centering
\includegraphics[width=0.32\textwidth]{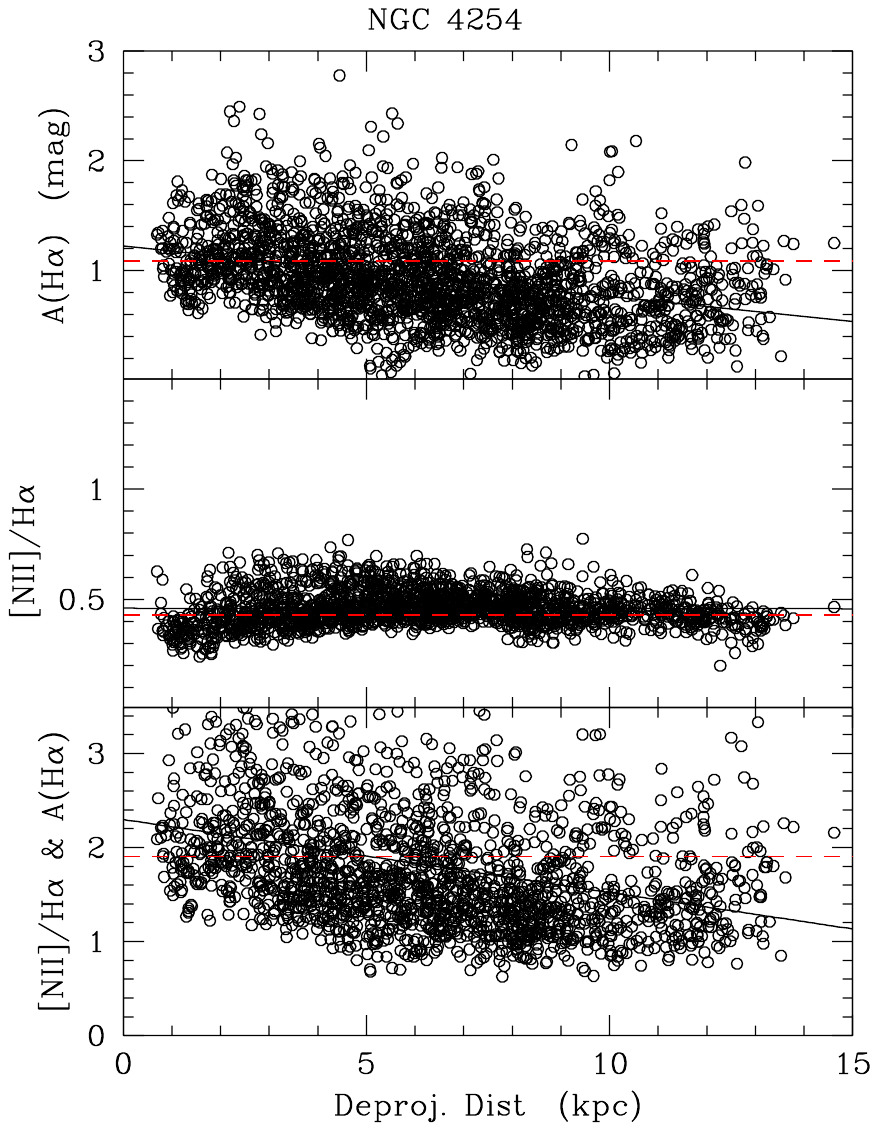}
\includegraphics[width=0.32\textwidth]{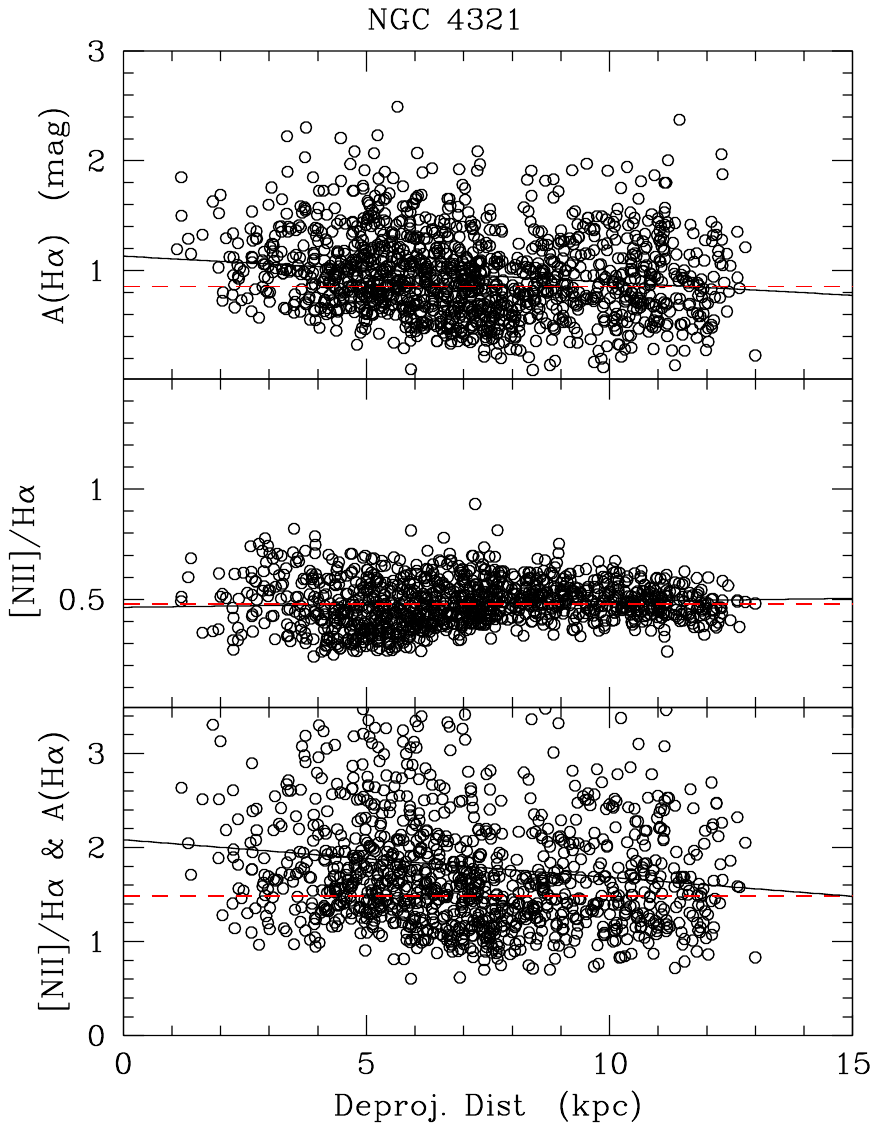}
\includegraphics[width=0.32\textwidth]{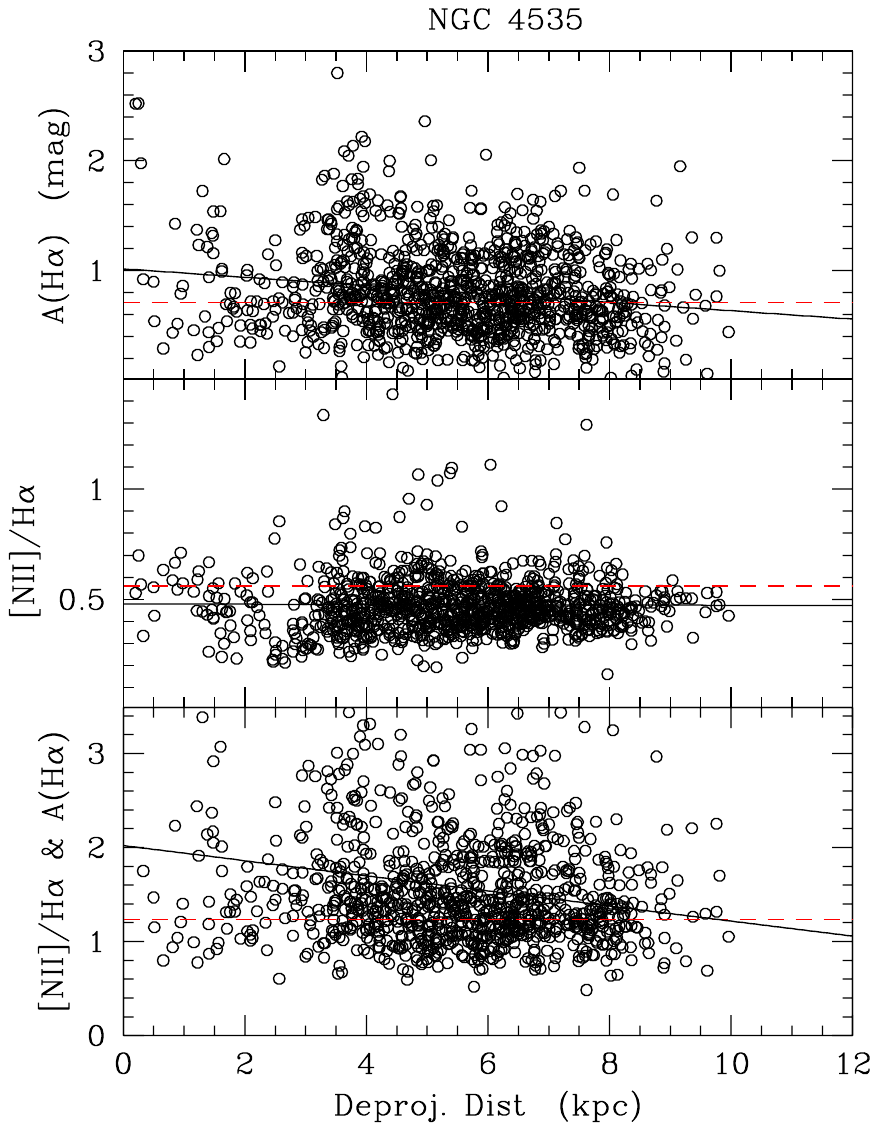}\\
\caption{Radial variation of the dust attenuation in the H$\alpha$ line derived using the Balmer decrement (upper panels),
of the [{\nii}] contamination (central panels), and their combined effect (lower panels; the Y-axis give the multiplicative factor which should
be used to correct luminosities) for the three galaxies of the MUSE/PHANGS 
sample in common, NGC 4254, NGC 4321, and NGC 4535. The red dashed line shows the values derived using integrated quantities and used in this work, 
the black solid line the linear best fit to the data.}
\label{dust}%
\end{figure*}

As described in Sec. 3, we apply a fixed correction to the total H$\alpha$ luminosity for [{\nii}] contamination and dust attenuation. This correction is 
derived for each individual galaxy using the available integrated spectroscopy and mid-infrared emission as described in Boselli et al. (2022a).
This simple zero-point correction, although optimised for each object, might introduce systematic effects in the total H$\alpha$ luminosities 
measured in this work and used through the analysis. To quantify this possible effect, we use the spectroscopic data gathered with MUSE at the VLT
for the three galaxies NGC 4254, 4321, and 4535 (see Sec. 3) and study the possible radial variation of the two main parameters used to correct the H$\alpha$ VESTIGE
data, i.e. the [{\nii}]/H$\alpha$ ratio, the dust attenuation measured using the Balmer decrement ($A(H\alpha)$), and their combined effect 
([{\nii}]/H$\alpha$ \& $A(H\alpha)$; Fig. \ref{dust}). For this exercise we use the spectroscopic data 
of Congiu et al. (2023). Figure \ref{dust} shows that the [{\nii}]/H$\alpha$ ratio is very constant all over the stellar disc (the Spearman correlation coefficient 
on the observed relations is in the range 0.03 $\leq$ $\rho$ $\leq$ 0.12, and any radial variation, if present, is $\lesssim$ 5\%\ ), with values similar to those 
derived using integrated spectroscopy and used in this work. The dispersion in the [{\nii}]/H$\alpha$ deprojected distance relation is fairly low, $\sigma$ $\simeq$ 0.1.
On the contrary, the dust attenuation $A(H\alpha)$ measured with the Balmer decrement has a significant larger dispersion ($\sigma$ $\simeq$ 0.4 mag), and slightly decreases 
at large radii (-0.38 $\leq$ $\rho$ $\leq$-0.14). The gradient is, however, low, with a radial variation of $A(H\alpha)$ $\simeq$ 0.2-0.3 mag. As for the [{\nii}],
the dust attenuation correction used in this work agrees fairly well with the one derived for individual {\hii} regions using the MUSE data. 
The lower panels of Fig. \ref{dust} shows the combined effect of the two variables. As for the Balmer decrement, there is a mild decrease of the combined effect of dust 
attenuation and [{\nii}] contamination with increasing deprojected distance from the galaxy 
centre (-0.38 $\leq$ $\rho$ $\leq$-0.14), but the scatter in the relation ($\sigma$ $\simeq$ 0.75) is comparable or larger than the observed gradient ($\simeq$ 0.4-0.7).  
The overall correction ([{\nii}]/H$\alpha$ \& $A(H\alpha)$) used in this work is within $\lesssim$ 15\%\ from the mean value derived using the PHANGS data, while the observed gradients 
within $\lesssim$ 35\%\ with respect to the mean correction. These systematic effects should be considered in the analysis. We notice, however, that they are
smaller than the bin sampled in the luminosity function, and generally smaller than the typical error in the variables analysed in the scaling relations. 
We also recall that these three galaxies are among the most massive spirals analysed in this work. Since dust attenuation and [{\nii}] contamination and their possible gradients become
less and less important with decreasing stellar mass, we expect that
the dust attenuation correction and [{\nii}] contamination adopted in this work do not introduce strong systematic effects in the data.

\FloatBarrier
\section{Properties of the sample galaxies}
\label{appB}

The galaxies analysed in this work (unperturbed sample) are listed in Table \ref{gal} along with their parameters.
Table \ref{Tabds9} gives the parameters of the elliptical apertures used for the identification of the {\hii} regions 
and for the extraction of the total H$\alpha$ flux. Table \ref{galLF} lists the best fit parameters for the luminosity function of individual objects.

\begin{landscape}
\begin{table}[h!]
\caption{Galaxies analysed in this work
}
\label{gal}
{\tiny 
\[
\resizebox{\columnwidth}{!}{\begin{tabular}{cccccccccccccccc}
\hline
\noalign{\smallskip}
\hline
Name		& NGC/IC  & RA(J2000)   & DEC(J2000)    & $E(B-V)$        &$2a_{B25.5}$ &$R_{eff,i}$&$PA(i)$    & $b/a$   &log $M_{star}$&SM     &dist        &$A(H\alpha)$&[{\nii}]/H$\alpha$&$HI-def$&N{\hii}$_{L(H\alpha)\geq 37}$\\
		&	  &  h m s      & $^o$ ' "      &                 & arcmin     & arcsec     & deg       &         & M$_{\odot}$  &       & Mpc	      & mag	   &	      &           &       \\
\hline
      VCC22	&    -	  & 12:10:24.39 & +13:10:14.2   &   0.0372	  & 0.18       & 4.69	    & 72        &0.69	  & 8.56   &   M	 &	  32  &     0.00   &	0.11  &     0.32  &	   6  \\
      VCC24	&    -	  & 12:10:35.68 & +11:45:38.9   &   0.0280	  & 0.78       & 4.35	    &140        &0.63	  & 8.38   &   A	 &	16.5  &     0.58   &	0.17  &     0.04  &	  16  \\
      VCC26	&    -	  & 12:10:40.89 & +14:38:46.5   &   0.0381	  & 0.54       & 5.44	    & 52        &0.71	  & 7.80   &   M	 &	  32  &     0.00   &	0.05  &     0.23  &	   7  \\
      VCC58	&   IC769 & 12:12:32.33 & +12:07:25.7   &   0.0319	  & 3.16       &25.68	    & 22        &0.51	  & 9.92   &   M	 &	  32  &     0.30   &	0.89  &     0.21  &	 237  \\
      VCC66	& NGC4178 & 12:12:46.33 & +10:51:54.9   &   0.0279	  & 6.61       &49.52	    & 30        &0.26	  & 9.76   &   A	 &	16.5  &     0.60   &	0.24  &     0.10  &	 719  \\
      VCC81	& NGC4186 & 12:13:26.32 & +14:46:19.3   &   0.0344	  & 1.82       &19.32	    & 78        &0.71	  & 9.07   &   M	 &	  32  &     0.07   &	0.20  &     0.28  &	 100  \\
      VCC89	& NGC4189 & 12:13:47.23 & +13:25:29.5   &   0.0329	  & 2.82       &30.83	    & 98        &0.51	  &10.38   &   M	 &	  32  &     1.02   &	0.51  &     0.16  &	 654  \\
      VCC92	& NGC4192 & 12:13:48.22 & +14:54:01.5   &   0.0351	  &12.02       &49.94	    &154        &0.49	  &10.67   &   A	 &	16.5  &     1.53   &	0.71  &     0.20  &	1899  \\
     VCC113	&    -	  & 12:14:32.98 & +12:06:11.1   &   0.0286	  & 0.54       & 9.24	    &160        &0.56	  & 8.13   &   M	 &	  32  &     1.21   &	0.13  &    -0.10  &	  16  \\
     VCC130	&    -	  & 12:15:04.00 & +09:45:13.1   &   0.0159	  & 0.45       & 4.33	    &152        &0.51	  & 7.65   &   A	 &	16.5  &     0.00   &	0.09  &     0.29  &	   6  \\
     VCC131	&  IC3061 & 12:15:04.44 & +14:01:44.1   &   0.0373	  & 3.24       &10.74	    &122        &0.28	  & 9.70   &   M	 &	  32  &     0.97   &	0.43  &     0.20  &	 195  \\
     VCC145	& NGC4206 & 12:15:16.77 & +13:01:26.5   &   0.0315	  & 6.31       &23.18	    &180        &0.37	  & 9.65   &   A	 &	16.5  &     0.97   &	0.40  &     0.27  &	 318  \\
     VCC169	&    -	  & 12:15:55.76 & +09:39:04.1   &   0.0185	  & 1.07       &24.41	    &100        &0.60	  & 7.29   &   A	 &	16.5  &     0.00   &	0.03  &     0.13  &	   6  \\
     VCC207	&    -	  & 12:16:47.81 & +08:02:57.0   &   0.0233	  & 0.25       & 1.94	    &121        &0.56	  & 7.05   & B  	 &	15.8  &     0.06   &	0.05  &     0.03  &	   8  \\
     VCC307	& NGC4254 & 12:18:49.57 & +14:24:59.6   &   0.0387	  & 7.59       &46.33	    &  8        &0.81	  &10.52   &   A	 &	16.5  &     1.09   &	0.43  &     0.06  &	2966  \\
     VCC334	&    -	  & 12:19:14.21 & +13:52:56.2   &   0.0460	  & 0.40       & 5.64	    & 18        &0.70	  & 8.21   & LVC	 &	16.5  &     0.45   &	0.12  &     0.07  &	  12  \\
     VCC340	&    -	  & 12:19:22.13 & +05:54:38.1   &   0.0189	  & 0.89       & 6.44	    &175        &0.54	  & 9.23   &   W	 &	  32  &     0.16   &	0.12  &     0.36  &	  23  \\
     VCC357	&    -	  & 12:19:35.14 & +06:27:11.2   &   0.0192	  & 0.18       & 4.82	    & 74        &0.61	  & 7.93   &   W	 &	  32  &     0.00   &	0.26  &     0.34  &	   1  \\
     VCC410	&    -	  & 12:20:21.89 & +12:11:15.4   &   0.0326	  & 0.21       & 2.94	    &139        &0.67	  & 7.31   &   A	 &	16.5  &     0.00   &	0.05  &     0.23  &	  12  \\
     VCC448	&    -	  & 12:21:00.18 & +12:43:34.1   &   0.0457	  & 0.27       & 4.88	    &175        &0.81	  & 7.64   &   A	 &	16.5  &     0.00   &	0.04  &    -0.02  &	   1  \\
     VCC465	& NGC4294 & 12:21:17.83 & +11:30:37.7   &   0.0339	  & 4.90       &22.73	    &155        &0.35	  & 9.47   &   A	 &	16.5  &     0.50   &	0.20  &     0.30  &	 304  \\
     VCC491	& NGC4299 & 12:21:40.80 & +11:30:10.2   &   0.0325	  & 2.34       &24.40	    &140        &0.89	  & 9.20   &   A	 &	16.5  &     0.25   &	0.17  &     0.21  &	 229  \\
     VCC566	&    -	  & 12:22:37.79 & +08:17:48.2   &   0.0259	  & 0.89       & 6.33	    & 55        &0.46	  & 7.67   & B  	 &	15.8  &     0.07   &	0.04  &     0.22  &	  12  \\
     VCC699	&  IC3268 & 12:24:07.55 & +06:36:27.4   &   0.0248	  & 1.82       &15.70	    &  3        &0.80	  & 9.38   & W' 	 &	  23  &     0.44   &	0.08  &     0.30  &	  69  \\
     VCC801	& NGC4383 & 12:25:25.46 & +16:28:11.8   &   0.0237	  & 3.24       &14.37	    & 28        &0.46	  & 9.69   &   A	 &	16.5  &     1.51   &	0.28  &    -0.00  &	 290  \\
     VCC890	&    -	  & 12:26:21.55 & +06:40:10.4   &   0.0224	  & 0.14       & 4.22	    &139        &0.90	  & 8.01   & W' 	 &	  23  &     0.04   &	0.10  &     0.21  &	   9  \\
     VCC905	& NGC4411 & 12:26:29.90 & +08:52:19.9   &   0.0251	  & 2.69       &35.69	    &122        &0.45	  & 9.16   & B  	 &	15.8  &     0.48   &	0.43  &     0.36  &	  93  \\
     VCC939	& NGC4411 & 12:26:47.23 & +08:53:04.6   &   0.0293	  & 3.24       &70.23	    & 96        &0.88	  & 9.44   & B  	 &	15.8  &     0.16   &	0.27  &     0.31  &	 162  \\
     VCC945	&  IC3355 & 12:26:50.83 & +13:10:36.9   &   0.0276	  & 1.62       &17.85	    &168        &0.32	  & 8.22   &   A	 &	16.5  &     0.00   &	0.08  &     0.40  &	   4  \\
     VCC950	&  IC3356 & 12:26:50.78 & +11:33:27.1   &   0.0307	  & 2.14       &27.76	    &132        &0.67	  & 8.41   &   A	 &	16.5  &     1.11   &	0.27  &     0.28  &	  14  \\
     VCC975	&    7557 & 12:27:11.29 & +07:15:46.9   &   0.0296	  & 4.90       &70.45	    & 33        &0.72	  & 9.49   & W' 	 &	  23  &     0.39   &	0.17  &     0.37  &	 241  \\
    VCC1091	&    7590 & 12:28:18.77 & +08:43:47.0   &   0.0217	  & 1.82       & 7.92	    &  4        &0.48	  & 8.68   & B  	 &	15.8  &     0.40   &	0.12  &     0.20  &	  77  \\
    VCC1141	&    -	  & 12:28:54.98 & +09:25:15.8   &   0.0201	  & 0.32       & 4.54	    & 31        &0.91	  & 8.05   & B  	 &	15.8  &     0.00   &	0.14  &     0.31  &	   7  \\
    VCC1313	&    -	  & 12:30:48.58 & +12:02:42.7   &   0.0270	  & 0.31       & 1.96	    & 61        &0.52	  & 7.02   &   A	 &	16.5  &     0.03   &	0.01  &     0.01  &	   6  \\
    VCC1356	&  IC3446 & 12:31:22.98 & +11:29:33.2   &   0.0454	  & 0.89       & 6.87	    &170        &0.55	  & 8.29   &   A	 &	16.5  &     0.17   &	0.08  &     0.19  &	  23  \\
    VCC1508	& NGC4519 & 12:33:30.25 & +08:39:17.0   &   0.0202	  & 4.47       &28.02	    & 81        &0.45	  & 9.47   & B  	 &	15.8  &     0.72   &	0.25  &     0.09  &	 486  \\
    VCC1524	& NGC4523 & 12:33:47.95 & +15:10:05.7   &   0.0398	  & 3.98       &43.10	    & 20        &0.48	  & 9.20   &   A	 &	16.5  &     0.08   &	0.20  &     0.28  &	 145  \\
    VCC1555	& NGC4535 & 12:34:20.31 & +08:11:52.6   &   0.0194	  & 8.91       &99.11	    & 27        &0.68	  &10.50   & B  	 &	15.8  &     0.71   &	0.56  &     0.18  &	1397  \\
    VCC1585	&  IC3522 & 12:34:45.69 & +15:13:14.9   &   0.0352	  & 2.09       &18.45	    & 82        &0.45	  & 8.39   &   A	 &	16.5  &     0.09   &	0.09  &     0.29  &	  27  \\
    VCC1673	& NGC4567 & 12:36:32.72 & +11:15:29.0   &   0.0326	  & 3.63       &34.08	    & 68        &0.72	  &10.07   &   A	 &	16.5  &     1.40   &	0.47  &     0.28  &	 401  \\
    VCC1750	&    -	  & 12:38:15.49 & +06:59:39.1   &   0.0202	  & 0.21       & 3.30	    &111        &0.72	  & 7.73   & B  	 &	15.8  &     0.15   &	0.04  &     0.26  &	   4  \\
    VCC1791	&  IC3617 & 12:39:25.01 & +07:57:58.4   &   0.0273	  & 1.62       &16.89	    & 68        &0.56	  & 8.45   & B  	 &	15.8  &     0.48   &	0.13  &     0.29  &	  85  \\
    VCC1987	& NGC4654 & 12:43:56.55 & +13:07:35.5   &   0.0260	  & 6.17       &43.15	    &118        &0.47	  &10.31   &   A	 &	16.5  &     1.10   &	0.41  &    -0.14  &	1015  \\
    VCC2062	&    -	  & 12:48:00.60 & +10:58:23.5   &   0.0395	  & 0.69       &19.69	    & 67        &0.49	  & 7.07   &   A	 &	16.5  &     0.00   &	0.02  &    -0.32  &	   6  \\
\hline
\end{tabular}}
\]
}
\end{table}
\end{landscape}

\begin{landscape}
\setcounter{table}{0}
\begin{table}
\caption{continued
}
\label{gal}
{\tiny 
\[
\resizebox{\columnwidth}{!}{\begin{tabular}{cccccccccccccccc}
\hline
\noalign{\smallskip}
\hline
Name		& NGC/IC  & RA(J2000)   & DEC(J2000)    & $E(B-V)$        &$2a_{B25.5}$ &$R_{eff,i}$&$PA(i)$    & $b/a$   &log $M_{star}$&SM     &dist        &$A(H\alpha)$&[{\nii}]/H$\alpha$&$HI-def$&N{\hii}$_{L(H\alpha)\geq 37}$\\
		&	  &  h m s      & $^o$ ' "      &                 & arcmin     & arcsec     & deg       &         & M$_{\odot}$  &       & Mpc	      & mag	   &	      &           &       \\
\hline
  AGC224696	&    -	  & 12:10:38.04 & +13:01:20.0   &   0.0387	  & 0.41       & 4.20	    &178        &0.43	  & 7.76   &   M	 &	  32  &     0.16   &	0.02  &     0.33  &	   5  \\
  AGC224807	&    -	  & 12:13:09.60 & +13:35:05.3   &   0.0312	  & 0.45       & 4.63	    & 66        &0.56	  & 7.85   &   M	 &	  32  &     0.00   &	0.05  &     0.21  &	   7  \\
  AGC224487	&    -	  & 12:14:12.05 & +12:46:59.1   &   0.0309	  & 0.39       & 4.07	    &165        &0.45	  & 7.41   &   A	 &	16.5  &     0.00   &	0.06  &     0.36  &	   4  \\
  AGC224705	&    -	  & 12:14:44.53 & +12:47:22.9   &   0.0347	  & 0.32       & 3.31	    & 93        &0.73	  & 7.73   &   M	 &	  32  &     0.00   &	0.07  &     0.16  &	   2  \\
  AGC224249	&    -	  & 12:15:27.29 & +10:30:44.5   &   0.0263	  & 0.27       & 2.85	    & 86        &0.69	  & 7.10   &   A	 &	16.5  &     0.00   &	0.02  &    -0.09  &	   1  \\
  AGC224251	&    -	  & 12:16:34.04 & +10:12:22.8   &   0.0240	  & 0.19       & 1.99	    &123        &0.71	  & 7.14   &   A	 &	16.5  &     0.00   &	0.04  &    -0.37  &	   5  \\
  AGC224385	&    -	  & 12:16:49.98 & +13:30:14.4   &   0.0327	  & 0.37       & 3.84	    &146        &0.89	  & 7.24   & LVC	 &	16.5  &     0.00   &	0.03  &     0.28  &	   2  \\
  AGC224489	&    -	  & 12:17:28.07 & +12:55:56.0   &   0.0376	  & 0.32       & 3.30	    &155        &0.37	  & 7.89   &   M	 &	  32  &     0.00   &	0.05  &     0.17  &	   8  \\
  AGC225058	&    -	  & 12:18:17.99 & +07:07:41.0   &   0.0210	  & 0.26       & 2.66	    &103        &0.53	  & 7.67   & B  	 &	15.8  &     0.00   &	0.14  &     0.14  &	   4  \\
  AGC226326	&    -	  & 12:23:58.24 & +07:27:01.0   &   0.0215	  & 0.34       & 3.54	    &149        &0.37	  & 7.04   & W' 	 &	  23  &     0.00   &	0.03  &     0.27  &	   4  \\
  AGC226357	&    -	  & 12:27:12.82 & +07:38:23.4   &   0.0231	  & 0.46       & 4.79	    &  1        &0.40	  & 6.81   & B  	 &	15.8  &     0.00   &	0.02  &     0.24  &	   2  \\
  AGC223724	&    -	  & 12:29:33.61 & +13:11:44.6   &   0.0243	  & 0.38       & 3.90	    & 59        &0.64	  & 6.91   &   A	 &	16.5  &     0.00   &	0.02  &    -0.37  &	   3  \\
  AGC224281	&    -	  & 12:30:42.00 & +05:52:43.8   &   0.0173	  & 0.18       & 1.88	    &147        &0.76	  & 7.24   & B  	 &	15.8  &     0.00   &	0.04  &    -0.22  &	   6  \\
  AGC225847	&    -	  & 12:38:56.82 & +13:33:05.3   &   0.0363	  & 0.50       & 5.17	    & 27        &0.64	  & 6.86   &   A	 &	16.5  &     0.00   &	0.02  &     0.17  &	   4  \\
  AGC742572	&    -	  & 12:43:54.12 & +16:32:50.6   &   0.0256	  & 0.46       & 4.81	    & 76        &0.64	  & 7.37   &   A	 &	16.5  &     0.00   &	0.03  &     0.36  &	   1  \\
  AGC223205	&    -	  & 12:51:06.83 & +12:03:35.8   &   0.0346	  & 0.78       & 8.09	    & 12        &0.77	  & 7.43   &   A	 &	16.5  &     0.00   &	0.03  &     0.16  &	   4  \\
  AGC223247	&    -	  & 12:53:11.59 & +12:38:06.3   &   0.0269	  & 1.04       &10.81	    &150        &0.76	  & 7.20   &   A	 &	16.5  &     0.00   &	0.02  &     0.30  &	   2  \\
    NGC4651	& NGC4651 & 12:43:42.63 & +16:23:36.2   &   0.0231	  & 3.90       &30.89	    & 81        &0.79	  &10.32   &   A	 &	16.5  &     0.59   &	0.56  &    -0.07  &	 725  \\
    NGC4746	& NGC4746 & 12:51:55.36 & +12:04:58.7   &   0.0358	  & 2.20       &12.20	    &122        &0.46	  & 9.59   &   A	 &	16.5  &     0.95   &	0.36  &     0.22  &	 178  \\
N12:20:30.91+06:38:23.9&-&12:20:30.91 & +06:38:23.9   &   0.0182          & 0.22       & 2.26       &103        &0.58     & 7.81   & W  	 &	 32   &     0.00   &	0.05  &     0.31  &	   2  \\
\hline
\end{tabular}}
\]
Column 1: galaxy name \\
Column 2: IC/NGC name \\
Column 3 and 4: right ascension and declination\\
Column 5: Galactic extinction $E(B-V)$, from Schlegel et al. (1998)\\
Column 6: $B$-band isophotal diameter at 25.5 mag arcsec$^{-2}$, from Binggeli et al. (1985) for all the VCC galaxies, from GoldMine (Gavazzi et al. 2003) for the NGC galaxies, 
or derived from the NGVS $g$-band effective radius using the relation $2 \times a_{B25.5}$ [arcsec] = 3.9886 $\times$ $R_{eff,g}$ [arcsec] for the remaining objects \\
Column 7: NGVS $i$-band effective radius $R_{eff,i}$ \\
Column 8: NGVS $i$-band position angle, measured from North counterclockwise\\
Column 9: $B$-band axial ratio, from Binggeli et al. (1985) whenever available, or from the NGVS $g$-band\\
Column 10: stellar mass, in solar units\\
Column 11: cluster subgroup membership\\
Column 12: distance, in Mpc\\
Column 13: $A(H\alpha)$, mag\\
Column 14: [{\nii}]$\lambda$6548,6583/H$\alpha$\\
Column 15: $HI-def$\\
Column 16: number of {\hii} regions brighter than $L(H\alpha)$$\geq$ 10$^{37}$ erg s$^{-1}$ (luminosity corrected for dust attenuation and {\nii} contamination)\\
}
\end{table}
\end{landscape}
\newpage

\begin{table*}[h!]
\caption{Apertures used for the flux extraction and for the identification of the {\hii} regions
}
\label{Tabds9}
{
\[
\begin{tabular}{cccccc}
\hline
\noalign{\smallskip}
\hline
Name	    & RA(J2000)   & DEC(J2000)  &Maj. axis& Min. axis   & PA   \\
	    & h m s	  & $^o$ ' "	& arcsec  & arcsec      & deg. \\
\hline
     VCC22 & 12:10:23.80 & +13:10:12.9	& 6.373   &   6.373	&     0\\
     VCC24 & 12:10:35.14 & +11:45:41.1	& 16.35   &   8.666	&   142\\
     VCC26 & 12:10:41.65 & +14:38:45.1	& 12.13   &   8.777	&    70\\
     VCC58 & 12:12:29.70 & +12:07:22.8	& 80.01   &   56.18	&    45\\
     VCC66 & 12:12:48.03 & +10:52:06.7	& 153.7   &   59.26	&    30\\
     VCC81 & 12:13:27.65 & +14:46:20.6	& 49.61   &   49.61	&     0\\
     VCC89 & 12:13:44.05 & +13:25:37.1	&  76.1   &    53.5	&    80\\
     VCC92 & 12:14:01.38 & +14:54:00.5	& 267.9   &   96.36	&   153\\
    VCC113 & 12:14:35.07 & +12:06:09.0	& 22.39   &   13.57	&   147\\
    VCC130 & 12:15:03.88 & +09:45:13.2	& 18.78   &   9.284	&   158\\
    VCC131 & 12:15:05.72 & +14:01:45.4	& 82.06   &   16.33	&   120\\
    VCC145 & 12:15:16.82 & +13:01:26.5	&   200   &   34.43	&     1\\
    VCC169 & 12:15:41.15 & +09:38:44.3	&  55.9   &    55.9	&     0\\
    VCC207 & 12:16:48.01 & +08:02:57.0	& 12.55   &   6.055	&   120\\
    VCC307 & 12:18:49.59 & +14:24:59.6	& 207.9   &   195.2	&    70\\
    VCC334 & 12:19:14.16 & +13:52:56.1	&    15   &	 15	&     0\\
    VCC340 & 12:19:22.02 & +05:54:39.2	& 12.27   &   7.416	&   170\\
    VCC357 & 12:19:35.42 & +06:27:11.4	& 4.212   &   2.543	&    65\\
    VCC410 & 12:20:21.89 & +12:11:15.3	&  22.8   &   17.23	&   160\\
    VCC448 & 12:21:00.15 & +12:43:34.1	& 15.84   &   13.01	&    15\\
    VCC465 & 12:21:17.98 & +11:30:37.6	& 110.3   &   50.58	&   155\\
    VCC491 & 12:21:40.73 & +11:30:10.2	& 50.77   &   50.72	&     0\\
    VCC566 & 12:22:37.91 & +08:17:45.8	& 26.19   &   14.81	&    35\\
    VCC699 & 12:24:05.85 & +06:36:25.2	& 51.39   &   24.89	&    70\\
    VCC801 & 12:25:27.87 & +16:28:14.8	& 132.3   &   84.23	&    20\\
    VCC890 & 12:26:21.77 & +06:40:10.6	&  10.7   &    10.7	&     0\\
    VCC905 & 12:26:29.71 & +08:52:20.2	& 96.71   &   87.18	&     0\\
    VCC939 & 12:26:47.17 & +08:53:04.5	&   100   &	100	&     0\\
    VCC945 & 12:26:51.36 & +13:10:36.6	& 48.69   &   28.74	&   177\\
    VCC950 & 12:26:54.73 & +11:33:28.6	& 57.99   &   49.52	&   165\\
    VCC975 & 12:27:02.43 & +07:16:03.9	&   112   &   93.05	&   135\\
    VCC1091 & 12:28:18.79 & +08:43:47.0	& 76.46   &   29.67	&   170\\
    VCC1141 & 12:28:55.00 & +09:25:15.8	& 12.61   &   10.23	&    30\\
    VCC1313 & 12:30:48.18 & +12:02:42.3	& 13.79   &   9.328	&    80\\
    VCC1356 & 12:31:21.58 & +11:29:30.5	& 22.93   &   17.89	&   170\\
    VCC1508 & 12:33:27.76 & +08:39:06.8	& 135.8   &   103.1	&   145\\
    VCC1524 & 12:33:47.88 & +15:10:05.7	& 127.8   &   93.85	&    45\\
    VCC1555 & 12:34:20.33 & +08:11:52.6	& 224.9   &   186.9	&    10\\
    VCC1585 & 12:34:42.93 & +15:13:12.8	& 55.69   &	 30	&    92\\
    VCC1673 & 12:36:25.64 & +11:15:30.4	& 52.44   &   36.36	&    60\\
    VCC1750 & 12:38:13.78 & +06:59:38.6	&  6.94   &    6.94	&     0\\
    VCC1791 & 12:39:15.74 & +07:57:47.7	& 50.59   &   23.96	&    65\\
    VCC1987 & 12:44:24.84 & +13:07:08.0	& 157.7   &   97.11	&   120\\
    VCC2062 & 12:47:52.02 & +10:58:11.6	&  34.7   &   26.18	&    91\\
  AGC224696 & 12:10:37.94 & +13:01:19.6	& 11.33   &   6.796  	&     0\\
  AGC224807 & 12:13:10.38 & +13:35:03.0	& 12.69   &   11.06	&    70\\
  AGC224487 & 12:14:10.98 & +12:47:00.7	& 10.52   &   5.956	&   160\\
  AGC224705 & 12:14:45.43 & +12:47:22.5	& 5.249   &   2.368	&   100\\
  AGC224249 & 12:15:27.10 & +10:30:46.0	& 6.973   &   3.658	&    90\\
  AGC224251 & 12:16:34.93 & +10:12:22.9	& 10.12   &   5.771	&    92\\
  AGC224385 & 12:16:49.68 & +13:30:15.4	& 6.492   &   6.492	&     0\\
  AGC224489 & 12:17:27.72 & +12:55:57.3	& 11.18   &    6.09	&   153\\
  AGC225058 & 12:18:18.84 & +07:07:40.2	& 8.021   &   5.313	&   118\\
  AGC226326 & 12:23:58.69 & +07:26:57.9	& 10.65   &   7.667	&   155\\
  AGC226357 & 12:27:13.48 & +07:38:22.0	& 10.78   &   5.218	&     0\\
  AGC223724 & 12:29:33.61 & +13:11:44.4	& 8.781   &   5.315	&    55\\
  AGC224281 & 12:30:42.04 & +05:52:43.9	& 7.263   &   4.703	&   140\\
  AGC225847 & 12:38:58.22 & +13:33:03.4	& 12.34   &   6.936	&     0\\
  AGC742572 & 12:43:53.20 & +16:32:49.4	& 3.877   &   1.959	&    58\\
  AGC223205 & 12:51:09.16 & +12:03:38.0	& 23.97   &   10.43	&    43\\
  AGC223247 & 12:53:10.35 & +12:38:11.0	& 23.66   &   13.31	&   105\\
    NGC4651 & 12:43:33.25 & +16:23:28.6	& 183.8   &   99.89	&    70\\
    NGC4746 & 12:51:55.51 & +12:04:57.5	& 84.77   &   20.48	&   117\\
N12:20:30.91+06:38:23.9 & 12:20:30.87 & +06:38:23.3 &   3.216	&  2.305 &	 90\\
\hline
\end{tabular}
\]
Column 1: galaxy name \\
Column 2 and 3: right ascension and declination of the centre of the elliptical aperture \\
Column 4 and 5: semi major and minor axis of the elliptical aperture, in arcsec \\
Column 6: position angle of the elliptical aperture, measured from north counter clockwise \\ 
}
\end{table*}
\newpage

\begin{table*}[h!]
\caption{Best fit parameters of the fit of the luminosity function for individual galaxies
}
\label{galLF}
{
\[
\begin{tabular}{cccccccccc}
\hline
\noalign{\smallskip}
\hline
Name		  &$\alpha_{16}$& $\alpha$&$\alpha_{84}$&log$L^*(H\alpha)_{16}$&log$L^*(H\alpha)$&log$L^*(H\alpha)_{84}$&$\Phi^*_{16}$&$\Phi^*$&$\Phi^*_{84}$	  \\
		  &	   &	     &  	   & erg s$^{-1}$ & erg s$^{-1}$ & erg s$^{-1}$ & N dex$^{-1}$ & N dex$^{-1}$ & N dex$^{-1}$ \\
\hline
VCC58	  &  	-2.32	 &  -2.24 &	-2.15	 &  39.46 &	40.10	 &  40.71 &	 0.01	 &   0.06  &	 0.42\\
VCC66	  &  	-1.58	 &  -1.53 &	-1.48	 &  39.07 &	39.18	 &  39.31 &	25.22	 &  35.46  &	46.95\\
VCC81	  &  	-2.17	 &  -2.04 &	-1.88	 &  38.80 &	39.52	 &  40.49 &	 0.02	 &   0.31  &	 2.71\\
VCC89	  &  	-1.73	 &  -1.69 &	-1.66	 &  40.13 &	40.42	 &  40.76 &	 1.19	 &   2.32  &	 4.19\\
VCC92	  &  	-1.87	 &  -1.84 &	-1.81	 &  39.47 &	39.63	 &  39.85 &	 7.28	 &  12.43  &	18.42\\
VCC131	  &  	-1.75	 &  -1.66 &	-1.56	 &  38.98 &	39.27	 &  39.73 &	 1.79	 &   5.33  &	11.30\\
VCC145	  &  	-2.07	 &  -1.99 &	-1.89	 &  38.91 &	39.29	 &  40.12 &	 0.23	 &   2.15  &	 7.41\\
VCC307	  &  	-1.38	 &  -1.36 &	-1.34	 &  39.25 &	39.29	 &  39.34 &    201.13	 & 222.31  &   247.82\\
VCC340	  &  	-1.12	 &  -0.96 &	-0.78	 &  39.41 &	39.71	 &  40.09 &	 2.31	 &   4.72  &	 8.19\\
VCC465	  &  	-1.30	 &  -1.23 &	-1.17	 &  38.98 &	39.09	 &  39.23 &	30.28	 &  41.07  &	53.47\\
VCC491	  &  	-1.41	 &  -1.35 &	-1.28	 &  39.32 &	39.48	 &  39.68 &	 9.77	 &  15.11  &	21.32\\
VCC699	  &  	-1.25	 &  -1.14 &	-1.04	 &  39.44 &	39.65	 &  39.95 &	 4.77	 &   8.17  &	12.63\\
VCC801	  &  	-1.58	 &  -1.54 &	-1.50	 &  40.60 &	40.80	 &  40.94 &	 1.15	 &   1.66  &	 2.43\\
VCC905	  &  	-2.49	 &  -2.34 &	-2.19	 &  39.32 &	40.01	 &  40.71 &	 0.00	 &   0.01  &	 0.18\\
VCC939	  &  	-2.43	 &  -2.32 &	-2.21	 &  39.26 &	39.99	 &  40.67 &	 0.00	 &   0.03  &	 0.38\\
VCC975	  &  	-2.11	 &  -2.04 &	-1.96	 &  39.42 &	40.01	 &  40.65 &	 0.05	 &   0.23  &	 1.21\\
VCC1091	  &  	-1.58	 &  -1.43 &	-1.27	 &  38.83 &	39.13	 &  39.68 &	 1.68	 &   5.48  &	11.88\\
VCC1356	  &  	-1.68	 &  -1.45 &	-1.18	 &  38.73 &	39.29	 &  40.26 &	 0.16	 &   1.24  &	 4.68\\
VCC1508	  &  	-1.75	 &  -1.69 &	-1.64	 &  39.36 &	39.58	 &  39.90 &	 3.26	 &   6.79  &	11.55\\
VCC1524	  &  	-2.05	 &  -1.94 &	-1.79	 &  38.75 &	39.27	 &  40.16 &	 0.12	 &   1.25  &	 5.97\\
VCC1555	  &  	-1.79	 &  -1.76 &	-1.73	 &  39.35 &	39.49	 &  39.66 &	11.15	 &  17.16  &	24.73\\
VCC1585	  &  	-2.85	 &  -2.50 &	-2.19	 &  38.65 &	39.60	 &  40.56 &	 0.00	 &   0.01  &	 0.31\\
VCC1673	  &  	-1.37	 &  -1.30 &	-1.24	 &  38.85 &	38.97	 &  39.10 &	37.85	 &  50.77  &	67.19\\
VCC1791	  &  	-1.87	 &  -1.76 &	-1.65	 &  39.37 &	39.90	 &  40.59 &	 0.10	 &   0.49  &	 1.74\\
VCC1987	  &  	-1.52	 &  -1.49 &	-1.46	 &  39.50 &	39.60	 &  39.72 &	25.33	 &  33.10  &	41.57\\
NGC4651   &     -1.55    &  -1.50 &	-1.46    &  39.07 &	39.18    &  39.31 &	28.49    &  38.80  &	50.81\\
NGC4746   &     -0.83    &  -0.74 &	-0.65    &  38.83 &	38.92    &  39.01 &	67.05    &  81.48  &	95.70\\
\hline
\end{tabular}
\]
Column 1: galaxy name \\
Column 2-4: $\alpha$ parameter of the best fit Schechter function with uncertainty estimates as the 16th and 84th percentiles
of the marginalised posterior distribution\\
Colunn 5-7: $L^*(H\alpha)$ parameter of the best fit Schechter function with uncertainty estimates as the 16th and 84th percentiles
of the marginalised posterior distribution\\
Column 8-10: $\Phi^*$ parameter of the best fit Schechter function with uncertainty estimates as the 16th and 84th percentiles
of the marginalised posterior distribution\\
}
\end{table*}
\newpage
\clearpage

\section{Scaling relations}
\label{appC}

\subsection{Main scaling relations }

In this Appendix we present all the scaling relations analysed in Sec. 4.2. 
These include the scaling relations between the total number of {\hii} regions, the brightest {\hii} region, 
their relative contribution to the total H$\alpha$ luminosity of the target galaxies, as well as the best fit parameters of the
luminosity function of the {\hii} regions and several physical parameters 
representing their hosting galaxies such as the star formation activity, the stellar mass, the mean star formation activity and stellar mass 
surface density, the metallicity, the molecular-to-atomic gas fraction, the gas-to-dust ratio, the specific star formation rate, and 
the morphological type.
The parameters of the best fit 
are shown in the figures and given in Table \ref{Tabscalingfit} 
only whenever the probability $P$ that the two variables are correlated with $P$ $\geq$ 95\%\ ($p$-value $<$0.05).

\begin{figure*}[h!]
\centering
\includegraphics[width=0.4\textwidth]{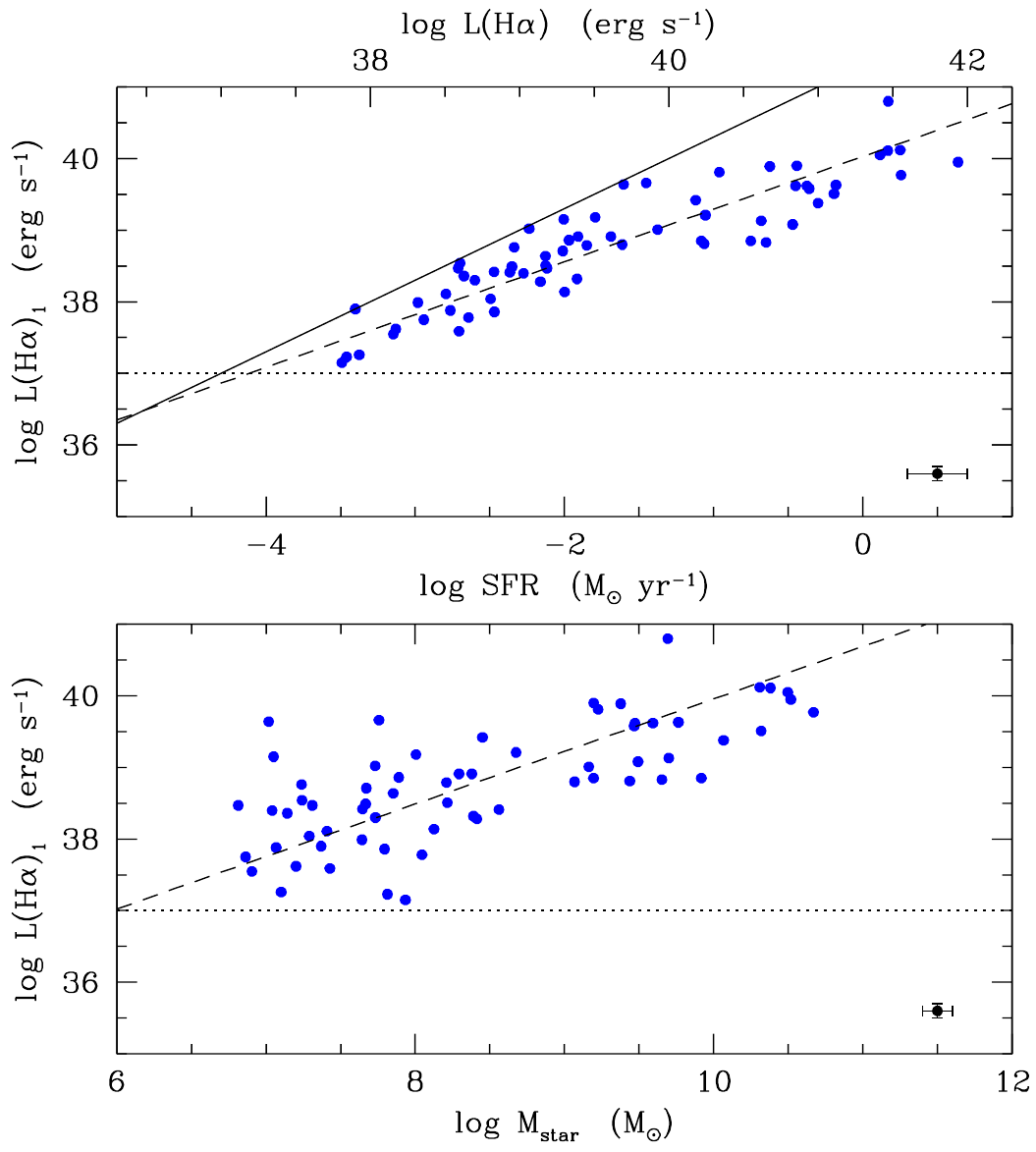}
\includegraphics[width=0.4\textwidth]{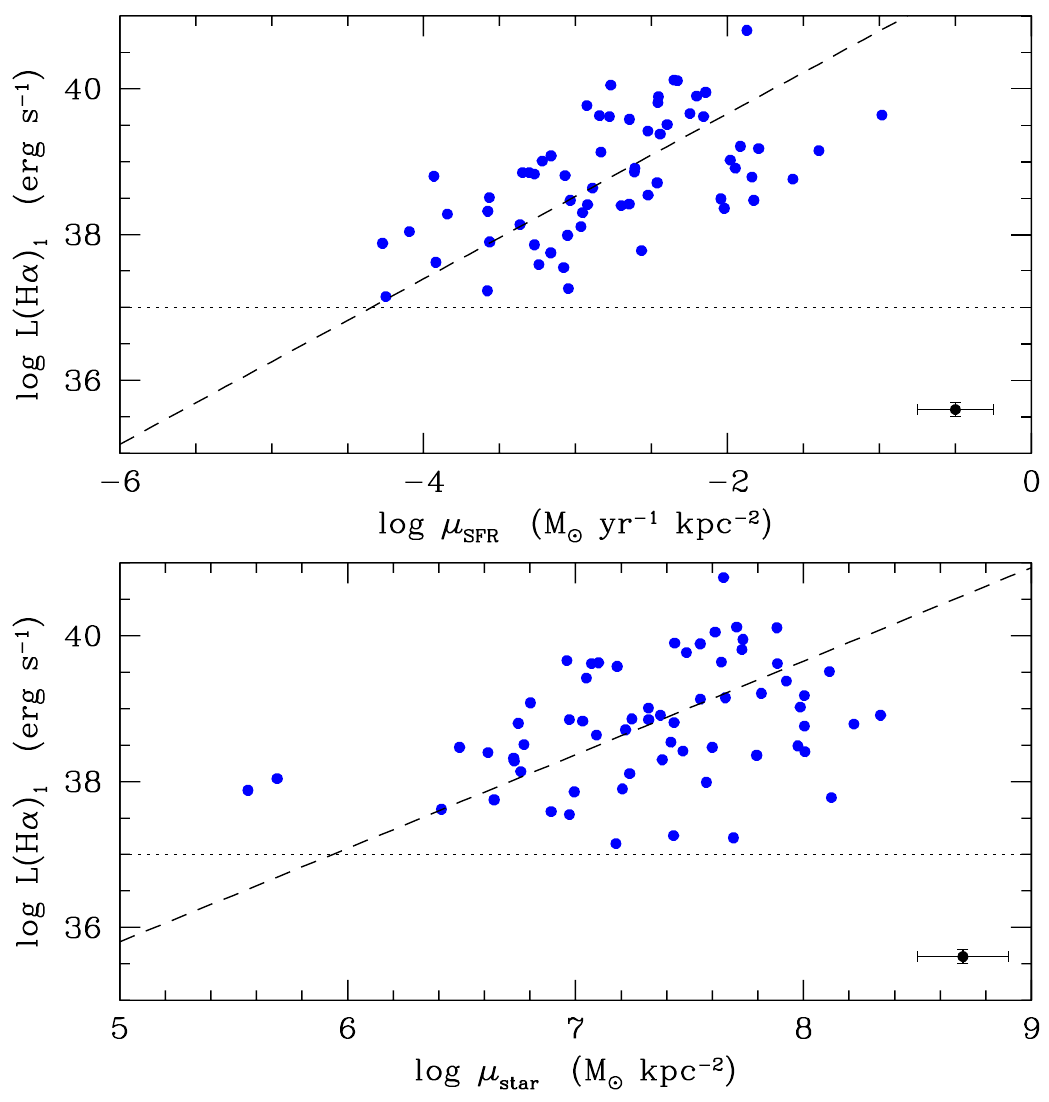}\\
\caption{Relation between the H$\alpha$ luminosity of the brightest {\hii} region and the total star formation rate  
(upper left panel), total stellar mass (lower left), mean star
formation rate surface density (upper right), and mean stellar mass surface density (lower right) of the host galaxies. The black dashed line shows the best fit to the data (bisector fit),
the dotted line the limiting H$\alpha$ luminosity of the selected sample ($L(H\alpha)$ $\geq$ 10$^{37}$ erg s$^{-1}$),
and the solid black line the $L(H\alpha)_{1}$ $\leq$ $L(H\alpha)$ physical limit in the relation. The dot in the lower right corner shows the typical error bar in the data.}
\label{scaling1Lup}%
\end{figure*}

\begin{figure*}[h!]
\centering
\includegraphics[width=0.4\textwidth]{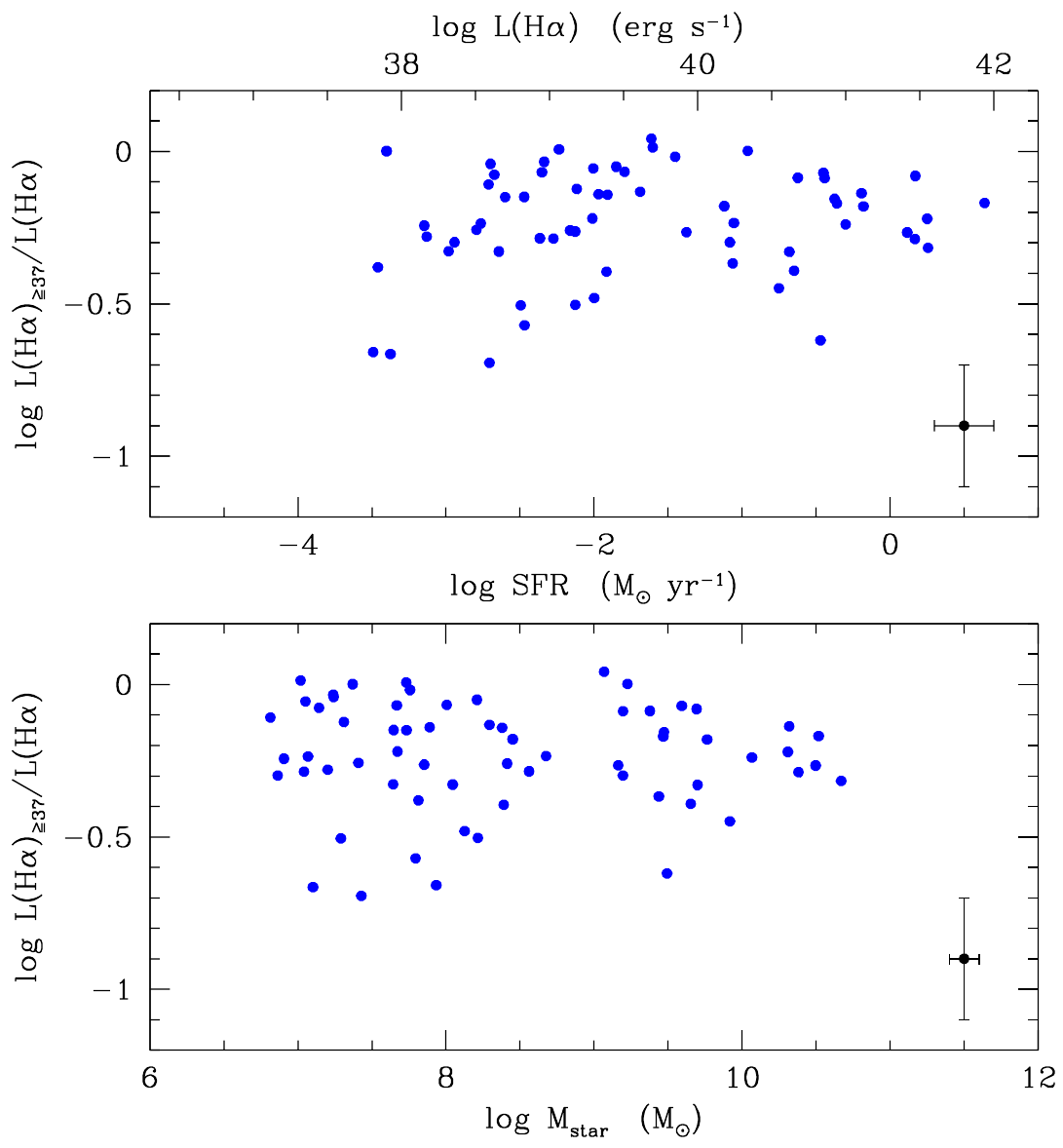}
\includegraphics[width=0.4\textwidth]{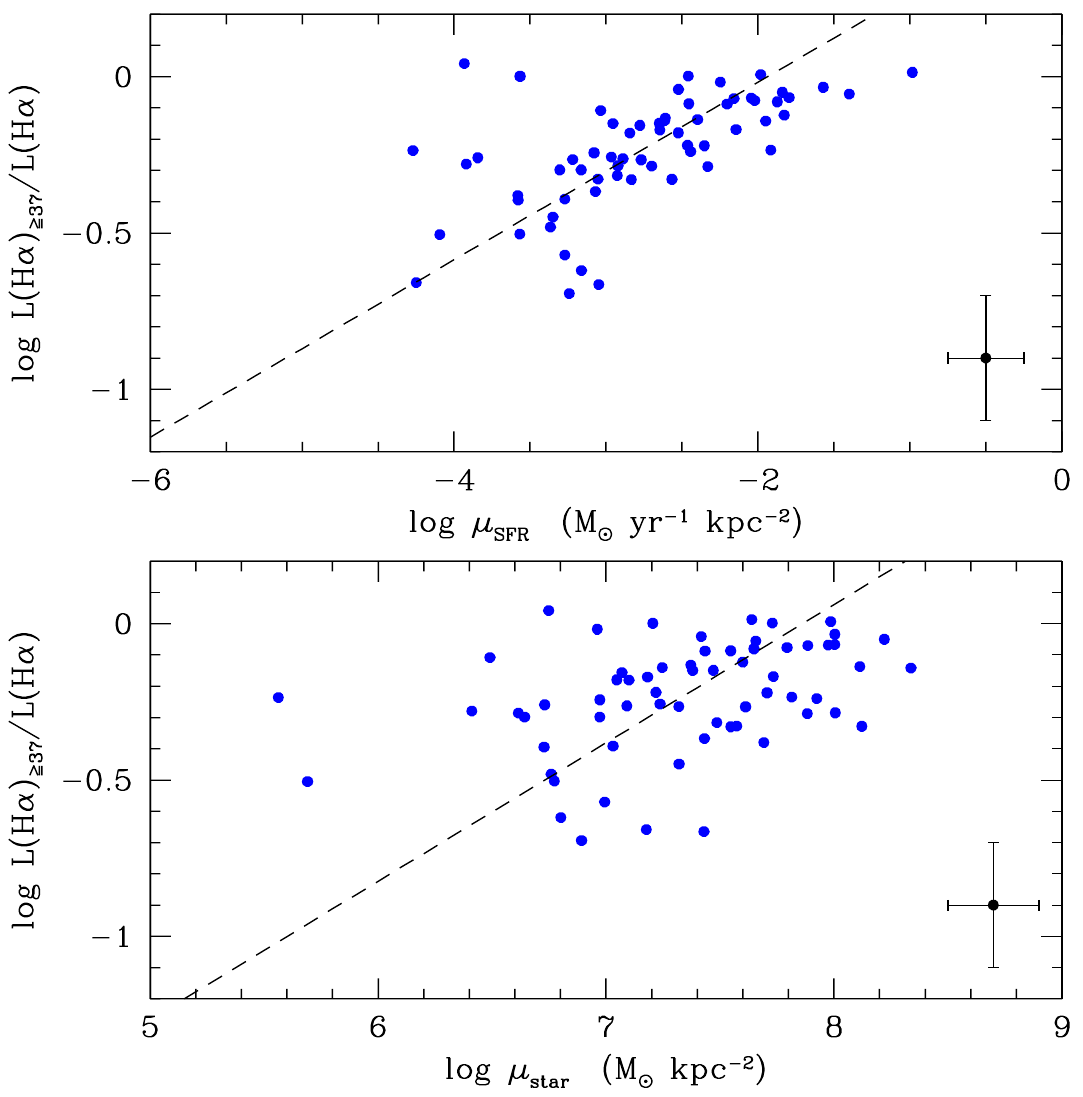}\\
\caption{Relation between $L(H\alpha)_{\geq 37}/L(H\alpha)$, defined as the ratio between the sum of the H$\alpha$ luminosity of all the {\hii} regions 
with luminosity $L(H\alpha)$ $\geq$ 10$^{37}$ erg s$^{-1}$ and the integrated H$\alpha$ luminosity of the galaxy, and the total star formation rate  
(upper left panel), total stellar mass (lower left), mean star
formation rate surface density (upper right), and mean stellar mass surface density (lower right) of the host galaxies. The dot in the lower right corner shows the typical error bar in the data.
}
\label{scaling1L37}%
\end{figure*}

\twocolumn

\begin{figure}[h!]
\centering
\includegraphics[width=0.41\textwidth]{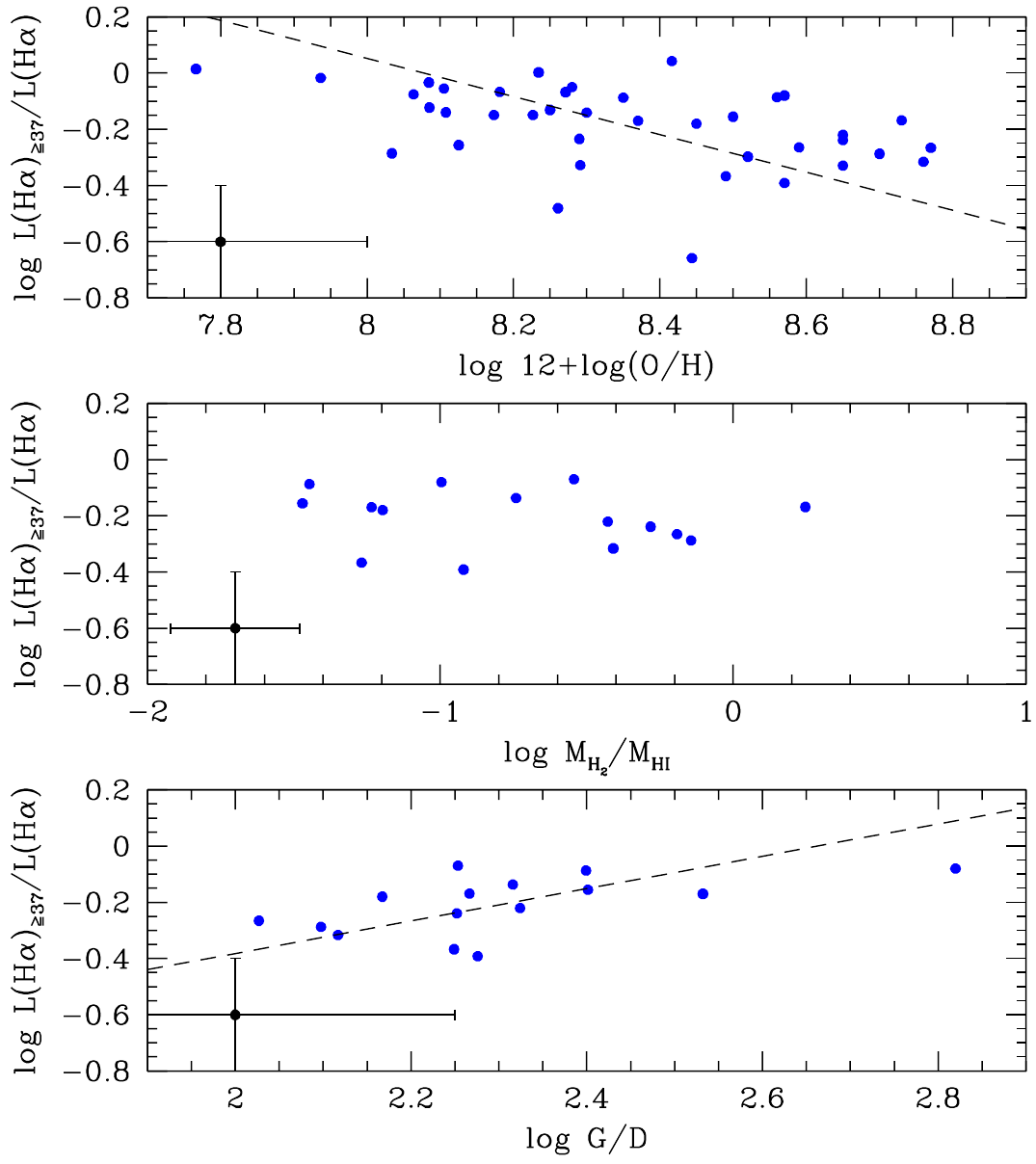}\\
\caption{Relation between $L(H\alpha)_{\geq 37}/L(H\alpha)$, defined as the ratio between the sum of the H$\alpha$ luminosity of all the {\hii} regions 
with luminosity $L(H\alpha)$ $\geq$ 10$^{37}$ erg s$^{-1}$ and the integrated H$\alpha$ luminosity of the galaxy, and the metallicity  
(upper panel), the molecular-to-atomic gas ratio (central panel), and the total gas-to-dust ratio (lower panel) of the host galaxies. 
The black dashed line shows the best fit to the data (bisector fit). The dot in the lower left corner shows the typical error bar in the data.}
\label{scaling1L37physical}%
\end{figure}

\begin{figure}[h!]
\centering
\includegraphics[width=0.41\textwidth]{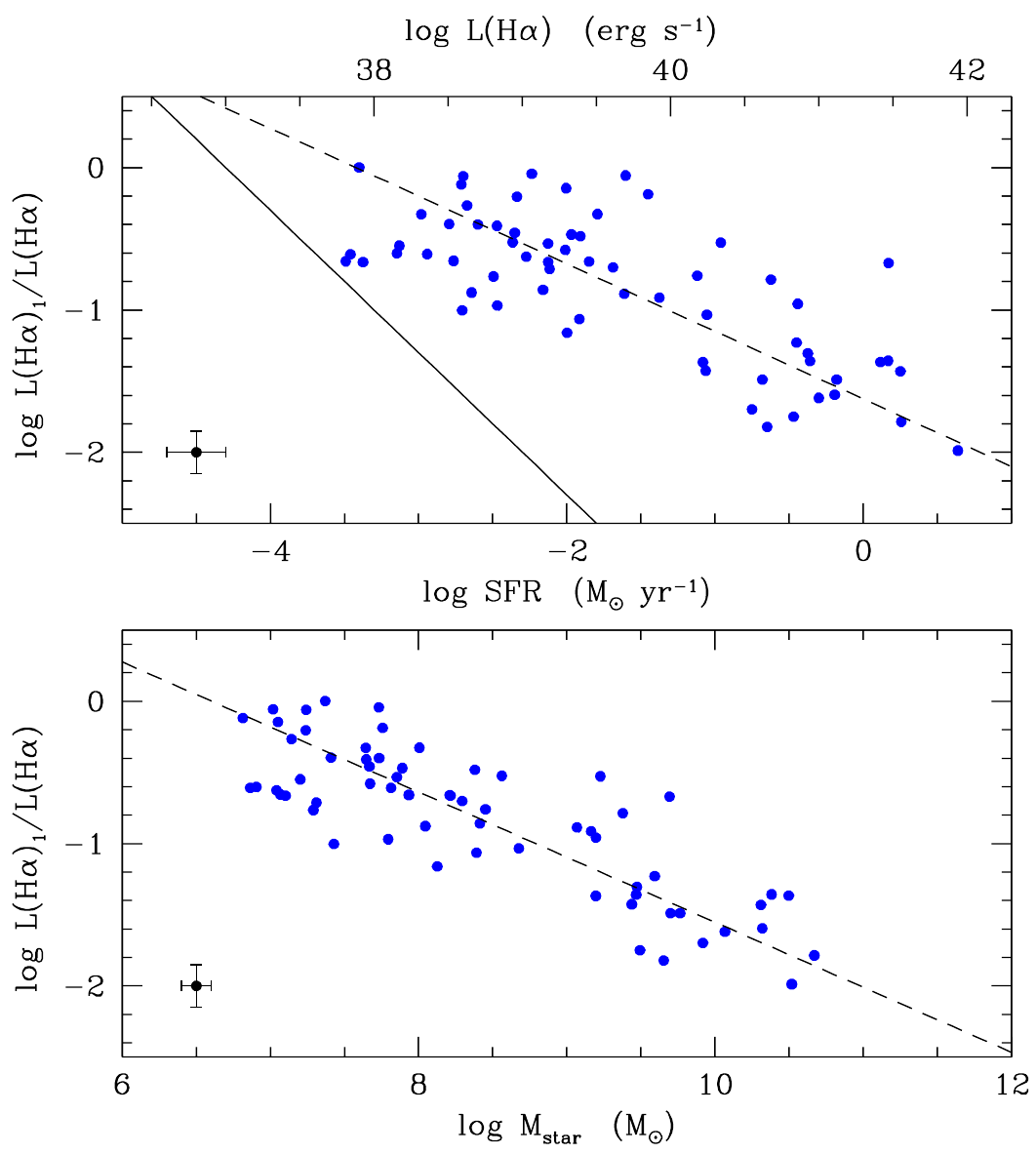}\\
\caption{Relation between $L(H\alpha)_{1}/L(H\alpha)$ and the total star formation rate  
(upper panel) and total stellar mass (lower panel) of the host galaxies. The black dashed line shows the best fit to the data (bisector fit)
and the solid black line the $L(H\alpha)_{1}$ $\leq$ 10$^{37}$ erg s$^{-1}$ physical limit in the relation.
The dot in the lower left corner shows the typical error bar in the data.}
\label{scaling2Lup}%
\end{figure}

\begin{figure}[h!]
\centering
\includegraphics[width=0.41\textwidth]{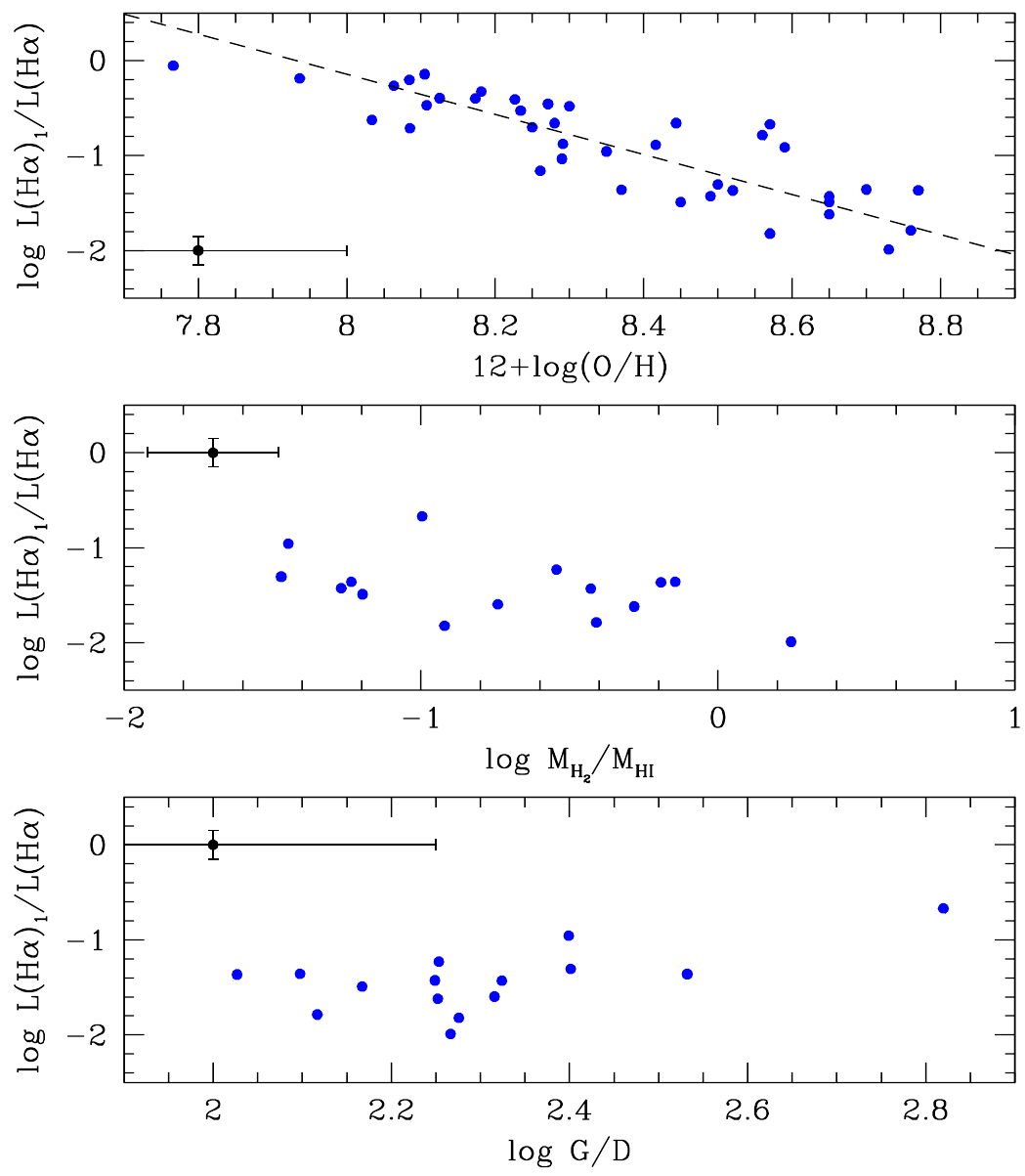}\\
\caption{Relation between $L(H\alpha)_{1}/L(H\alpha)$ and and the metallicity  
(upper panel), the molecular-to-atomic gas ratio (central panel), and the total gas-to-dust ratio (lower panel) of the host galaxies. 
The black dashed line shows the best fit to the data (bisector fit).
The dot in the lower left corner shows the typical error bar in the data.}
\label{scaling1Lupphysical}%
\end{figure}

\begin{figure}[h!]
\centering
\includegraphics[width=0.41\textwidth]{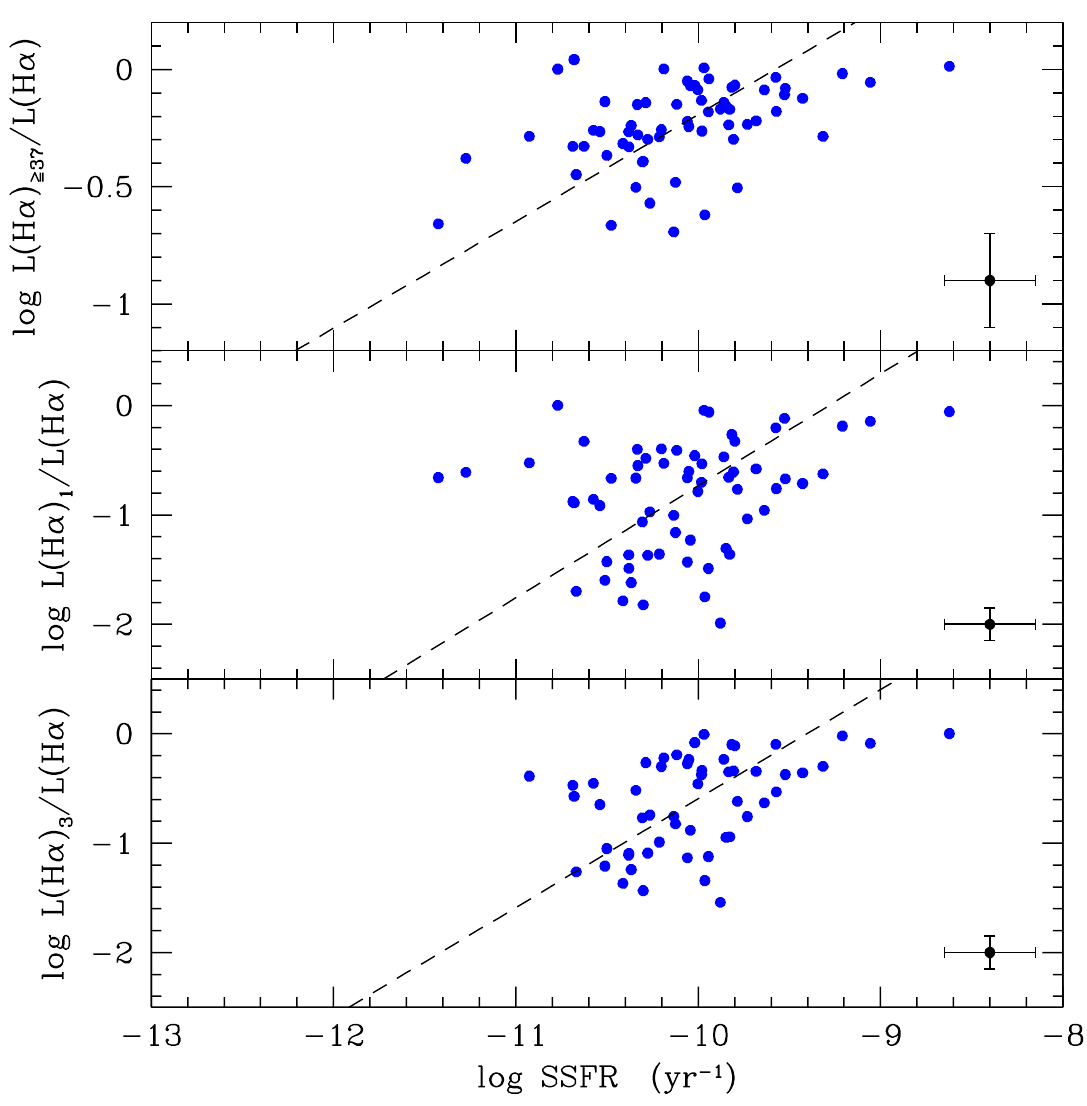}\\
\caption{Relation between the ratio of the total H$\alpha$ luminosity of {\hii} regions with $L(H\alpha)$ $\geq$ 10$^{37}$ erg s$^{-1}$  
(upper panel), of the brightest {\hii} region
(central panel), and of the sum of the H$\alpha$ luminosity of the brightest three {\hii} regions and the integrated H$\alpha$ luminosity 
and the specific star formation rate of the parent galaxies. 
The black dashed line shows the best fit to the data (bisector fit). The dot in the lower right corner shows the typical error bar in the data. }
\label{SSFR}%
\end{figure}

\onecolumn

\begin{figure}[h!]
\centering
\includegraphics[width=0.4\textwidth]{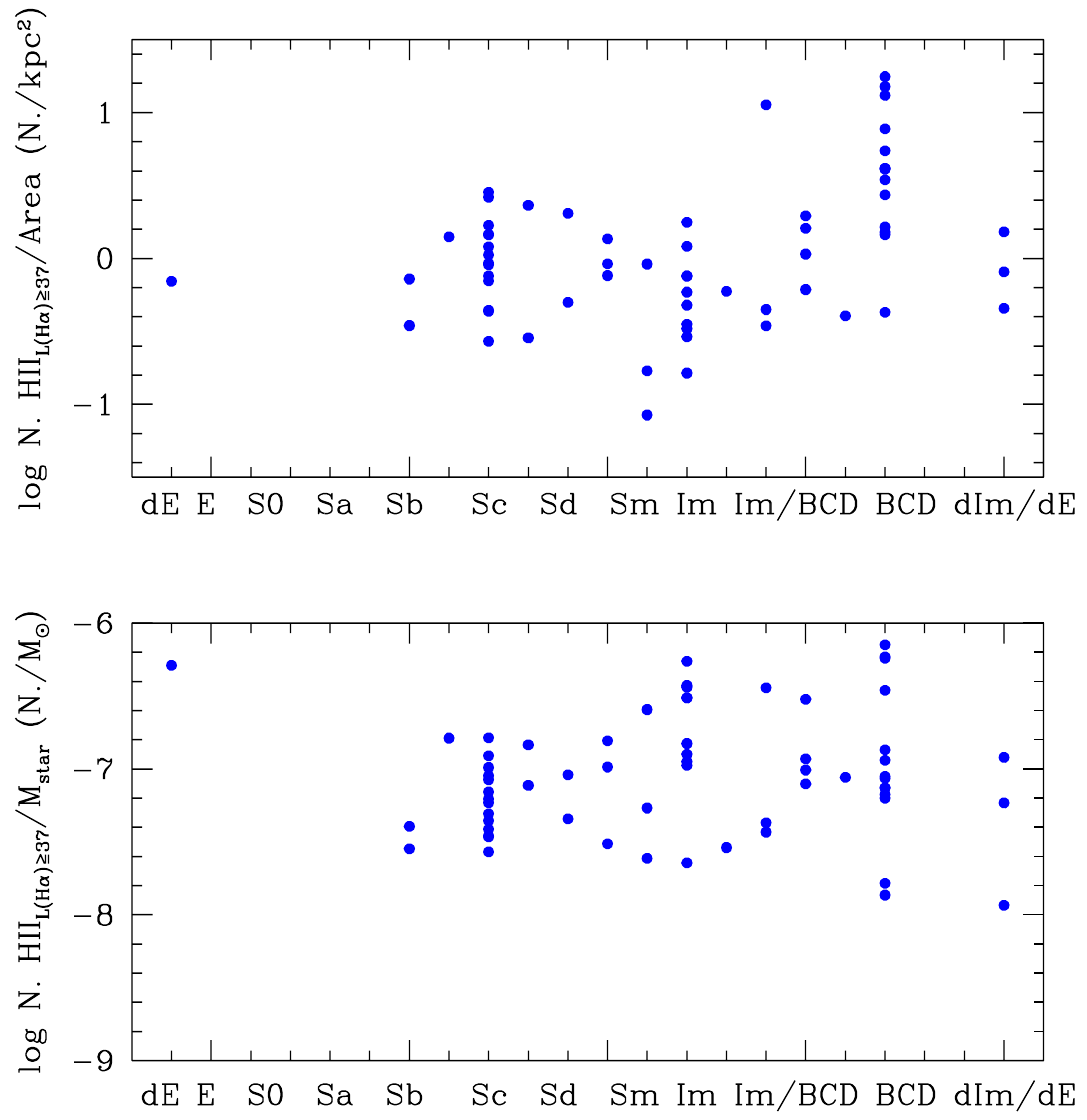}\\
\caption{Relation between the number of {\hii} regions of luminosity $L(H\alpha)$ $\geq$ 10$^{37}$ erg s$^{-1}$ per unit disc size (upper panel) 
and stellar mass (lower panel) and the morphological type.}
\label{density37type}%
\end{figure}

\begin{figure*}[h!]
\centering
\includegraphics[width=0.4\textwidth]{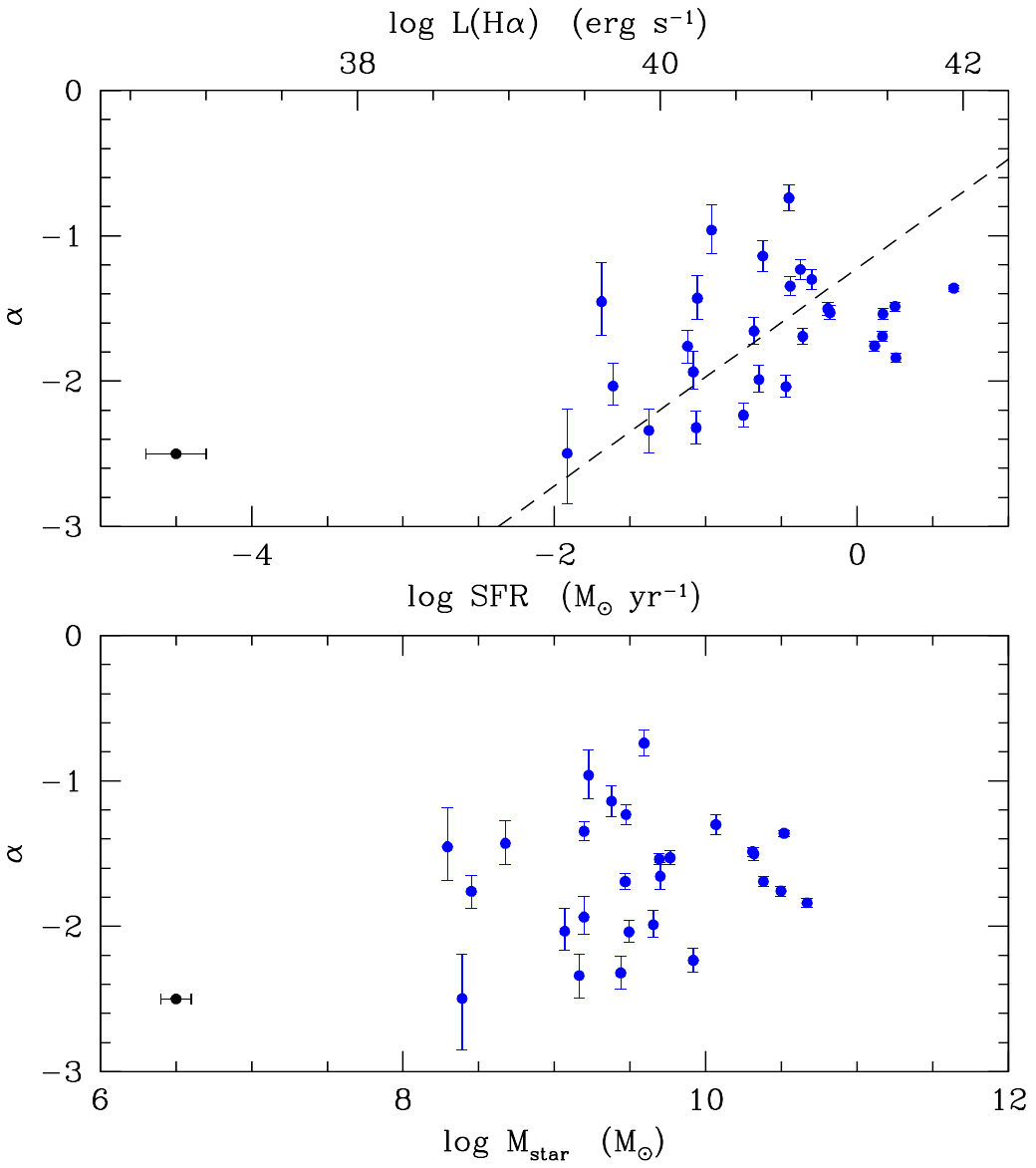}
\includegraphics[width=0.4\textwidth]{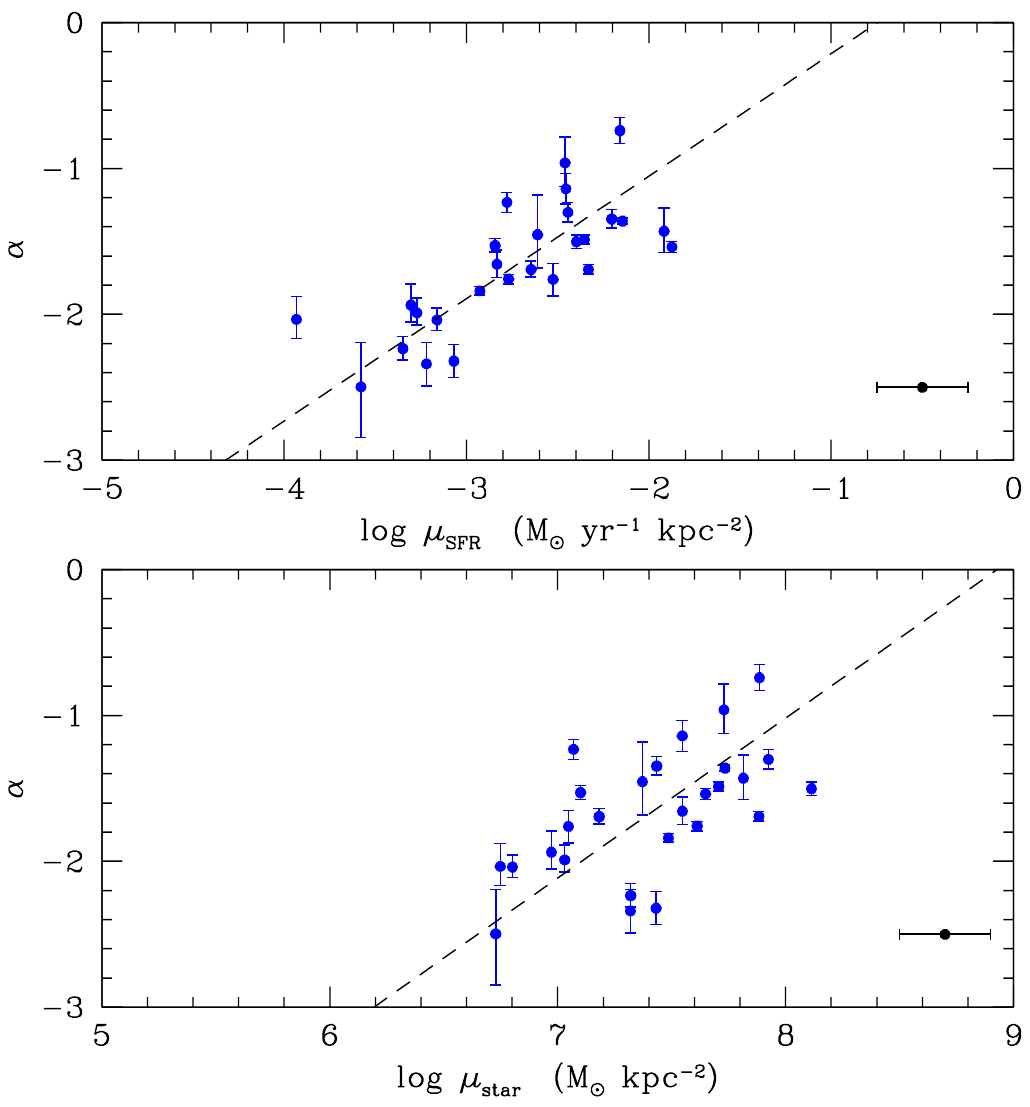}\\
\caption{Relation between the faint end slope of the luminosity function $\alpha$ and the total star formation rate  
(upper left panel), total stellar mass (lower left), mean star
formation rate surface density (upper right), and mean stellar mass surface density (lower right) of the host galaxies. 
The black dashed line shows the best fit to the data (bisector fit). The dots in the lower left and right corners show the typical error bar in the data.
}
\label{scaling1alfa}%
\end{figure*}

\begin{figure*}[h!]
\centering
\includegraphics[width=0.4\textwidth]{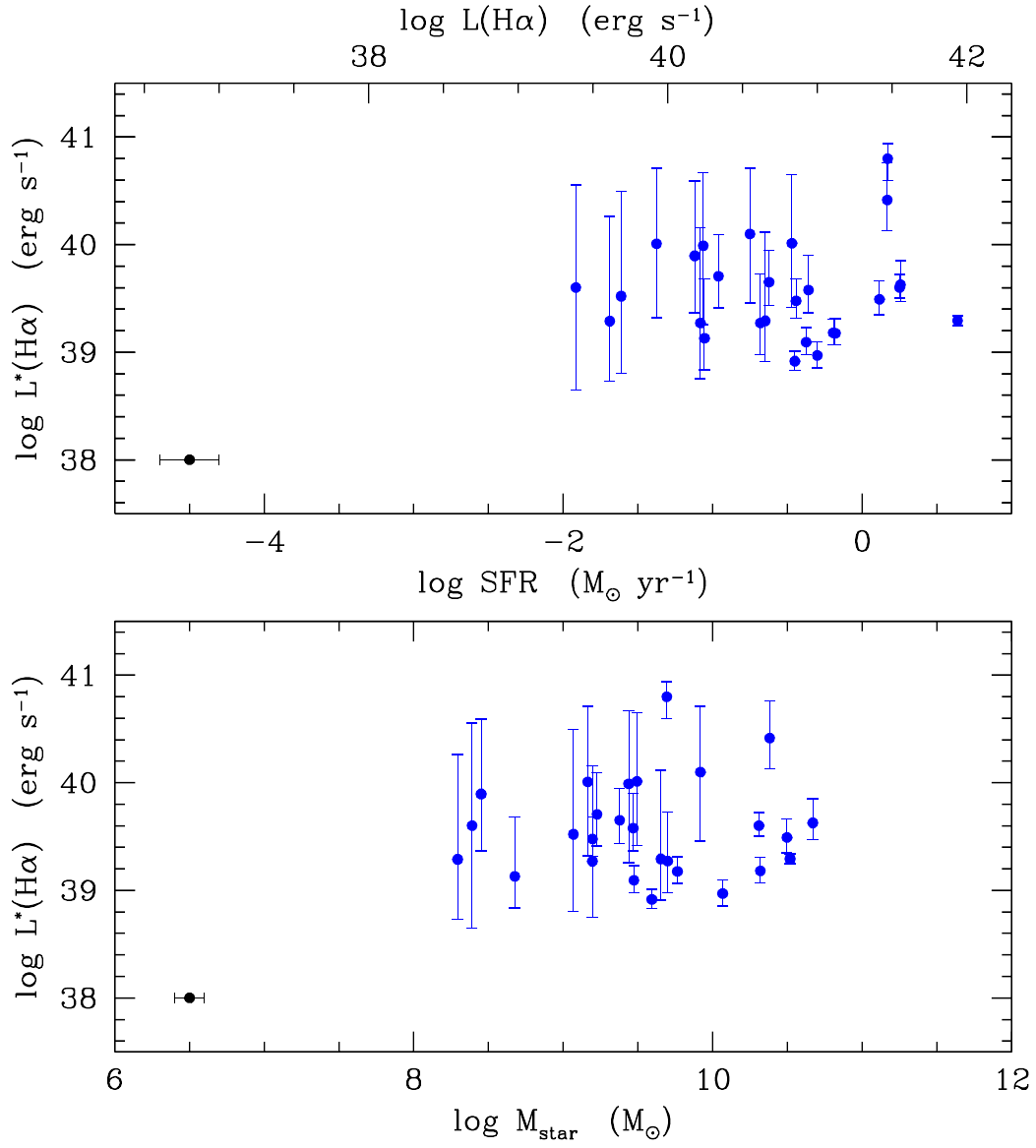}
\includegraphics[width=0.4\textwidth]{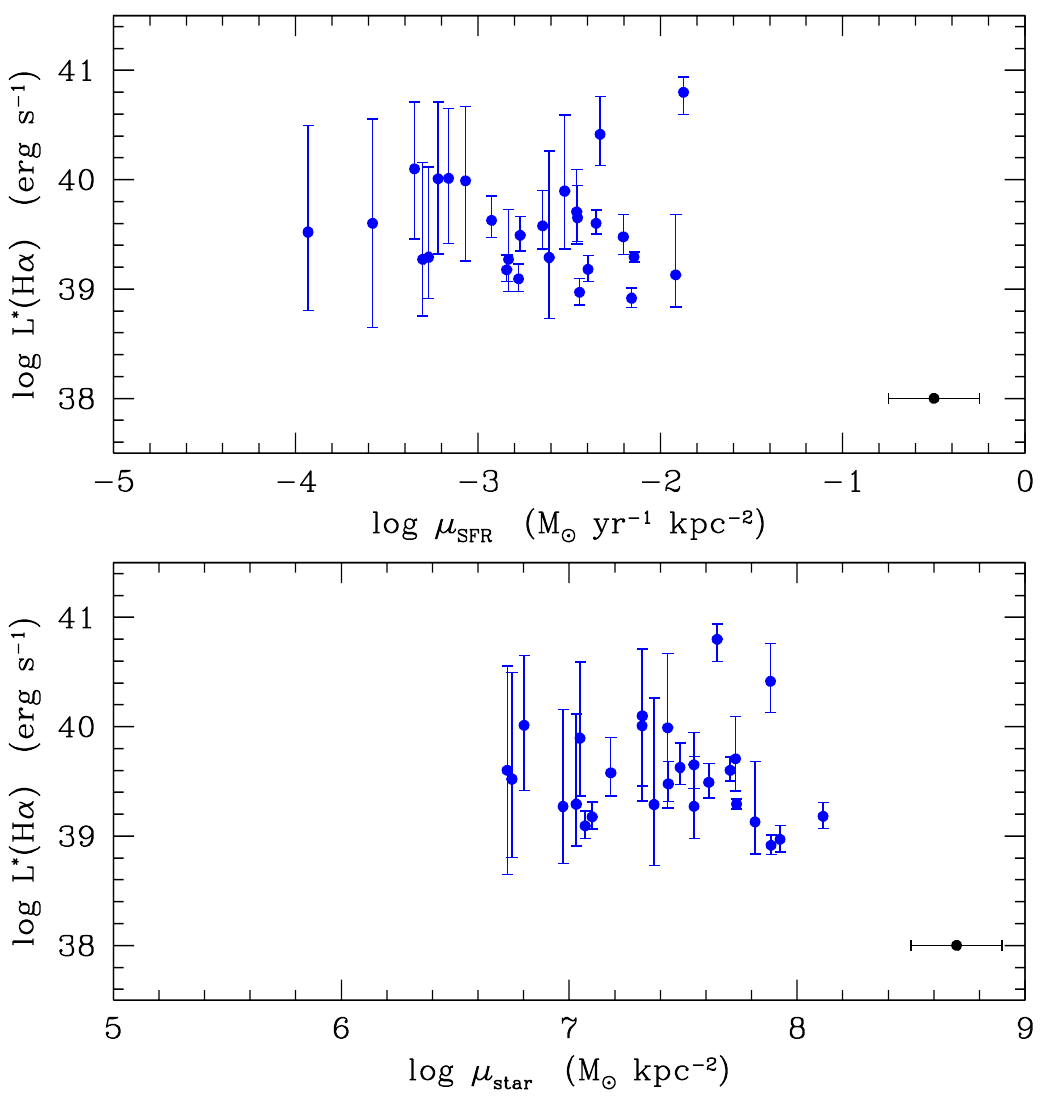}\\
\caption{Relation between the characteristic luminosity of the luminosity function $L^*$ and the total star formation rate  
(upper left panel), total stellar mass (lower left), mean star
formation rate surface density (upper right), and mean stellar mass surface density (lower right) of the host galaxies. 
The dots in the lower left and right corners show the typical error bar in the data.}
\label{scaling1Lstar}%
\end{figure*}

\begin{figure*}[h!]
\centering
\includegraphics[width=0.4\textwidth]{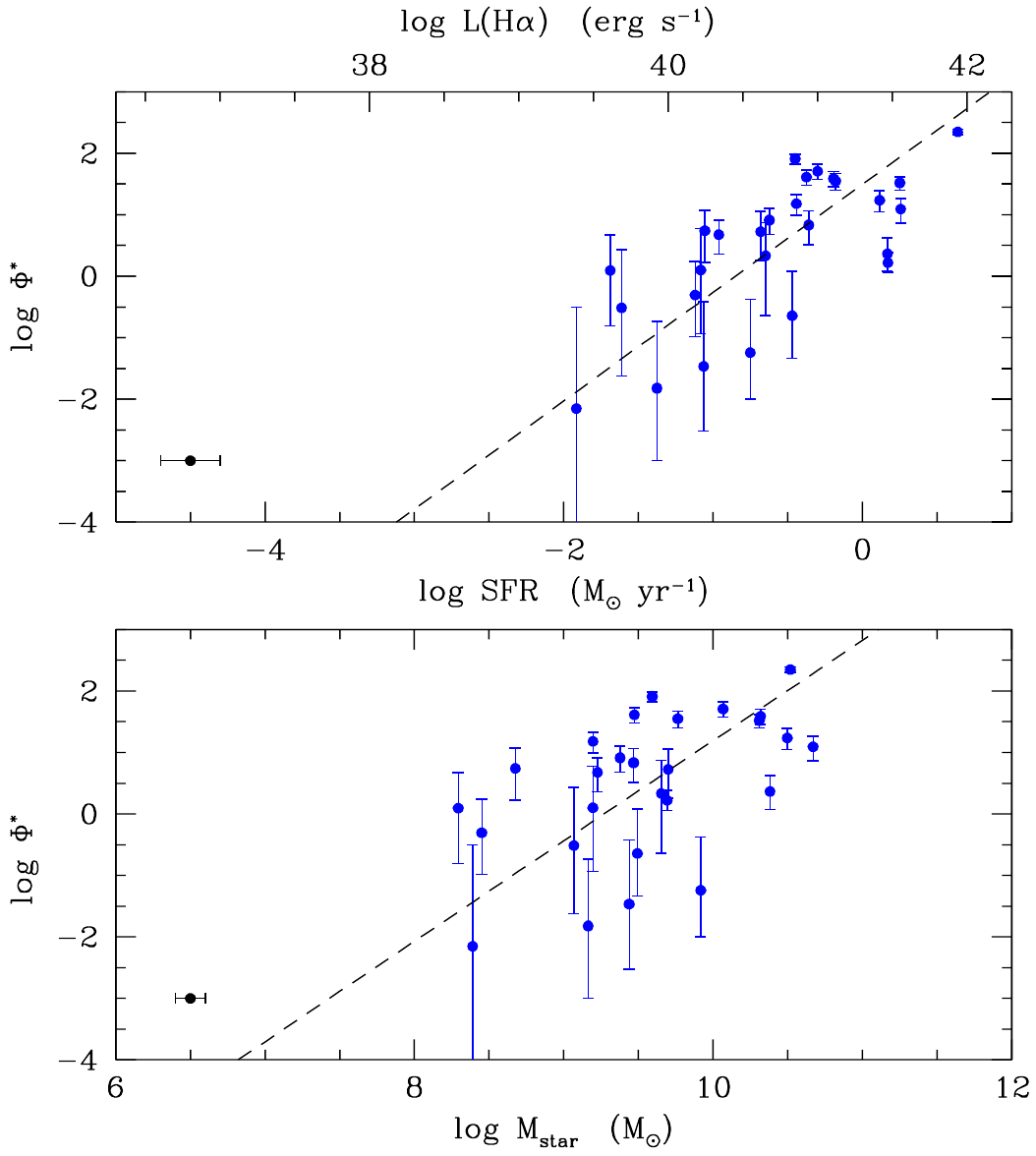}
\includegraphics[width=0.4\textwidth]{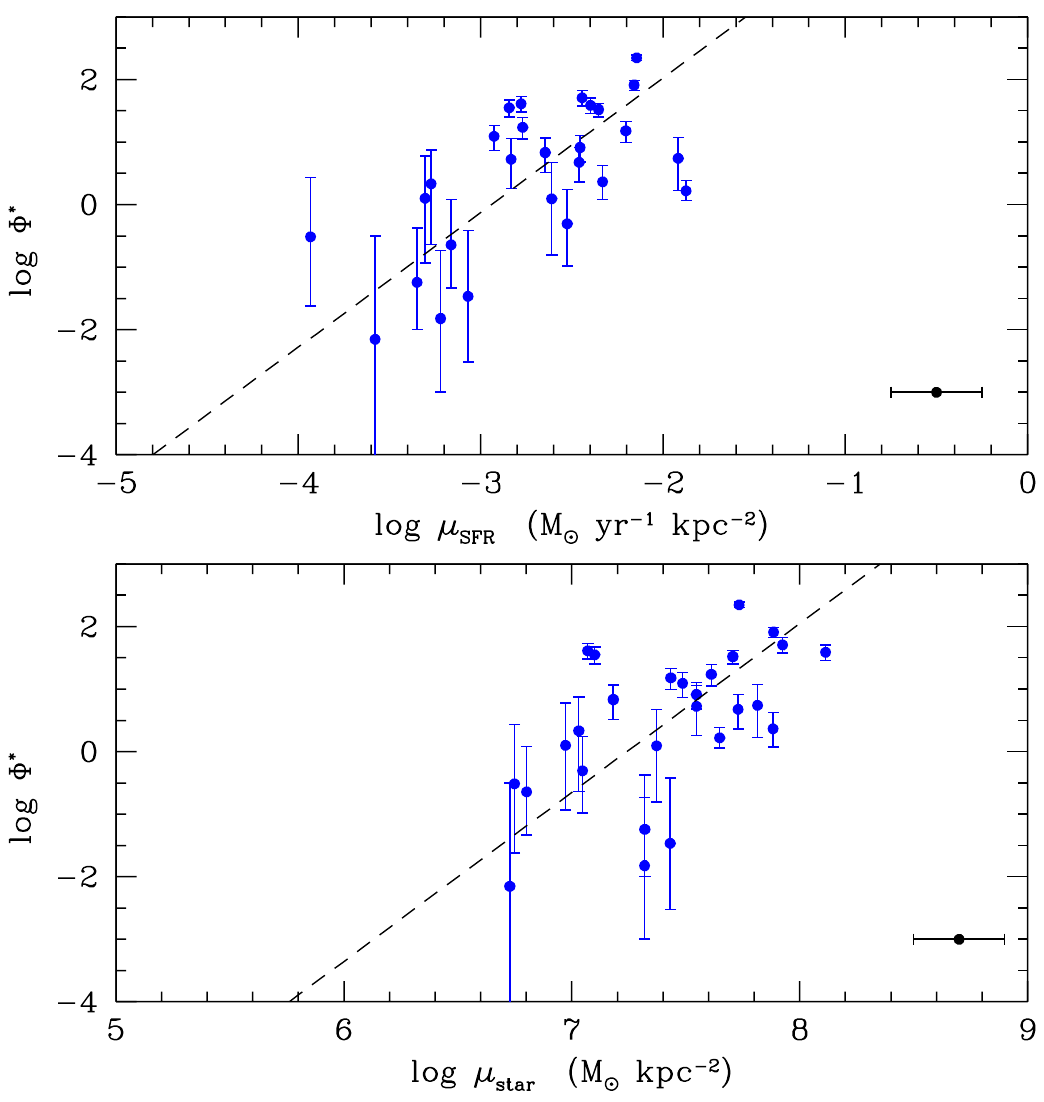}\\
\caption{Relation between the characteristic number density of the luminosity $\phi^*$ and the total star formation rate  
(upper left panel), total stellar mass (lower left), mean star
formation rate surface density (upper right), and mean stellar mass surface density (lower right) of the host galaxies. 
The black dashed line shows the best fit to the data (bisector fit). The dots in the lower left and right corners show the typical error bar in the data.}
\label{scaling1phistar}%
\end{figure*}

\begin{figure}[h!]
\centering
\includegraphics[width=0.4\textwidth]{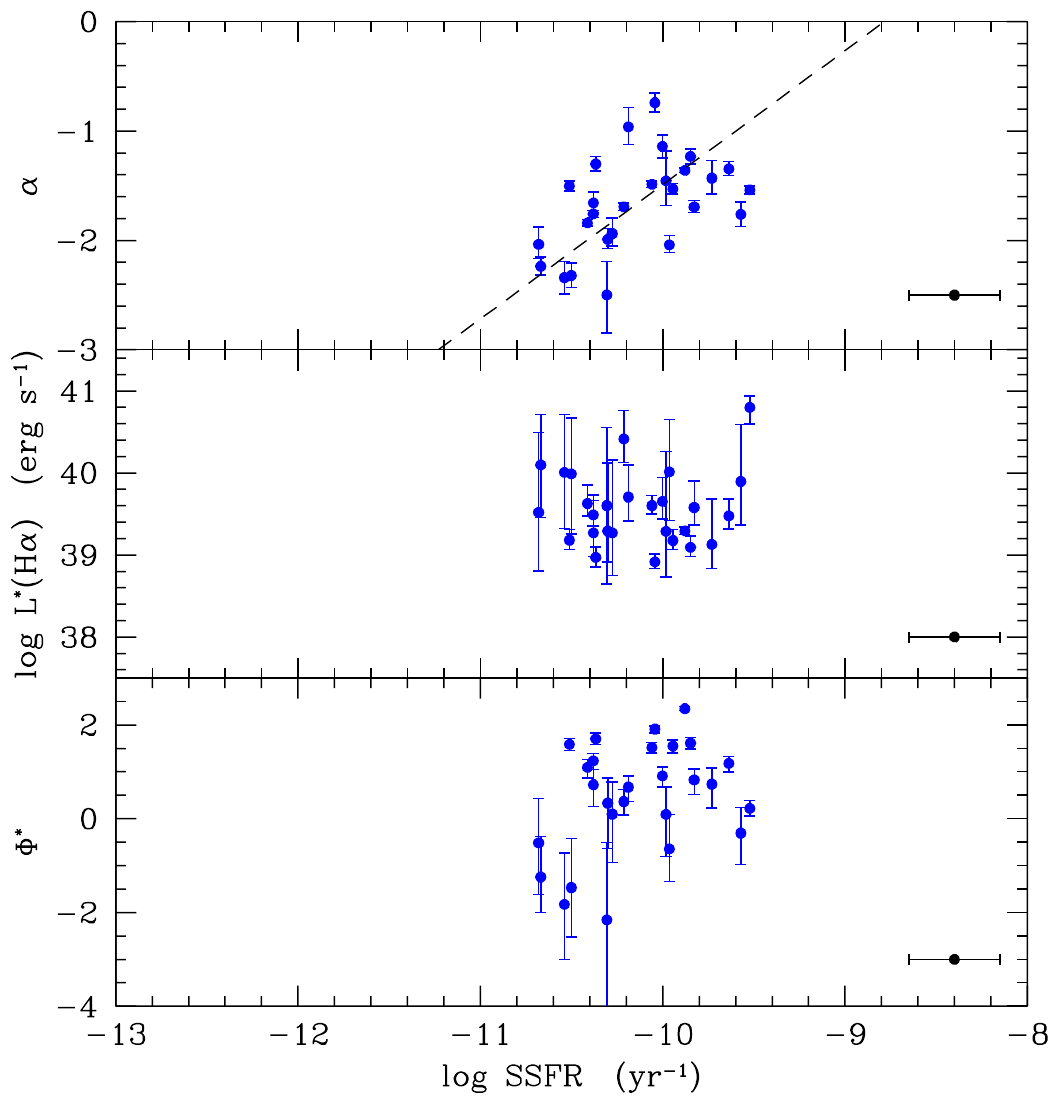}\\
\caption{Relation between the faint end slope $\alpha$ (upper panel), the characteristic luminosity $L^*$ (central panel), and $\phi^*$ (lower panel)
of the luminosity function
and the specific star formation rate of the parent galaxies. 
The black dashed line shows the best fit to the data (bisector fit). The dot in the lower right corner shows the typical error bar in the data.}
\label{SSFRLF}%
\end{figure}

\subsection{Scaling relations with the first three ranked {\hii} regions}

Often used in the literature are also the scaling relations with the mean ($\langle L(H\alpha)_{3}\rangle$) or the sum ($L(H\alpha)_{3}$) of the first three ranked {\hii} regions. 
These relations generally follow those found for the first ranked {\hii} region presented in Sec. 4 and discussed in Sec. 5.
For completeness, we include these relations in this Appendix. The best fit parameters of these relations are given in Table \ref{Tabscalingfit}.

\begin{figure*}[h!]
\centering
\includegraphics[width=0.49\textwidth]{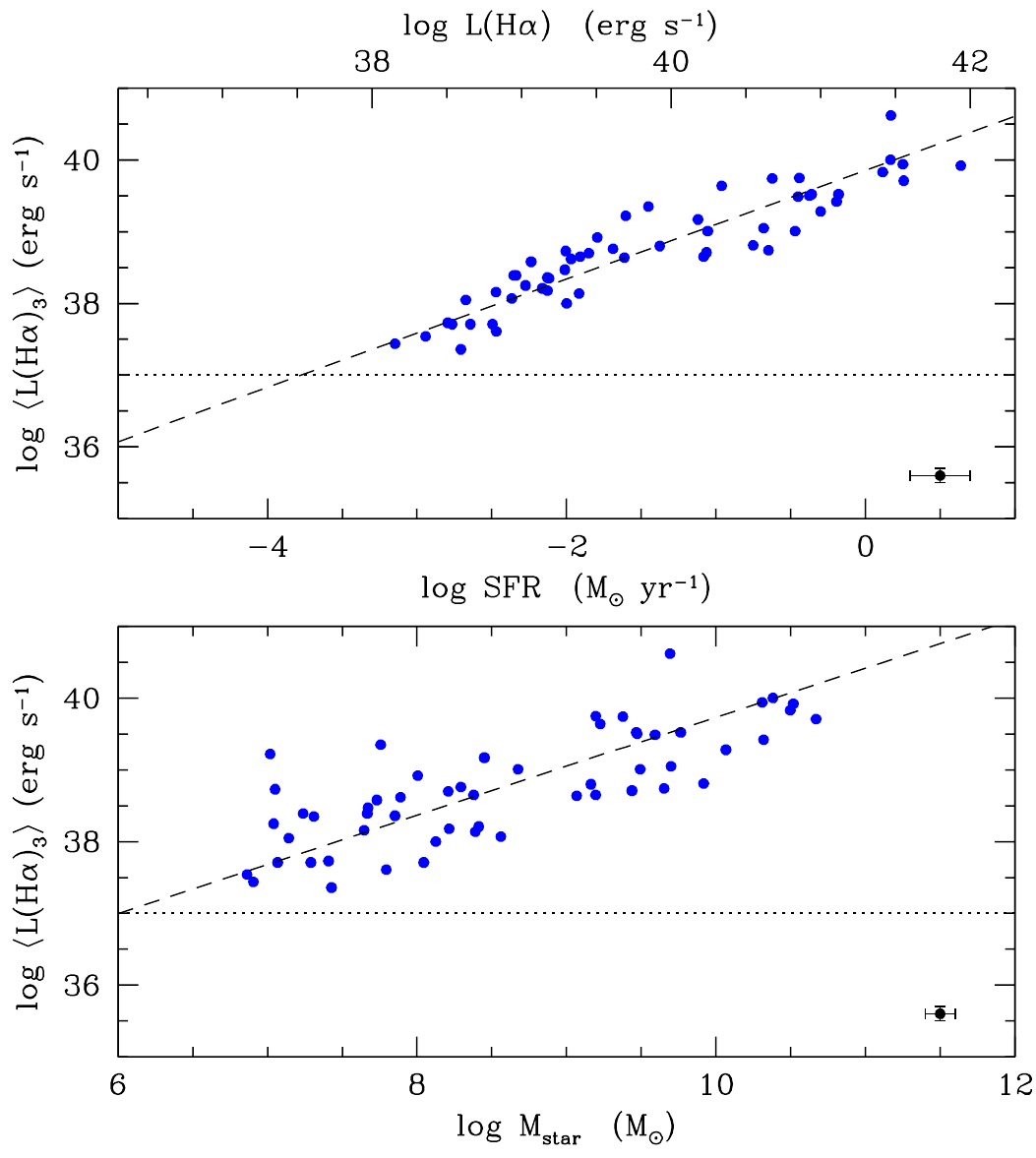}
\includegraphics[width=0.49\textwidth]{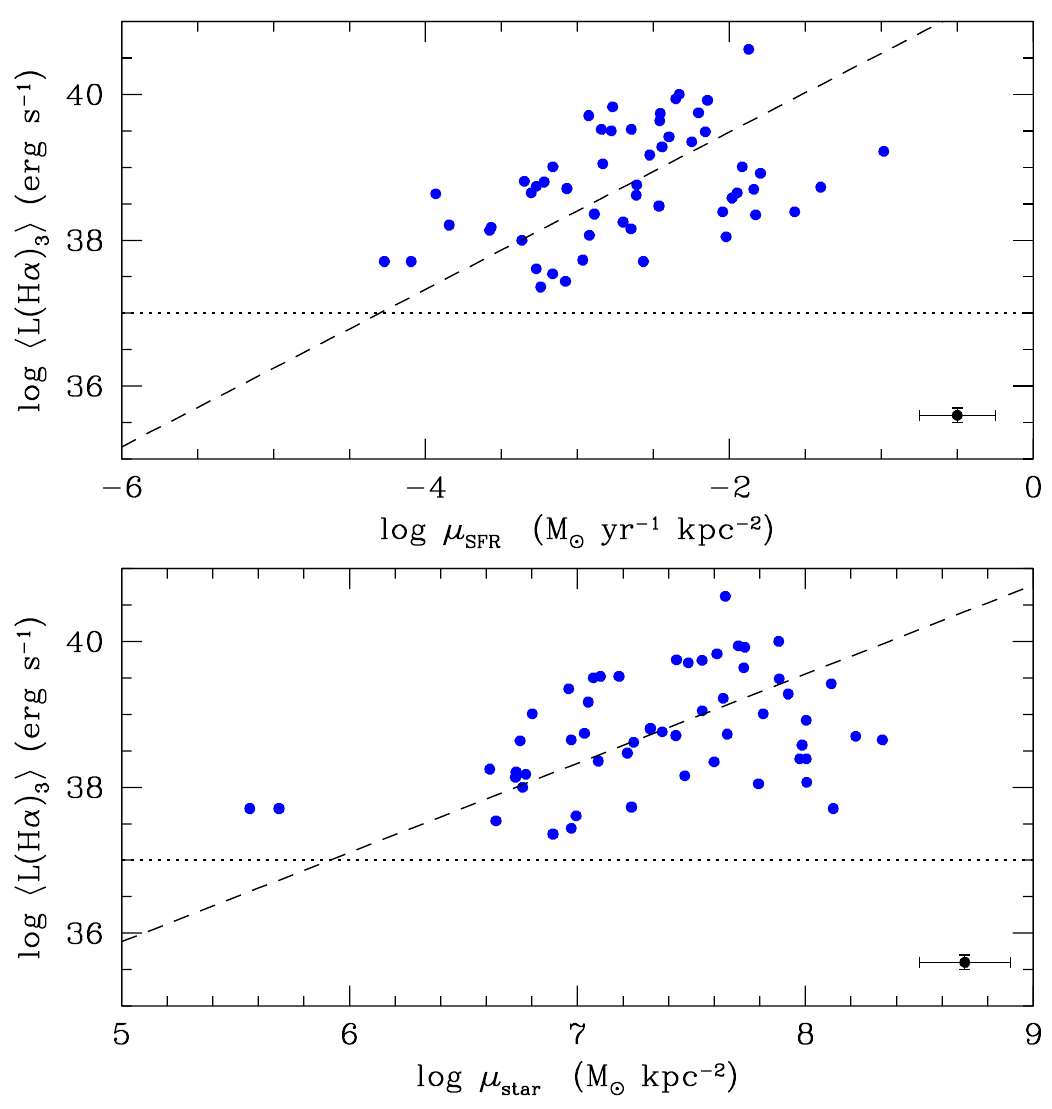}\\
\caption{Relation between the mean H$\alpha$ luminosity of the three brightest {\hii} regions and the total star formation rate  
(upper left panel), total stellar mass (lower left), mean star
formation rate surface density (upper right), and mean stellar mass surface density (lower right) of the host galaxies. The black dashed line shows the best fit to the data (bisector fit),
the dotted line the limiting H$\alpha$ luminosity of the selected sample ($L(H\alpha)$ $\geq$ 10$^{37}$ erg s$^{-1}$). The dot in the lower right corner shows the typical error bar in the data.}
\label{scaling1Lup3}%
\end{figure*}

Figure \ref{scaling1Lup3} shows the relations between the mean H$\alpha$ 
luminosity of the three brightest {\hii} regions and the total star formation rate, 
stellar mass (and their surface densities) of the parent galaxies. 
The mean H$\alpha$ luminosity of the three brightest {\hii} region is
tightly related to the star formation rate, stellar mass, mean star formation rate, and stellar mass surface density of their host.

\begin{figure}[h!]
\centering
\includegraphics[width=0.44\textwidth]{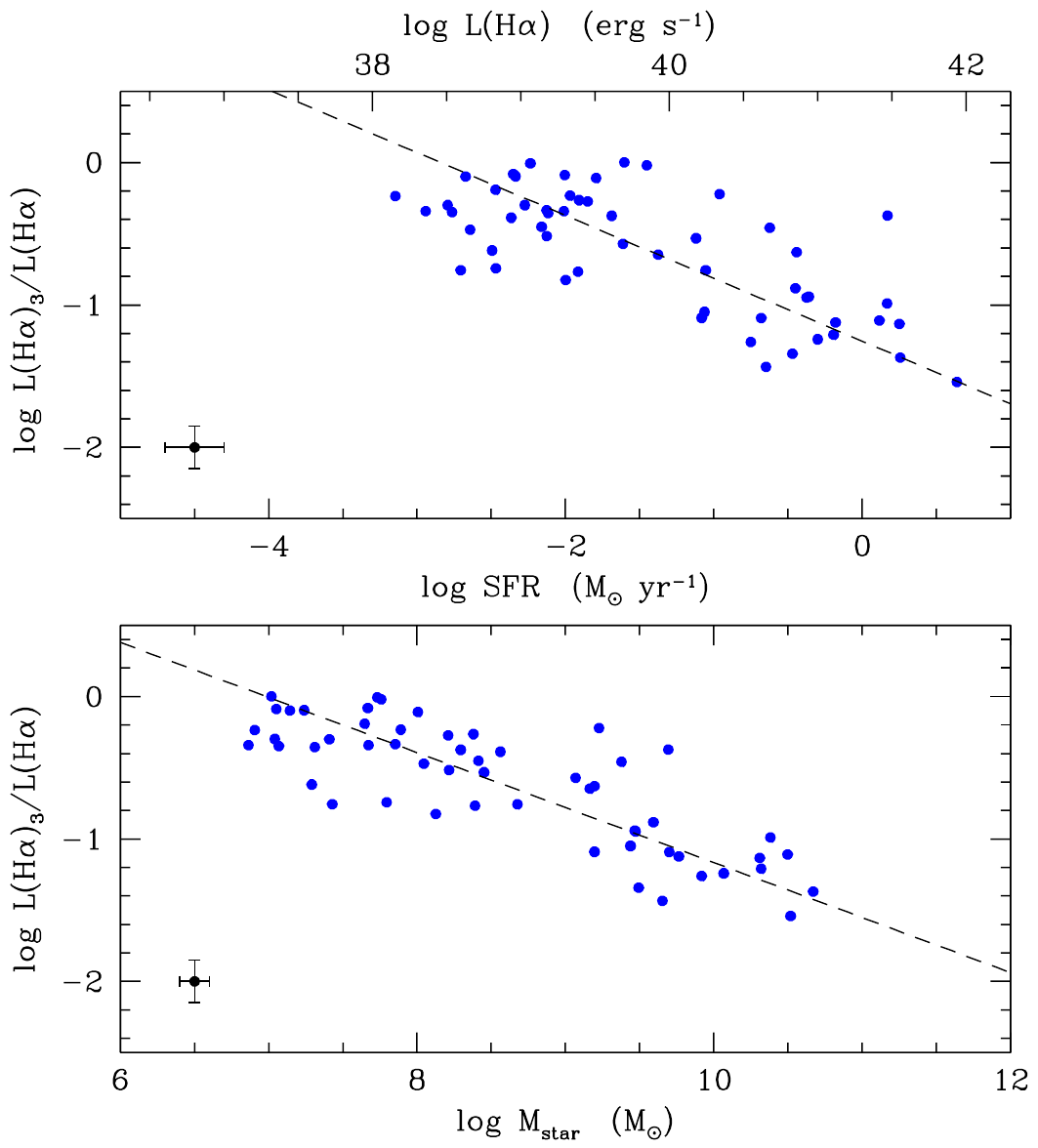}\\
\caption{Relation between $L(H\alpha)_{3}/L(H\alpha)$ (contribution of the brightest three {\hii} regions to the total H$\alpha$ luminosity of the galaxy)
and the total star formation rate  
(upper panel) and total stellar mass (lower panel) of the host galaxies. The black dashed line shows the best fit to the data (bisector fit).
The dot in the lower left corner shows the typical error bar in the data.}
\label{scaling2Lup3}%
\end{figure}

Figure \ref{scaling2Lup3} shows the relation between $L(H\alpha)_{3}/L(H\alpha)$, the contribution of the three brightest
{\hii} region to the total H$\alpha$ luminosity of the galaxy, and the integrated 
star formation rate and stellar mass of the host. 
Figure \ref{scaling2Lup3} indicates that the contribution of the three brightest 
{\hii} regions to the total H$\alpha$ emission of galaxies strongly increases from $\simeq$ 1-10\%\ in star forming ($SFR$ $\simeq$ 1 M$_{\odot}$ yr$^{-1}$)
and massive ($M_{star}$ $\simeq$ 10$^{10}$ M$_{\odot}$) objects to $\simeq$ 50-100\%\ in dwarf systems ($SFR$ $\simeq$ 10$^{-3}$ M$_{\odot}$ yr$^{-1}$;
$M_{star}$ $\simeq$ 10$^{7}$ M$_{\odot}$).

\begin{figure}[h!]
\centering
\includegraphics[width=0.44\textwidth]{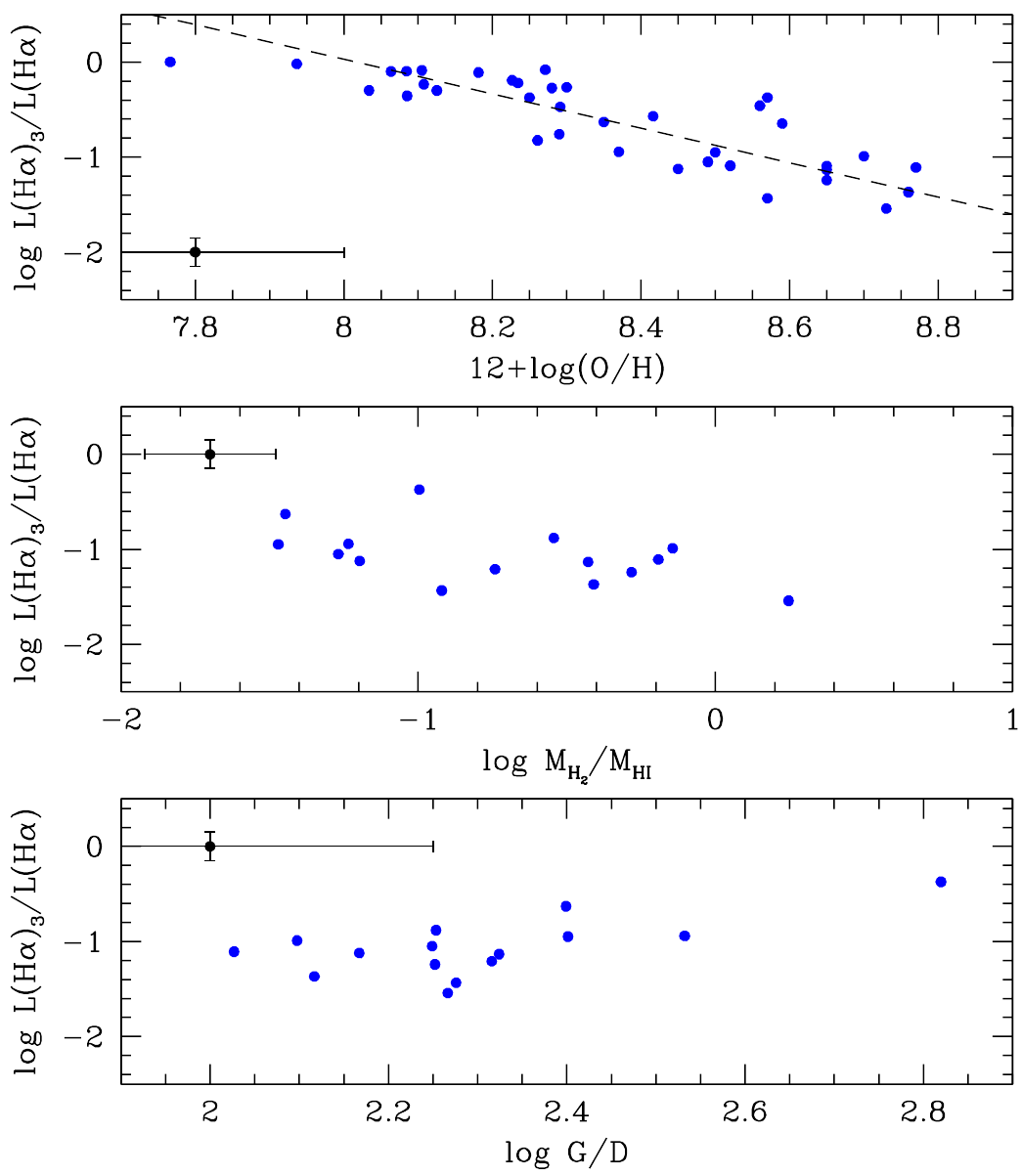}\\
\caption{Relation between $L(H\alpha)_{3}/L(H\alpha)$ (contribution of the brightest three {\hii} regions to the total H$\alpha$ luminosity of the galaxy)
and the total star formation rate  
(upper panel) and total stellar mass (lower panel) of the host galaxies. The black dashed line shows the best fit to the data (bisector fit), while the black dot the typical error in the data.}
\label{scaling1Lup3physical}%
\end{figure}

Figure \ref{scaling1Lup3physical} shows the relation $L(H\alpha)_{3}/L(H\alpha)$ and the metallicity, the molecular-to-atomic gas ratio, and the
total gas-to-dust ratio. The poor statistic (15 objects) prevents us to see whether the two variables are related to $M_{H_2}/M_{HI}$ and $M_{gas}/M_{dust}$,
but clearly show an anti-correlation with the gas metallicity.

\twocolumn

\onecolumn


\section{Relation between the star formation rate and stellar mass surface densities}
\label{appD}

\begin{figure}[h!]
\centering
\includegraphics[width=0.49\textwidth]{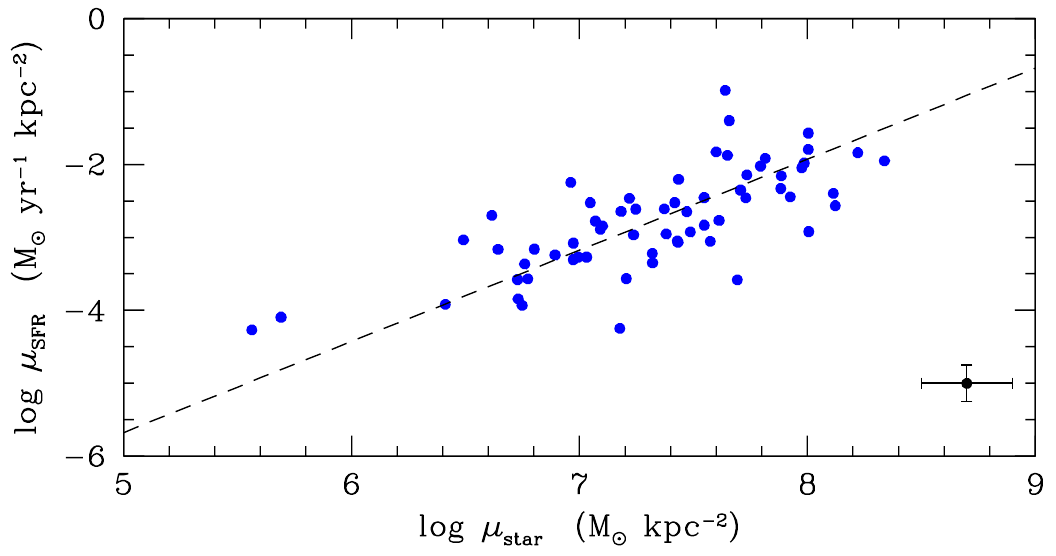}\\
\caption{Relation between the mean star formation rate surface density and the mean stellar mass surface density. 
The black dashed line shows the best fit to the data (bisector fit). The dot in the lower right corner shows the typical error bar in the data.}
\label{sfrmstarsurf}%
\end{figure}

Two of the scaling parameters used in Sec. 4.2 to characterise the typical properties of the selected galaxies are the mean stellar mass and star formation rates 
surface densities as defined in eq. 6 and 5, respectively. These two quantities are tightly related, as shown in Fig. \ref{sfrmstarsurf}. The bisector fit of the relation
is: log $\mu_{SFR}$ \rm{[M$_{\odot}$ yr$^{-1}$ kpc$^{-2}$]} = 1.25$\pm$0.08 $\times$ log $\mu_{star}$ \rm{[M$_{\odot}$ kpc$^{-2}$]} -11.92$\pm$0.50 ($\rho$=0.74, $\sigma$=0.32).
The strong relation between these two variables suggests that the star formation process is tightly connected to the stellar gravity of the disc (Shi et al. 2018).
This relation is expected given that the two variables $M_{star}$ and $L(H\alpha)$ are strongly correlated (main sequence relation, see Boselli et al. (2023a)
for details on this relation in the VESTIGE sample). While the main sequence relation is an obvious scaling relation where "bigger galaxies have more of everything" (Kennicutt 1990),
the one using normalised entities is more indicated to compare objects of different size and nature.

\section{Luminosity function on individual objects}
\label{appE}

\begin{figure*}[h!]
\centering
\includegraphics[width=0.99\textwidth]{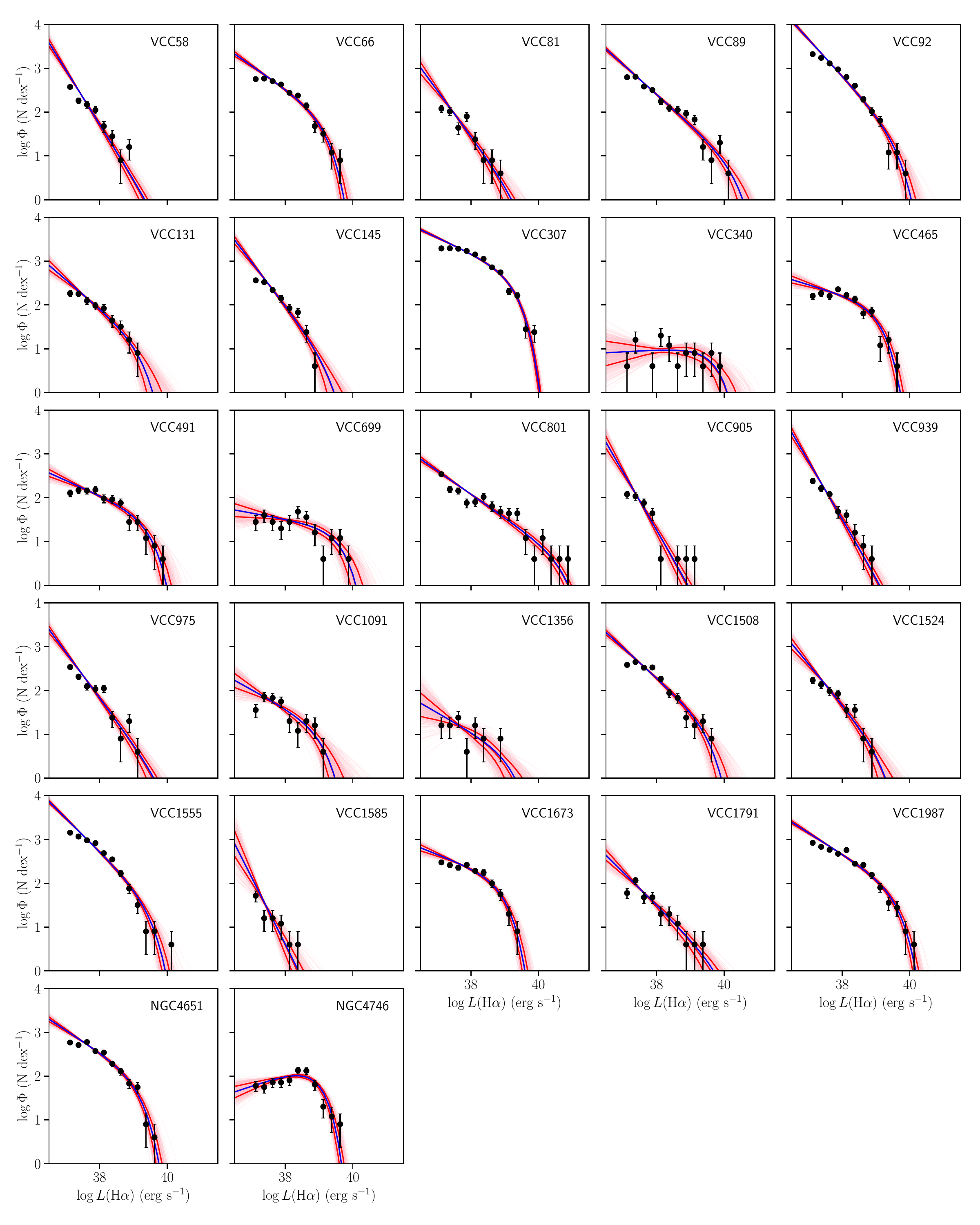}\\
\caption{Luminosity function of the {\hii} regions detected by \textsc{HIIphot} in individual galaxies. The H$\alpha$ luminosity of 
individual {\hii} regions is corrected for dust attenuation and [{\nii}] contamination as described in Sec. 3.2. The solid  blue and red lines
indicate the best fit and 1$\sigma$ confidence regions for the Schechter luminosity function parametrisation. Black solid dots indicate the number of {\hii} regions
in 0.25 dex bins of H$\alpha$ luminosity above the adopted completeness of the survey ($L(H\alpha)$ $\geq$ 10$^{37}$ erg s$^{-1}$).}
\label{LFindividual}%
\end{figure*}

Figure \ref{LFindividual} shows the luminosity function of {\hii} regions derived for
the 27 galaxies of the sample with at least 20 {\hii} regions detected by \textsc{HIIphot}
above the completeness limit of the survey ($L(H\alpha)$ $\geq$ 10$^{37}$ erg s$^{-1}$).

\end{appendix}
\end{document}